\documentclass[12pt,fleqn,a4paper]{article} 
\usepackage{epsfig,graphics,graphicx}
\usepackage{clshan-math,clshan-dm-dd}
\newcommand{\imageswitch} [2] {#2}
\def \lsim {\:\raisebox{-0.7ex}{$\stackrel{\textstyle<}{\sim}$}\:}
\def \gsim {\:\raisebox{-0.7ex}{$\stackrel{\textstyle>}{\sim}$}\:}
\begin{document}
\thispagestyle{empty}
\begin{flushright}
 March 2010
\end{flushright}
\begin{center}
{\large\bf
 Effects of Residue Background Events
 in Direct Dark Matter \\ \vspace{-0.04cm} Detection Experiments on
 the Reconstruction of \\ \vspace{0.1cm} the Velocity Distribution Function of Halo WIMPs} \\
\vspace*{0.7cm}
 {\sc Chung-Lin Shan} \\
\vspace*{0.5cm}
 {\it Department of Physics, National Cheng Kung University \\
      No.~1, University Road,
      Tainan City 70101, Taiwan, R.O.C.}                    \\~\\
%
 {\it Physics Division,
      National Center for Theoretical Sciences              \\
      No.~101, Sec.~2, Kuang-Fu Road,
      Hsinchu City 30013, Taiwan, R.O.C.}                   \\~\\
%
 {\it E-mail:} {\tt clshan@mail.ncku.edu.tw}                \\
\end{center}
\vspace{1cm}
\begin{abstract}
 In our earlier work on the development of
 a model--independent data analysis method
 for reconstructing the (moments of the)
 time--averaged one--dimensional
 velocity distribution function of
 Weakly Interacting Massive Particles (WIMPs)
 by using measured recoil energies
 from direct Dark Matter detection experiments directly,
 it was assumed that
 the analyzed data sets are background--free,
 i.e., all events are WIMP signals.
 In this article,
 as a more realistic study,
 we take into account
 a fraction of possible residue background events,
 which pass all discrimination criteria and
 then mix with other real WIMP--induced events
 in our data sets.
 Our simulations show that,
 for the reconstruction of the one--dimensional
 WIMP velocity distribution,
 the maximal acceptable fraction of residue background events
 in the analyzed data set(s)
 of ${\cal O}(500)$ total events
 is $\sim$ 10\% -- 20\%.
 For a WIMP mass of 50 GeV
 with a negligible uncertainty
 and 20\% residue background events,
 the deviation of the reconstructed velocity distribution
 would in principle be $\sim$ 7.5\%
 with a statistical uncertainty of $\sim$ 18\%
 ($\sim$ 19\% for a background--free data set).
\end{abstract}
\clearpage
\section{Introduction}
 Currently,
 direct Dark Matter detection experiments
 searching for Weakly Interacting Massive Particles (WIMPs)
 are one of the promising methods
 for understanding the nature of Dark Matter
 and identifying them among new particles produced at colliders
 as well as reconstructing the (sub)structure of our Galactic halo
 \cite{Smith90, Lewin96, SUSYDM96, Bertone05}.
 In our earlier work
 \cite{DMDDf1v},
 we developed methods
 for reconstructing the (moments of the)
 time--averaged one--dimensional velocity distribution of halo WIMPs
 by using (a functional form of) the recoil spectrum
 as well as the measured recoil energies directly.
 This analysis requires
 {\em no} prior knowledge about
 the WIMP density near the Earth
 {\em nor} about their scattering cross section on nucleus,
 the unique required information
 is the mass of incident WIMPs.
 We therefore turned to also develop the method for
 determining the WIMP mass model--independently
 by combining two experimental data sets
 with two different target nuclei
 \cite{DMDDmchi-SUSY07, DMDDmchi}.

 In the work on the development of
 these model--independent data analysis procedures
 by using measured recoil energies
 from direct detection experiments directly,
 it was assumed that
 the analyzed data sets are background--free,
 i.e., all events are WIMP signals.
 Active background discrimination techniques
 should make this condition possible.
 For example,
 the ratio of the ionization to recoil energy,
 the so--called ``ionization yield'',
 used in the CDMS-II experiment
 provides an event--by--event rejection
 of electron recoil events
 to be better than $10^{-4}$ misidentification
 \cite{Ahmed09b}.
 By combining the ``phonon pulse timing parameter'',
 the rejection ability of
 the misidentified electron recoils
 (most of them are ``surface events''
  with sufficiently reduced ionization energies)
 can be improved to be $< 10^{-6}$ for electron recoils
 \cite{Ahmed09b}.
 Moreover,
 as demonstrated by the CRESST collaboration
 \cite{CRESST-bg}, 
 by means of inserting a scintillating foil,
 which causes some additional scintillation light
 for events induced by $\alpha$-decay of $\rmXA{Po}{210}$
 and thus shifts the pulse shapes of these events
 faster than pulses induced by WIMP interactions in the crystal,
 the pulse shape discrimination (PSD) technique
 can then easily distinguish WIMP--induced nuclear recoils
 from those induced by backgrounds%
\footnote{
 For more details
 about background discrimination techniques and status
 in currently running and projected direct detection experiments,
 see e.g.,
 Refs.~\cite{Aprile09a,
             EDELWEISS-bg, 
             Lang09b} 
}.

 However,
 as the most important issue in all underground experiments,
 the signal identification ability and
 possible residue background events
 which pass all discrimination criteria and
 then mix with other real WIMP--induced events in our data sets
 should also be considered.
 Therefore,
 in this article,
 as a more realistic study,
 we follow our first work
 on the effects of residue background events
 on the determination of the WIMP mass
 \cite{DMDDbg-mchi}
 and want to study
 how well we could reconstruct
 the WIMP velocity distribution model--independently
 by using ``impure'' data sets
 and how ``dirty'' these data sets could be
 to be still useful.

 The remainder of this article is organized as follows.
 In Sec.~2
 I review the model--independent method
 for reconstructing the time--averaged
 one--dimensional velocity distribution function of halo WIMPs
 by using data from direct detection experiments directly.
 In Sec.~3
 the effects of residue background events
 in the analyzed data sets
 on the measured energy spectrum
 as well as on the reconstructed WIMP mass
 will briefly be discussed.
 In Sec.~4
 I show numerical results of
 the reconstructed WIMP velocity distribution
 by using mixed data sets
 with different fractions of residue background events
 based on Monte Carlo simulations.
 I conclude in Sec.~5.
 Some technical details will be given in an appendix.
\section{Methods for reconstructing
         the one--dimensional WIMP velocity distribution function}
 In this section
 I review briefly the methods for reconstructing
 the one--dimensional WIMP velocity distribution function
 from the recoil spectrum as well as
 from experimental data directly.
 Detailed derivations and discussions
 can be found in Refs.~\cite{DMDDf1v, PhD}.
\subsection{From the recoil spectrum}
 The basic expression for the differential event rate
 for elastic WIMP--nucleus scattering is given by \cite{SUSYDM96}:
\beq
   \dRdQ
 = \calA \FQ \int_{\vmin}^{\vmax} \bfrac{f_1(v)}{v} dv
\~.
\label{eqn:dRdQ}
\eeq
 Here $R$ is the direct detection event rate,
 i.e., the number of events
 per unit time and unit mass of detector material,
 $Q$ is the energy deposited in the detector,
 $F(Q)$ is the elastic nuclear form factor,
 $f_1(v)$ is the one--dimensional velocity distribution function
 of the WIMPs impinging on the detector,
 $v$ is the absolute value of the WIMP velocity
 in the laboratory frame.
 The constant coefficient $\calA$ is defined as
\beq
        \calA
 \equiv \frac{\rho_0 \sigma_0}{2 \mchi \mrN^2}
\~,
\label{eqn:calA}
\eeq
 where $\rho_0$ is the WIMP density near the Earth
 and $\sigma_0$ is the total cross section
 ignoring the form factor suppression.
 The reduced mass $\mrN$ is defined by
\beq
        \mrN
 \equiv \frac{\mchi \mN}{\mchi + \mN}
\~,
\label{eqn:mrN}
\eeq
 where $\mchi$ is the WIMP mass and
 $\mN$ that of the target nucleus.
 Finally,
 $\vmin$ is the minimal incoming velocity of incident WIMPs
 that can deposit the energy $Q$ in the detector:
\beq
   \vmin
 = \alpha \sqrt{Q}
\label{eqn:vmin}
\eeq
 with the transformation constant
\beq
        \alpha
 \equiv \sfrac{\mN}{2 \mrN^2}
\~,
\label{eqn:alpha}
\eeq
 and $\vmax$ is the maximal WIMP velocity
 in the Earth's reference frame,
 which is related to
 the escape velocity from our Galaxy
 at the position of the Solar system,
 $\vesc~\gsim~600$ km/s.

 In our earlier work \cite{DMDDf1v},
 it was found that,
 by using a time--averaged recoil spectrum $dR / dQ$
 and assuming that no directional information exists,
 the normalized one--dimensional velocity distribution function
 of incident WIMPs, $f_1(v)$, can be solved
 from Eq.~(\ref{eqn:dRdQ}) directly as
\beq
   f_1(v)
 = \calN
   \cbrac{ -2 Q \cdot \dd{Q} \bbrac{ \frac{1}{\FQ} \aDd{R}{Q} } }\Qva
\~,
\label{eqn:f1v_dRdQ}
\eeq 
 where the normalization constant $\calN$ is given by
\beq
   \calN
 = \frac{2}{\alpha}
   \cbrac{\intz \frac{1}{\sqrt{Q}}
                \bbrac{ \frac{1}{\FQ} \aDd{R}{Q} } dQ}^{-1}
\~.
\label{eqn:calN_int}
\eeq
 Here the integral goes over
 the entire physically allowed range of recoil energies:
 starting at $Q = 0$,
 and the upper limit of the integral has been written as $\infty$.
 Note that,
 because $f_1(v)$ in Eq.~(\ref{eqn:f1v_dRdQ})
 is the normalized velocity distribution,
 the normalization constant $\cal N$ here is independent of
 the constant coefficient $\cal A$ defined in Eq.~(\ref{eqn:calA}).
 Moreover,
 as the most important consequence,
 the velocity distribution function of halo WIMPs
 reconstructed by Eq.~(\ref{eqn:f1v_dRdQ})
 is independent of the local WIMP density $\rho_0$
 as well as of the WIMP--nucleus cross section $\sigma_0$.
 However,
 as we will see later,
 not only the overall normalization constant $\calN$
 given in Eq.~(\ref{eqn:calN_int}),
 but also the shape of the velocity distribution,
 through the transformation $Q = v^2 / \alpha^2$ in Eq.~(\ref{eqn:f1v_dRdQ}),
 depends on the WIMP mass $\mchi$
 involved in the coefficient $\alpha$ defined in Eq.~(\ref{eqn:alpha}).
\subsection{From experimental data directly}
 In order to use the expressions
 (\ref{eqn:f1v_dRdQ}) and (\ref{eqn:calN_int})
 for reconstructing $f_1(v)$,
 one needs a functional form for the recoil spectrum $dR / dQ$.
 In practice
 this requires usually a fit to experimental data.
 However,
 data fitting will re--introduce some model dependence
 and make the error analysis more complicated.
 Hence,
 expressions that allow to reconstruct $f_1(v)$
 directly from experimental data
 (i.e., measured recoil energies)
 have also been developed \cite{DMDDf1v}.
 We started by considering experimental data described by
\beq
     {\T Q_n - \frac{b_n}{2}}
 \le \Qni
 \le {\T Q_n + \frac{b_n}{2}}
\~,
     ~~~~~~~~~~~~ 
     i
 =   1,~2,~\cdots,~N_n,~
     n
 =   1,~2,~\cdots,~B.
\label{eqn:Qni}
\eeq
 Here the entire experimental possible
 energy range between $\Qmin$ and $\Qmax$
 has been divided into $B$ bins
 with central points $Q_n$ and widths $b_n$.
 In each bin,
 $N_n$ events will be recorded.

 As argued in Ref.~\cite{DMDDf1v},
 the statistical uncertainty on
 the ``slope of the recoil spectrum'',
\beqN
 \bbrac{\dd{Q} \aDd{R}{Q}}_{Q = Q_n}
\~,
\eeqN
 appearing in the expression (\ref{eqn:f1v_dRdQ}),
 scales like the bin width to the power $-1.5$.
 In addition,
 the wider the bin width,
 the more the recorded events in this bin,
 and thus the smaller the statistical uncertainty
 on the estimator of $\bbrac{d/dQ \~ (dR / dQ)}_{Q = Q_n}$.
 Hence,
 since the recoil spectrum $dR / dQ$ is expected
 to be approximately exponential,
 in order to approximate the spectrum
 in a rather wider range,
 instead of the conventional standard linear approximation,
 the following {\em exponential} ansatz
 for the {\em measured} recoil spectrum
 ({\em before} normalized by the exposure $\calE$)
 in the $n$th bin has been introduced \cite{DMDDf1v}:
\beq
        \adRdQ_{{\rm expt}, \~ n}
 \equiv \adRdQ_{{\rm expt}, \~ Q \simeq Q_n}
 \equiv \rn  \~ e^{k_n (Q - Q_{s, n})}
\~.
\label{eqn:dRdQn}
\eeq
 Here $r_n$ is the standard estimator
 for $(dR / dQ)_{\rm expt}$ at $Q = Q_n$:
\beq
   r_n
 = \frac{N_n}{b_n}
\~,
\label{eqn:rn}
\eeq
 $k_n$ is the logarithmic slope of
 the recoil spectrum in the $n$th $Q-$bin,
 which can be computed numerically
 from the average value of the measured recoil energies
 in this bin:
\beq
   \bQn
 = \afrac{b_n}{2} \coth\afrac{k_n b_n}{2}-\frac{1}{k_n}
\~,
\label{eqn:bQn}
\eeq
 where
\beq
        \bQxn{\lambda}
 \equiv \frac{1}{N_n} \sumiNn \abrac{\Qni - Q_n}^{\lambda}
\~.
\label{eqn:bQn_lambda}
\eeq
 Then the shifted point $Q_{s, n}$
 in the ansatz (\ref{eqn:dRdQn}),
 at which the leading systematic error
 due to the ansatz
 is minimal \cite{DMDDf1v},
 can be estimated by
\beq
   Q_{s, n}
 = Q_n + \frac{1}{k_n} \ln\bfrac{\sinh(k_n b_n/2)}{k_n b_n/2}
\~.
\label{eqn:Qsn}
\eeq
 Note that $Q_{s, n}$ differs from the central point of the $n$th bin, $Q_n$.

 Now,
 substituting the ansatz (\ref{eqn:dRdQn})
 into Eq.~(\ref{eqn:f1v_dRdQ})
 and then letting $Q = Q_{s, n}$,
 we can obtain that
\beq
   f_{1, {\rm rec}}(v_{s, n})
 = \calN
   \bBigg{\frac{2 Q_{s, n} r_n}{F^2(Q_{s, n})}}
   \bbrac{\dd{Q} \ln \FQ \bigg|_{Q = Q_{s, n}} - k_n}
\~.
\label{eqn:f1v_Qsn}
\eeq
 Here
\beq
   v_{s, n}
 = \alpha \sqrt{Q_{s, n}}
\~,
\label{eqn:vsn}
\eeq
 and the normalization constant $\calN$
 given in Eq.~(\ref{eqn:calN_int})
 can be estimated directly from the data:
\beq
   \calN
 = \frac{2}{\alpha}
   \bbrac{\sum_{a} \frac{1}{\sqrt{Q_a} \~ F^2(Q_a)}}^{-1}
\~,
\label{eqn:calN_sum}
\eeq
 where the sum runs over all events in the sample.
\subsection{Windowing the data set}
 As mentioned above,
 the statistical uncertainty on
 the slope of the recoil spectrum
 around the central point $Q_n$,
 $\bbrac{d/dQ \~ (dR / dQ)}_{Q \simeq Q_n}$,
 is approximately proportional to $b_n^{-1.5}$.
 Thus,
 in order to reduce the statistical uncertainty
 on the reconstructed velocity distribution function
 by Eq.~(\ref{eqn:f1v_Qsn}),
 it seems to be better
 to use large bin width.
 However,
 neither the conventional linear approximation:
\beq
   \adRdQ_{{\rm expt}, \~ Q = Q_n}
 = \frac{N_n}{b_n}
\eeq
 nor the exponential ansatz
 given in Eq.~(\ref{eqn:dRdQn})
 can describe the real (but as yet unknown)
 recoil spectrum exactly.
 The neglected terms of higher powers of $Q - Q_n$
 could therefore induce some uncontrolled systematic errors
 which increase with increasing bin width.
 Moreover,
 since the number of bins scales inversely with their size,
 by using large bins we would be able to estimate $f_1(v)$
 only at a small number of velocities.
 Additionally,
 once a quite large bin width is used,
 it would correspondingly lead to a quite large value
 of the first reconstructible point of $f_1(v)$,
 i.e., $f_{1, {\rm rec}}(v_{s, 1})$,
 since the central point $Q_1$
 as well as the shifted point $Q_{s, 1}$
 of the first bin would be quite large.
 Finally,
 choosing a fixed bin size,
 as one conventionally does,
 would let errors on the estimated logarithmic slopes,
 and hence also on the estimates of $f_1(v)$,
 increase quickly with increasing $Q$ or $v$.
 This is due to the essentially exponential form
 of the expected recoil spectrum,
 which would lead to a quickly falling number of events
 in equal--sized bins.
 By some trial--and--error analyses
 it was found that
 the errors are roughly equal in all bins
 if the bin widths increase linearly \cite{DMDDf1v}.

 Therefore,
 it has been introduced in Ref.~\cite{DMDDf1v} that
 one can first collect experimental data
 in relatively small bins
 and then combining varying numbers of bins
 into overlapping ``windows''.
 In particular,
 the first window would be identical with the first bin.
 One starts by binning the data,
 as in Eq.~(\ref{eqn:Qni}),
 where the bin widths satisfy
\beq
   b_n
 = b_1 + (n - 1) \delta
\~,
\label{eqn:bn_delta}
\eeq
 i.e.,
\beq
   Q_n
 = \Qmin + \abrac{n - \frac{1}{2}} b_1 + \bfrac{(n - 1)^2}{2} \delta
\~.
\label{eqn:Qn_delta}
\eeq
 Here the increment $\delta$ satisfies
\beq
   \delta
 = \frac{2}{B (B - 1)} \aBig{\Qmax - \Qmin - B b_1}
\~,
\label{eqn:delta_B}
\eeq
 $B$ being the total number of bins,
 and $Q_{\rm (min, max)}$ are
 the experimental minimal and maximal cut--off energies.
 Assume up to $n_W$ bins are collected into a window,
 with smaller windows at the borders of the range of $Q$.

 In order to distinguish the numbers of bins and windows,
 hereafter Latin indices $n,~m,~\cdots$ are used to label bins,
 and Greek indices $\mu,~\nu,~\cdots$ to label windows.
 For $1 \leq \mu \leq n_W$,
 the $\mu$th window simply consists of the first $\mu$ bins;
 for $n_W \leq \mu \leq B$,
 the $\mu$th window consists of bins
 $\mu-n_W + 1,~\mu-n_W + 2,~\cdots,~\mu$;
 and for $B \leq \mu \leq B+n_W-1$,
 the $\mu$th window consists of the last $n_W - (\mu - B)$ bins.
 This can also be described by introducing
 the indices $n_{\mu-}$ and $n_{\mu+}$
 which label the first and last bin
 contributing to the $\mu$th window,
 with
\cheqna
\beq
\renewcommand{\arraystretch}{1.3}
   n_{\mu-}
 = \cleft{\begin{array}{l c l}
           1,         & ~~~~~~ & {\rm for}~\mu \leq n_W, \\
           \mu-n_W+1, &        & {\rm for}~\mu \geq n_W,
          \end{array}}
\label{eqn:n_mu_minus}
\eeq
 and
\cheqnb
\beq
\renewcommand{\arraystretch}{1.3}
   n_{\mu+}
 = \cleft{\begin{array}{l c l}
           \mu, & ~~~~~~ & {\rm for}~\mu \leq B, \\
           B,   &        & {\rm for}~\mu \geq B.
          \end{array}}
\label{eqn:n_mu_plus}
\eeq
\cheqn
 The total number of windows
 defined through Eqs.~(\ref{eqn:n_mu_minus}) and (\ref{eqn:n_mu_plus})
 is evidently $W = B + n_W - 1$,
 i.e., $1 \leq \mu \leq B + n_W - 1$.

 As shown in the previous subsection,
 the basic observables needed
 for the reconstruction of $f_1(v)$ by Eq.~(\ref{eqn:f1v_Qsn})
 are the number of events in the $n$th $Q-$bin, $N_n$,
 as well as the average value of the measured recoil energies
 in this bin, $\bQn$.
 For a ``windowed'' data set,
 one can easily calculate
 the number of events per window as
\beq
   N_{\mu}
 = \sum_{n = n_{\mu-}}^{n_{\mu+}} N_n
\~,
\label{eqn:N_mu}
\eeq
 as well as the average value of
 the measured recoil energies
\beq
   \Bar{Q - Q_{\mu}}|_{\mu}
 = \frac{1}{N_{\mu}}
   \abrac{\sum_{n = n_{\mu-}}^{n_{\mu+}} N_n \Bar{Q}|_{n}} - Q_{\mu}
\~,
\label{eqn:wQ_mu}
\eeq
 where $Q_{\mu}$ is the central point of the $\mu$th window.
 The exponential ansatz in Eq.~(\ref{eqn:dRdQn})
 is now assumed to hold over an entire window.
 We can then estimate the prefactor as
\beq
   r_{\mu}
 = \frac{N_{\mu}}{w_{\mu}}
\~,
\label{eqn:r_mu}
\eeq
 $w_{\mu}$ being the width of the $\mu$th window.
 The logarithmic slope of the recoil spectrum
 in the $\mu$th window, $k_{\mu}$,
 as well as the shifted point $Q_{s, \mu}$
 (from the central point of each ``window'', $Q_{\mu}$)
 can be calculated as in Eqs.~(\ref{eqn:bQn}) and (\ref{eqn:Qsn})
 with ``bin'' quantities replaced by ``window'' quantities.
 Note that,
 due to the combination of bins
 into overlapping windows,
 these quantities are all correlated (for $n_W \neq 1$).
 The expressions for estimating the covariance matrices
 are given in the appendix.

 Finally,
 the covariance matrix of
 the estimates of $f_1(v)$
 at adjacent values of $v_{s, \mu} = \alpha \sqrt{Q_{s, \mu}}$
 is given by
\beqn
 \conti {\rm cov}\aBig{f_{1, {\rm rec}}(v_{s, \mu}), f_{1, {\rm rec}}(v_{s, \nu})}
        \non\\
 \=     \bfrac{f_{1, {\rm rec}}(v_{s, \mu}) f_{1, {\rm rec}}(v_{s, \nu})}
              {r_{\mu} r_{\nu}}
        {\rm cov}\abrac{r_{\mu}, r_{\nu}}
       +\abrac{2 \calN}^2
        \bfrac{Q_{s, \mu} Q_{s, \nu} r_{\mu} r_{\nu}}{F^2(Q_{s, \mu}) F^2(Q_{s, \nu})}
        {\rm cov}\abrac{k_{\mu}, k_{\nu}}
        \non\\
\conti ~~~~~~~~~~~~ 
       -\calN
        \cbrac{ \bfrac{f_{1, {\rm rec}}(v_{s, \mu})}{r_{\mu}}
                \bfrac{2 Q_{s, \nu} r_{\nu} }{F^2(Q_{s, \nu})}
                {\rm cov}\abrac{r_{\mu}, k_{\nu}}
               +\aBig{\mu \lgetsto \nu}}
\~.
\label{eqn:cov_f1v_Qs_mu}
\eeqn
 Note that
 Eq.~(\ref{eqn:cov_f1v_Qs_mu}) should in principle
 also include contributions involving
 the statistical error on the estimator for $\calN$
 in Eq.~(\ref{eqn:calN_sum}).
 However,
 this error and its correlations
 with the errors on the $r_{\mu}$ and $k_{\mu}$
 have been found to be negligible
 compared to the errors included in Eq.~(\ref{eqn:cov_f1v_Qs_mu})
 \cite{DMDDf1v}.
\section{Effects of residue background events}
 In this section
 I first show some numerical results of
 the energy spectrum of WIMP recoil signals
 mixed with a few background events.
 Then
 I review the effects of residue background events
 in the analyzed data sets
 on the reconstruction of the WIMP mass $\mchi$.

 For generating WIMP--induced signals,
 we use the shifted Maxwellian velocity distribution
 \cite{Lewin96, SUSYDM96, DMDDf1v}:
\beq
   f_{1, \sh}(v)
 = \frac{1}{\sqrt{\pi}} \afrac{v}{\ve v_0}
   \bbigg{ e^{-(v - \ve)^2 / v_0^2} - e^{-(v + \ve)^2 / v_0^2} }
\~.
\label{eqn:f1v_sh}
\eeq
 Here $v_0 \simeq 220~{\rm km/s}$
 is the orbital velocity of the Sun
 in the Galactic frame,
 and $\ve$ is the Earth's velocity in the Galactic frame
 \cite{Freese88, SUSYDM96, Bertone05}%
\footnote{
 The time dependence of the Earth's velocity
 in the Galactic frame,
 the second term of $\ve(t)$,
 will be ignored in our simulations,
 i.e., $\ve = 1.05 \~ v_0$ will be used.
}:
\beq
   v_{\rm e}(t)
 = v_0 \bbrac{1.05 + 0.07 \cos\afrac{2 \pi (t - t_{\rm p})}{1~{\rm yr}}}
\~;
\label{eqn:ve}
\eeq
 with $t_{\rm p} \simeq$ June 2nd is the date
 on which the velocity of the Earth
 relative to the WIMP halo is maximal.
 As a maximal cut--off
 of the velocity distribution function,
 the escape velocity has been set as $\vesc = 700$ km/s.
 The Woods--Saxon elastic nuclear form factor
 \cite{Engel91, SUSYDM96, Bertone05}
 for the SI WIMP--nucleus cross section
 will also be used%
\footnote{
 Other commonly used analytic forms
 for the one--dimensional WIMP velocity distribution
 as well as for the elastic nuclear form factor
 for the SI WIMP--nucleus cross section
 can be found in Refs.~\cite{DMDDf1v, DMDDmchi-NJP}.
}.

 Meanwhile,
 we use the target--dependent exponential form
 introduced in Ref.~\cite{DMDDbg-mchi}
 for the residue background spectrum:
\beq
   \aDd{R}{Q}_{\rm bg, ex}
 = \exp\abrac{-\frac{Q /{\rm keV}}{A^{0.6}}}
\~.
\label{eqn:dRdQ_bg_ex}
\eeq
 Here $Q$ is the recoil energy,
 $A$ is the atomic mass number of the target nucleus.
 The power index of $A$, 0.6, is an empirical constant,
 which has been chosen so that
 the exponential background spectrum is
 somehow similar to,
 but still different from
 the expected recoil spectrum of the target nuclei;
 otherwise,
 there is in practice no difference between
 the WIMP scattering and background spectra.
 Note that,
 among different possible choices,
 we use in our simulations the atomic mass number $A$
 as the simplest, unique characteristic parameter
 in the general analytic form (\ref{eqn:dRdQ_bg_ex})
 for defining the residue background spectrum
 for different target nuclei.
 However,
 it does not mean that
 the (superposition of the real) background spectra
 would depend simply/primarily on $A$ or
 on the mass of the target nucleus, $\mN$.
 In other words,
 it is practically equivalent to
 use expression (\ref{eqn:dRdQ_bg_ex})
 or $(dR / dQ)_{\rm bg, ex} = e^{-Q / 13.5~{\rm keV}}$ directly
 for a $\rmXA{Ge}{76}$ target.

 Note also that,
 firstly,
 as argued in Ref.~\cite{DMDDbg-mchi},
 the exponential form of background spectrum
 is rather naive;
 but,
 since we consider here
 only {\em a few (tens) residue} background events
 induced by perhaps {\em two or more} different sources,
 pass all discrimination criteria,
 and then mix with other WIMP--induced events
 in our data sets of a few hundreds {\em total} events,
 exact forms of different background spectra
 are actually not very important and
 this exponential form should practically not be unrealistic%
\footnote{
 Other (more realistic) forms for background spectrum
 (perhaps also for some specified targets/experiments)
 can be tested on the \amidas\ website
 \cite{AMIDAS-web, AMIDAS-eprints}. 
}.
 Secondly,
 as demonstrated in Ref.~\cite{DMDDf1v}
 and reviewed in the previous section,
 the model--independent data analysis procedure
 for reconstructing the WIMP velocity distribution function
 requires only measured recoil energies
 (induced mostly by WIMPs and
  occasionally by background sources)
 from direct detection experiments.
 Therefore,
 for applying this method to future real data,
 a prior knowledge about (different) background source(s)
 is {\em not required at all}.

 Moreover,
 for our numerical simulations
 presented here as well as in the next section,
 the actual numbers of signal and background events
 in each simulated experiment
 are Poisson--distributed around their expectation values
 {\em independently};
 and the total event number recorded in one experiment
 is then the sum of these two numbers.
 Additionally,
 we assumed that
 all experimental systematic uncertainties
 as well as the uncertainty on
 the measurement of the recoil energy
 could be ignored.
 The energy resolution of most existing detectors
 is so good that its error can be neglected
 compared to the statistical uncertainty
 for the foreseeable future
 with pretty few events.
\subsection{On the measured energy spectrum}
\begin{figure}[p!]
\begin{center}
\vspace{-0.75cm}
\imageswitch{
\begin{picture}(16.5,20.5)
\put(0  ,14  ){\framebox(8,6.5){dRdQ-bg-ex-Ge-000-100-20-010}}
\put(8.5,14  ){\framebox(8,6.5){dRdQ-bg-ex-Ge-000-100-20-025}}
\put(0  , 7  ){\framebox(8,6.5){dRdQ-bg-ex-Ge-000-100-20-050}}
\put(8.5, 7  ){\framebox(8,6.5){dRdQ-bg-ex-Ge-000-100-20-100}}
\put(0  , 0  ){\framebox(8,6.5){dRdQ-bg-ex-Ge-000-100-20-250}}
\put(8.5, 0  ){\framebox(8,6.5){dRdQ-bg-ex-Ge-000-100-20-500}}
\end{picture}}
{\hspace*{-1.6cm}
 \includegraphics[width=9.8cm]{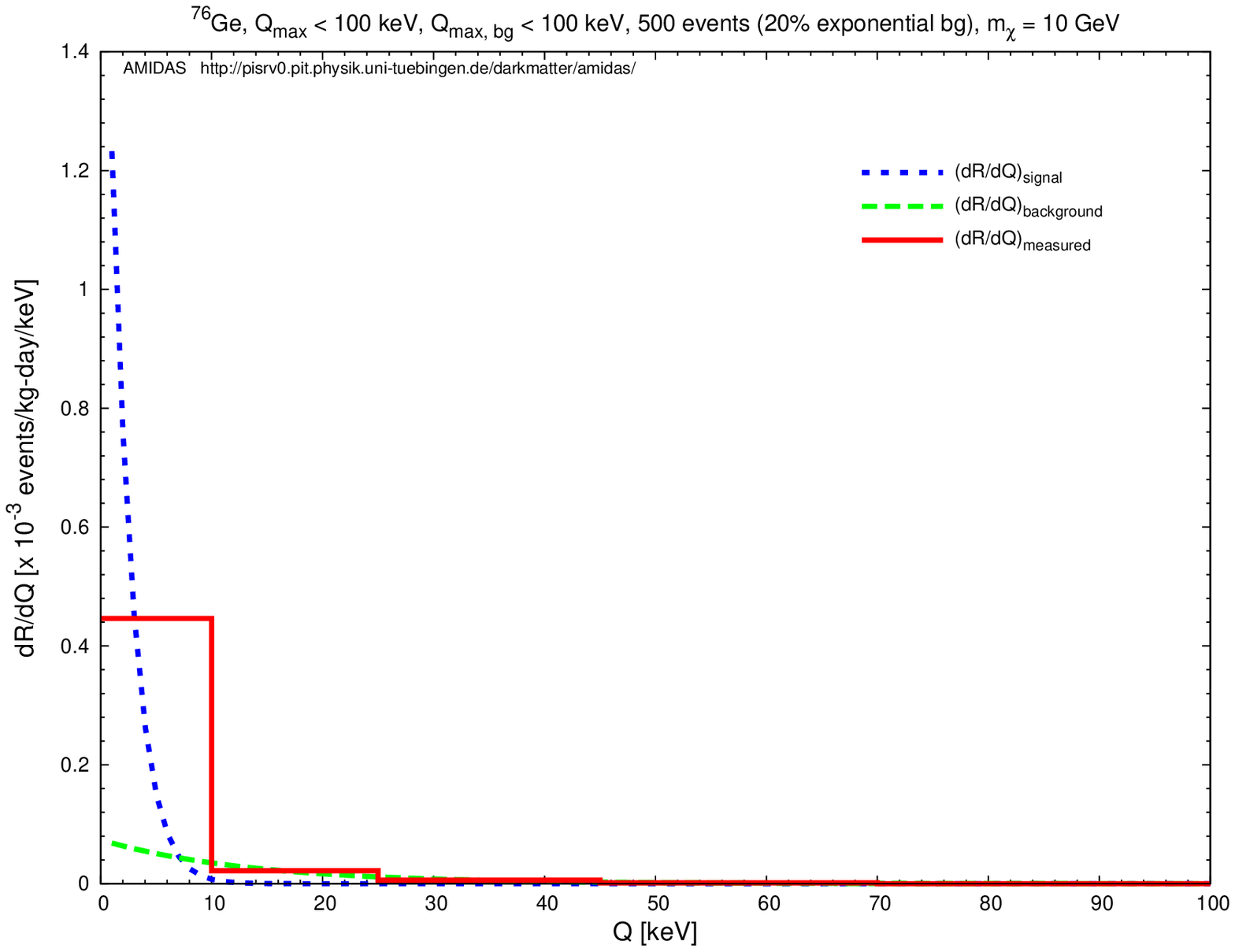} \hspace{-1.1cm}
 \includegraphics[width=9.8cm]{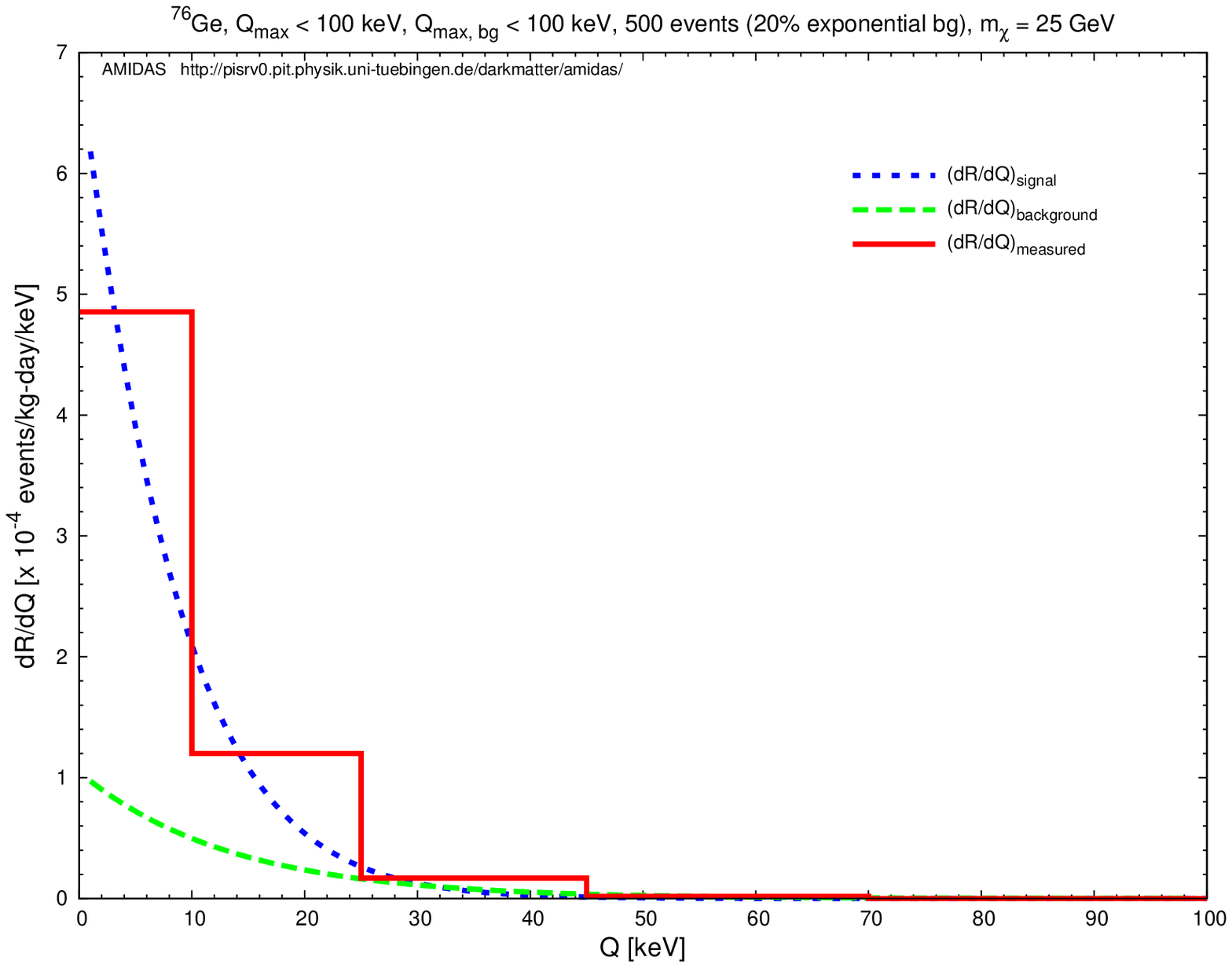} \hspace*{-1.6cm} \\
 \vspace{0.5cm}
 \hspace*{-1.6cm}
 \includegraphics[width=9.8cm]{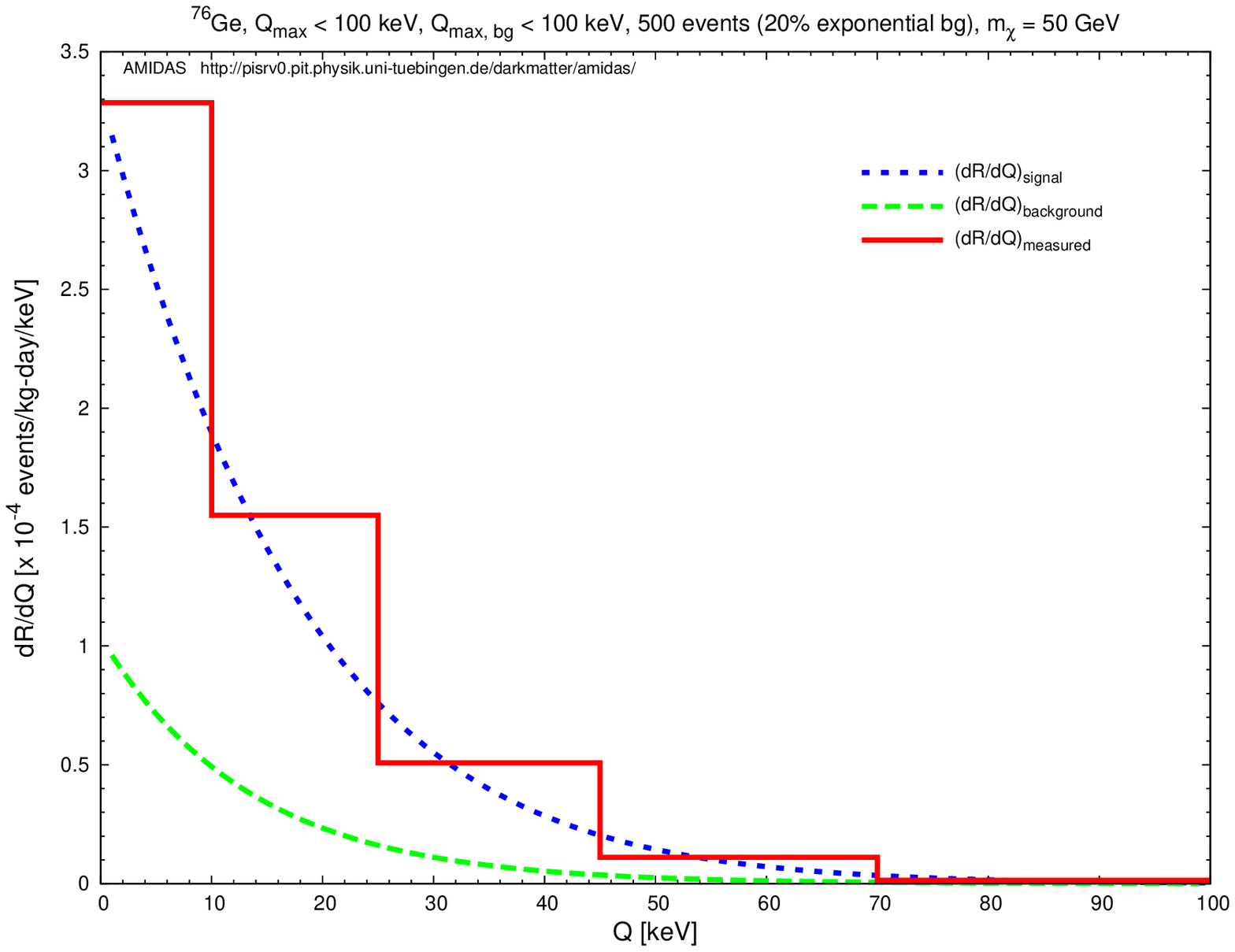} \hspace{-1.1cm}
 \includegraphics[width=9.8cm]{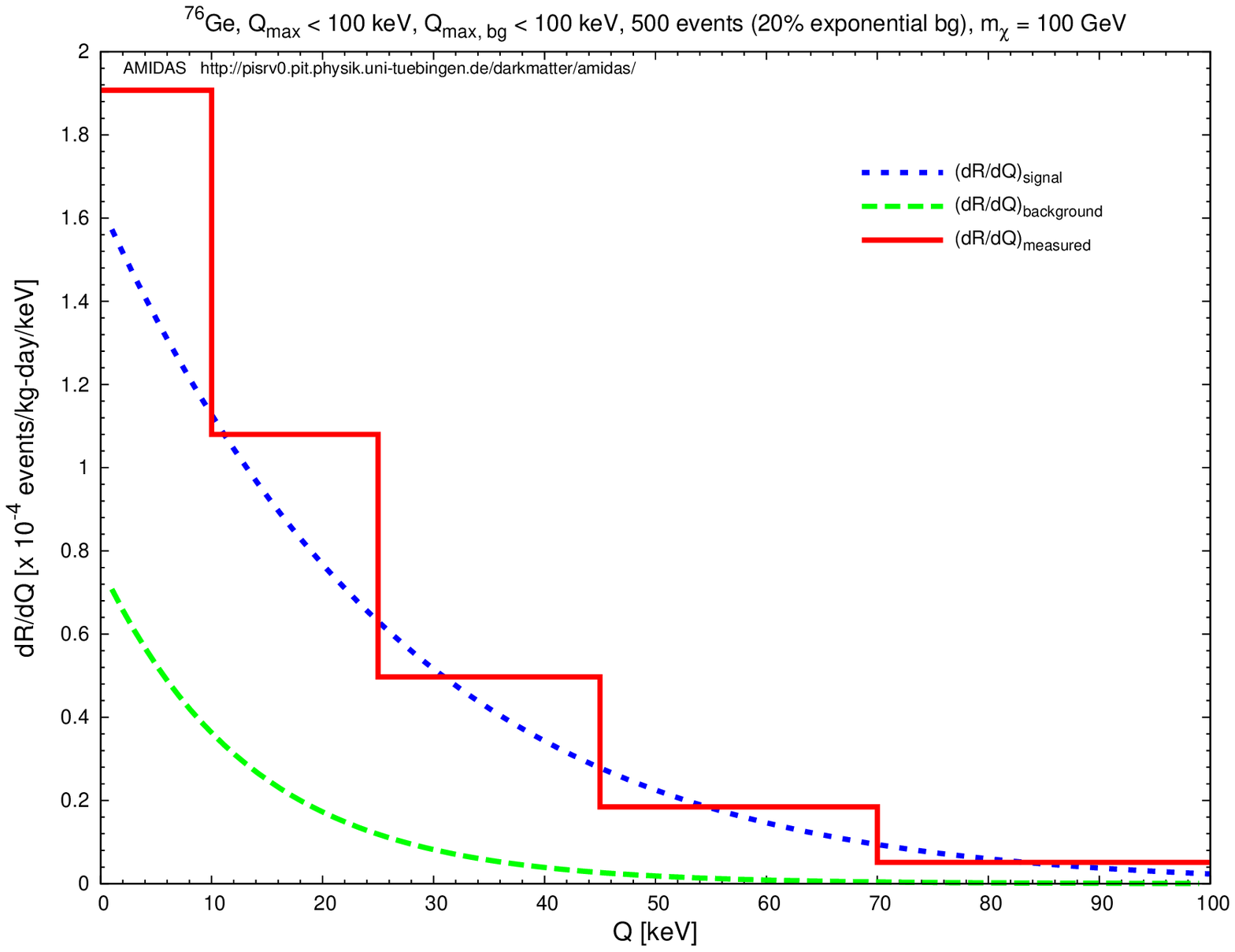} \hspace*{-1.6cm} \\
 \vspace{0.5cm}
 \hspace*{-1.6cm}
 \includegraphics[width=9.8cm]{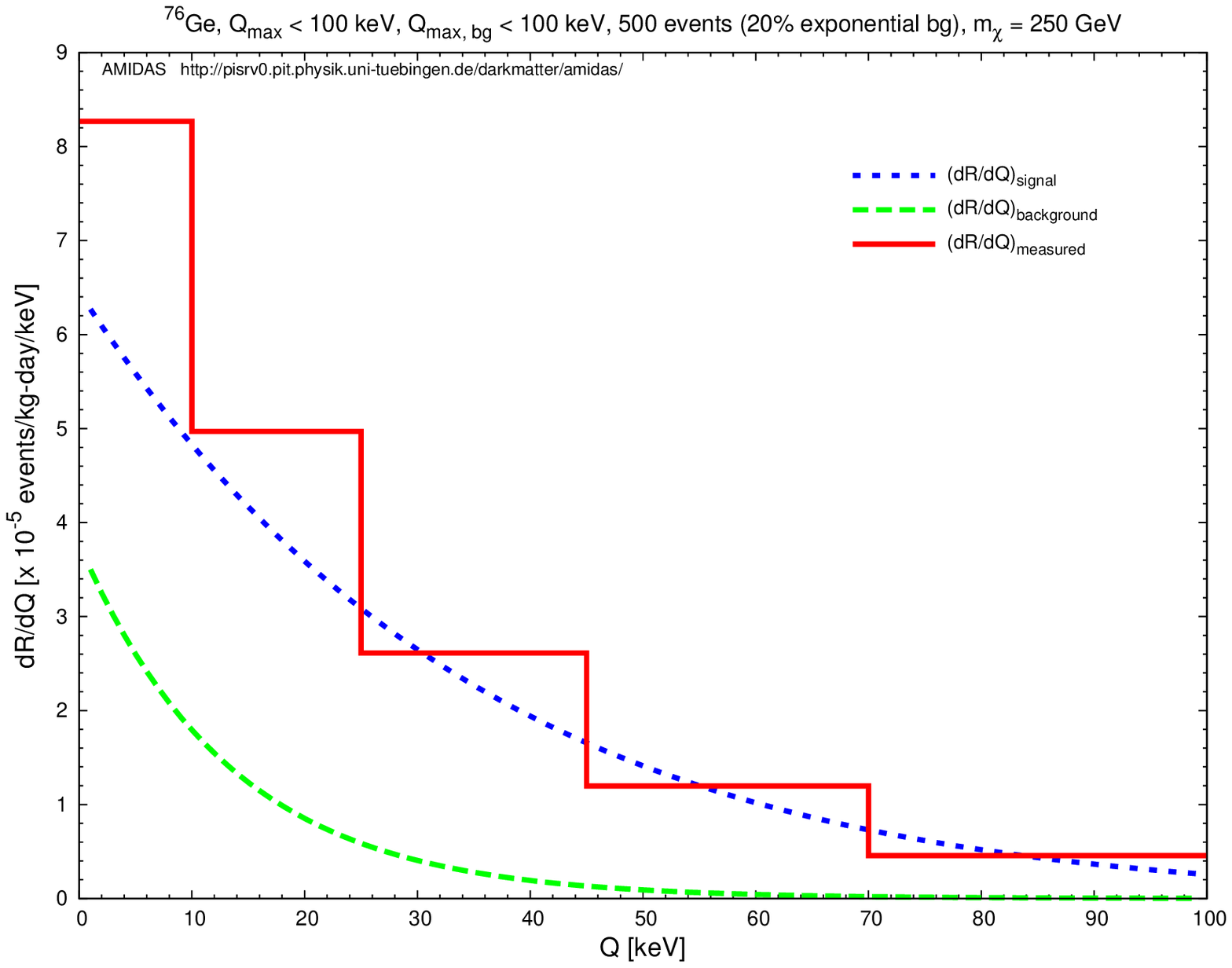} \hspace{-1.1cm}
 \includegraphics[width=9.8cm]{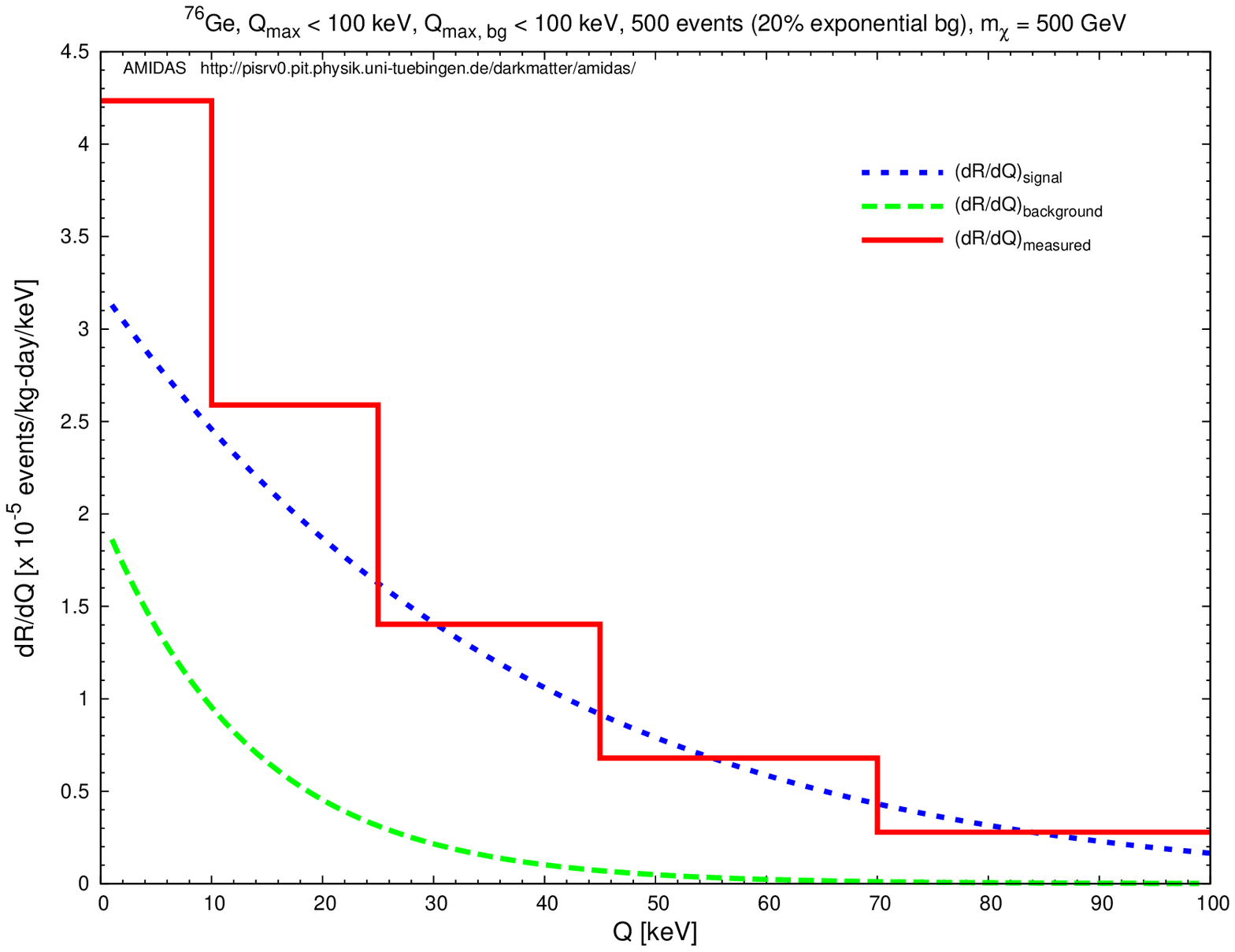} \hspace*{-1.6cm} \\
}
\vspace{-0.75cm}
\end{center}
\caption{
 Measured energy spectra (solid red histograms)
 for a $\rmXA{Ge}{76}$ target
 with six different WIMP masses:
 10, 25, 50, 100, 250, and 500 GeV.
 The dotted blue curves are
 the elastic WIMP--nucleus scattering spectra,
 whereas
 the dashed green curves are
 the exponential background spectra
 normalized to fit to the chosen background ratio,
 which has been set as 20\% here.
 The experimental threshold energy
 has been assumed to be negligible
 and the maximal cut--off energy
 is set as 100 keV.
 The background windows
 have been assumed to be the same as
 the experimental possible energy ranges.
 5,000 experiments with 500 total events on average
 in each experiment have been simulated.
 See the text for further details
 (plots from Ref.~\cite{DMDDbg-mchi}).
}
\label{fig:dRdQ-bg-ex-Ge-000-100-20}
\end{figure}
\begin{figure}[p!]
\begin{center}
\vspace{-0.75cm}
\imageswitch{
\begin{picture}(15,21.25)
\put(0,10.75){\framebox(15,10.5){mchi-SiGe-ex-000-100-050}}
\put(0, 0   ){\framebox(15,10.5){mchi-SiGe-ex-000-100-500}}
\end{picture}}
{\includegraphics[width=15cm]{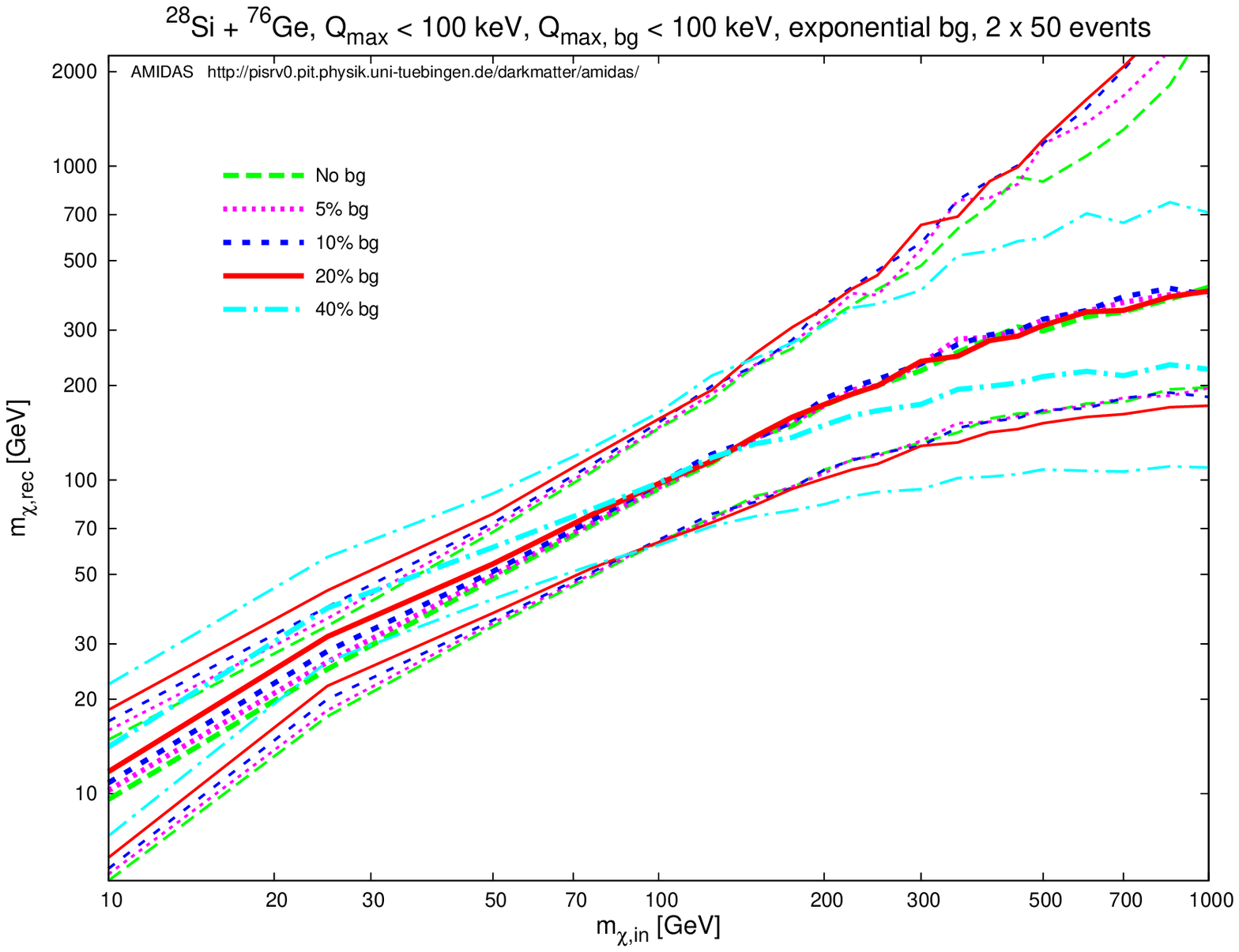}  \\ \vspace{0.25cm}
 \includegraphics[width=15cm]{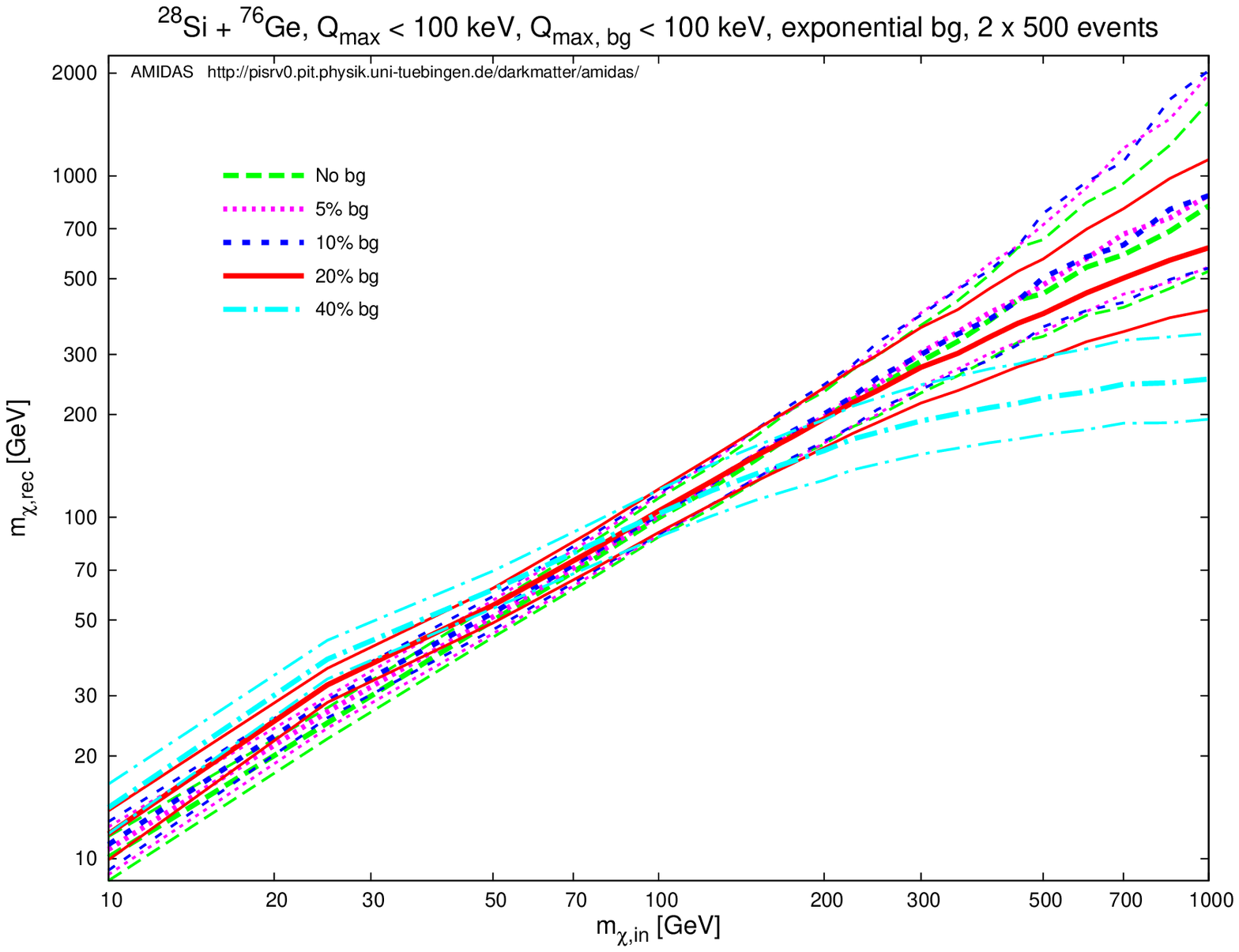}  \\
}
\vspace{-0.5cm}
\end{center}
\caption{
 The reconstructed WIMP mass
 and the lower and upper bounds of
 the 1$\sigma$ statistical uncertainty
 by using mixed data sets
 from WIMP--induced and background events
 as functions of the input WIMP mass.
 $\rmXA{Si}{28}$ and $\rmXA{Ge}{76}$
 have been chosen as two target nuclei.
 The background ratios shown here
 are no background (dashed green curves),
  5\% (dotted magenta curves),
 10\% (long--dotted blue curves),
 20\% (solid red curves),
 and 40\% (dash--dotted cyan curves)
 background events in the analyzed data sets
 in the experimental energy ranges
 between 0 and \mbox{100 keV}.
 Each experiment contains 50 (upper)
 and 500 (lower) total events
 on average.
 Other parameters are as
 in Figs.~\ref{fig:dRdQ-bg-ex-Ge-000-100-20}.
 See the text for further details.
}
\label{fig:mchi-SiGe-ex-000-100}
\end{figure}

 In Figs.~\ref{fig:dRdQ-bg-ex-Ge-000-100-20}
 I show measured energy spectra (solid red histograms)
 for a $\rmXA{Ge}{76}$ target
 with six different WIMP masses:
 10, 25, 50, 100, 250, and 500 GeV
 based on Monte Carlo simulations.
 The dotted blue curves are
 the elastic WIMP--nucleus scattering spectra,
 whereas
 the dashed green curves are
 the exponential background spectra
 given in Eq.~(\ref{eqn:dRdQ_bg_ex}),
 which have been normalized so that
 the ratios of the areas under these background spectra
 to those under the (dotted blue) WIMP scattering spectra
 are equal to the background--signal ratio
 in the whole data sets
 (e.g.,~20\% backgrounds to 80\% signals
  shown in Figs.~\ref{fig:dRdQ-bg-ex-Ge-000-100-20}).
 The experimental threshold energy
 has been assumed to be negligible
 and the maximal cut--off energy
 is set as 100 keV.
 The background windows
 (the possible energy ranges
  in which residue background events exist)
 have been assumed to be the same as
 the experimental possible energy ranges.
 5,000 experiments with 500 total events on average
 in each experiment have been simulated.
 Remind that
 the measured energy spectra shown here
 are averaged over the simulated experiments.
 Five bins with linear increased bin widths
 have been used for binning
 generated signal and background events.
 As argued in Sec.~2.3,
 for reconstructing the one--dimensional
 WIMP velocity distribution function,
 this unusual, particular binning has been chosen
 in order to accumulate more events
 in high energy ranges
 and thus to reduce the statistical uncertainties
 in high velocity ranges.

 It can be found
 in Figs.~\ref{fig:dRdQ-bg-ex-Ge-000-100-20} that,
 the shape of the WIMP scattering spectrum
 depends highly on the WIMP mass:
 for light WIMPs ($\mchi~\lsim~50$ GeV),
 the recoil spectra drop sharply with increasing recoil energies,
 while for heavy WIMPs ($\mchi~\gsim~100$ GeV),
 the spectra become flatter.
 In contrast,
 the exponential background spectra shown here
 depend only on the target mass
 and are rather flatter (sharper)
 for light (heavy) WIMP masses
 compared to the WIMP scattering spectra.
 This means that,
 once input WIMPs are light (heavy),
 background events would contribute relatively more to
 high (low) energy ranges,
 and, consequently,
 the measured energy spectra
 would mimic scattering spectra
 induced by heavier (lighter) WIMPs.

 More detailed illustrations and discussions
 about the effects of residue background events
 on the measured energy spectrum
 can be found in Ref.~\cite{DMDDbg-mchi}.
\subsection{On the reconstructed WIMP mass}
 Figs.~\ref{fig:mchi-SiGe-ex-000-100}
 show the reconstructed WIMP mass
 and the lower and upper bounds of
 the 1$\sigma$ statistical uncertainty
 by means of the model--independent procedure introduced
 in Refs.~\cite{DMDDmchi-SUSY07, DMDDmchi}
 with mixed data sets
 from WIMP--induced and background events
 as functions of the input WIMP mass.
 As in Ref.~\cite{DMDDmchi},
 $\rmXA{Si}{28}$ and $\rmXA{Ge}{76}$
 have been chosen as two target nuclei.
 The experimental threshold energies of two experiments
 have been assumed to be negligible
 and the maximal cut--off energies
 are set the same as 100 keV.
 The background windows are set as
 the same as the experimental possible energy ranges
 for both experiments.
 The background ratios shown here
 are no background (dashed green curves),
  5\% (dotted magenta curves),
 10\% (long--dotted blue curves),
 20\% (solid red curves),
 and 40\% (dash--dotted cyan curves)
 background events in the analyzed data sets.
 2 $\times$ 5,000 experiments have been simulated.
 Each experiment contains 50 (upper)
 and 500 (lower) total events
 on average.
 Note that
 {\em all} events recorded in our data sets
 are treated as WIMP signals in the analysis,
 although statistically we know that
 a fraction of these events could be backgrounds.

 It can be seen clearly that,
 for light WIMP masses ($\mchi~\lsim~100$ GeV),
 due to the relatively flatter background spectrum
 (compared to the scattering spectrum induced by light WIMPs),
 the energy spectrum of all recorded events
 would mimic a scattering spectrum induced
 by WIMPs with a relatively heavier mass,
 and, consequently,
 the reconstructed WIMP masses
 as well as the statistical uncertainty intervals
 could be overestimated.
 In contrast,
 for heavy WIMP masses ($\mchi~\gsim~100$ GeV),
 due to the relatively sharper background spectrum,
 relatively more background events
 contribute to low energy ranges,
 and the energy spectrum of all recorded events
 would mimic a scattering spectrum induced
 by WIMPs with a relatively lighter mass.
 Hence,
 the reconstructed WIMP masses
 as well as the statistical uncertainty intervals
 could be underestimated.
 Moreover,
 Figs.~\ref{fig:mchi-SiGe-ex-000-100}
 show that
 the larger the fraction of background events in the data sets,
 the more strongly over-/underestimated
 the reconstructed WIMP masses
 as well as the statistical uncertainty intervals.
 Nevertheless,
 from Figs.~\ref{fig:mchi-SiGe-ex-000-100}
 it can be found that,
 with $\sim$ 10\% residue background events
 in the analyzed data sets
 of $\sim$ 500 total events,
 one could still estimate the WIMP mass pretty well;
 if WIMPs are light ($\mchi~\lsim~200$ GeV),
 the maximal acceptable fraction of
 residue background events
 could even be as large as $\sim$ 20\%.

 More detailed illustrations and discussions
 about the effects of residue background events
 on the determination of the WIMP mass
 can be found in Ref.~\cite{DMDDbg-mchi}.
\section{Results of the reconstructed
         one--dimensional WIMP velocity distribution function}
 In this section
 I present simulation results
 of the reconstructed one--dimensional
 velocity distribution function of halo WIMPs
 with mixed data sets
 from WIMP--induced and background events
 by means of the model--independent method
 described in Sec.~2.%
\footnote{
 Note that,
 rather than the mean values,
 the (bounds on the) reconstructed $f_{1, {\rm rec}}(v_{s, \mu})$
 are always the median values
 of the simulated results.
}
 The WIMP mass $\mchi$
 involved in the coefficient $\alpha$
 in Eqs.~(\ref{eqn:vsn}) and (\ref{eqn:calN_sum})
 for estimating the reconstructed points $v_{s, n}$
 (or $v_{s, \mu}$ for a windowed data set)
 as well as the normalization constant $\calN$
 has been assumed to be known precisely
 with a negligible uncertainty
 from other (e.g., collider) experiments
 or can be determined from
 {\em other} direct detection experiments.
 As in Ref.~\cite{DMDDf1v},
 a $\rmXA{Ge}{76}$ nucleus has been chosen
 as our detector target for reconstructing $f_1(v)$;
 while a $\rmXA{Si}{28}$ target
 and a {\em second} $\rmXA{Ge}{76}$ target
 have been used for determining $\mchi$.
 The experimental threshold energy of each experiment
 has been assumed to be negligible
 and the maximal cut--off energies
 are set the same as 100 keV.
 The exponential background spectrum
 given in Eq.~(\ref{eqn:dRdQ_bg_ex})
 has been used for generating background events
 in windows of the entire experimental possible ranges
 As in Figs.~\ref{fig:dRdQ-bg-ex-Ge-000-100-20},
 five bins have been used%
\footnote{
 For the input WIMP masses of 10 (25) GeV,
 the widths of the first bin
 have been modified to 1.5 (5) keV
 due to a kinematic maximal cut--off energy
 discussed later.
} and
 up to three bins have been combined to a window.
 (3 $\times$) 5,000 experiments
 have been simulated.
\subsection{With a precisely known WIMP mass}
 In this subsection
 we first assume that
 the required WIMP mass
 for determining the shape of the reconstructed velocity distribution
 through the transformation (\ref{eqn:vsn})
 from $Q_{s, n}$ to $v_{s, n}$
 (or from $Q_{s, \mu}$ to $v_{s, \mu}$)
 and for estimating the normalization constant $\calN$
 by Eq.~(\ref{eqn:calN_sum})
 has been known precisely
 with a negligible uncertainty.

\begin{figure}[p!]
\begin{center}
\vspace{-0.75cm}
\imageswitch{
\begin{picture}(16.5,20)
\put(0  ,13.5 ){\framebox(8,6.5){f1v-Ge-ex-000-100-010-0500}}
\put(8.5,13.5 ){\framebox(8,6.5){f1v-Ge-ex-000-100-025-0500}}
\put(0  , 6.75){\framebox(8,6.5){f1v-Ge-ex-000-100-050-0500}}
\put(8.5, 6.75){\framebox(8,6.5){f1v-Ge-ex-000-100-100-0500}}
\put(0  , 0   ){\framebox(8,6.5){f1v-Ge-ex-000-100-250-0500}}
\put(8.5, 0   ){\framebox(8,6.5){f1v-Ge-ex-000-100-500-0500}}
\end{picture}}
{\hspace*{-1.6cm}
 \includegraphics[width=9.8cm]{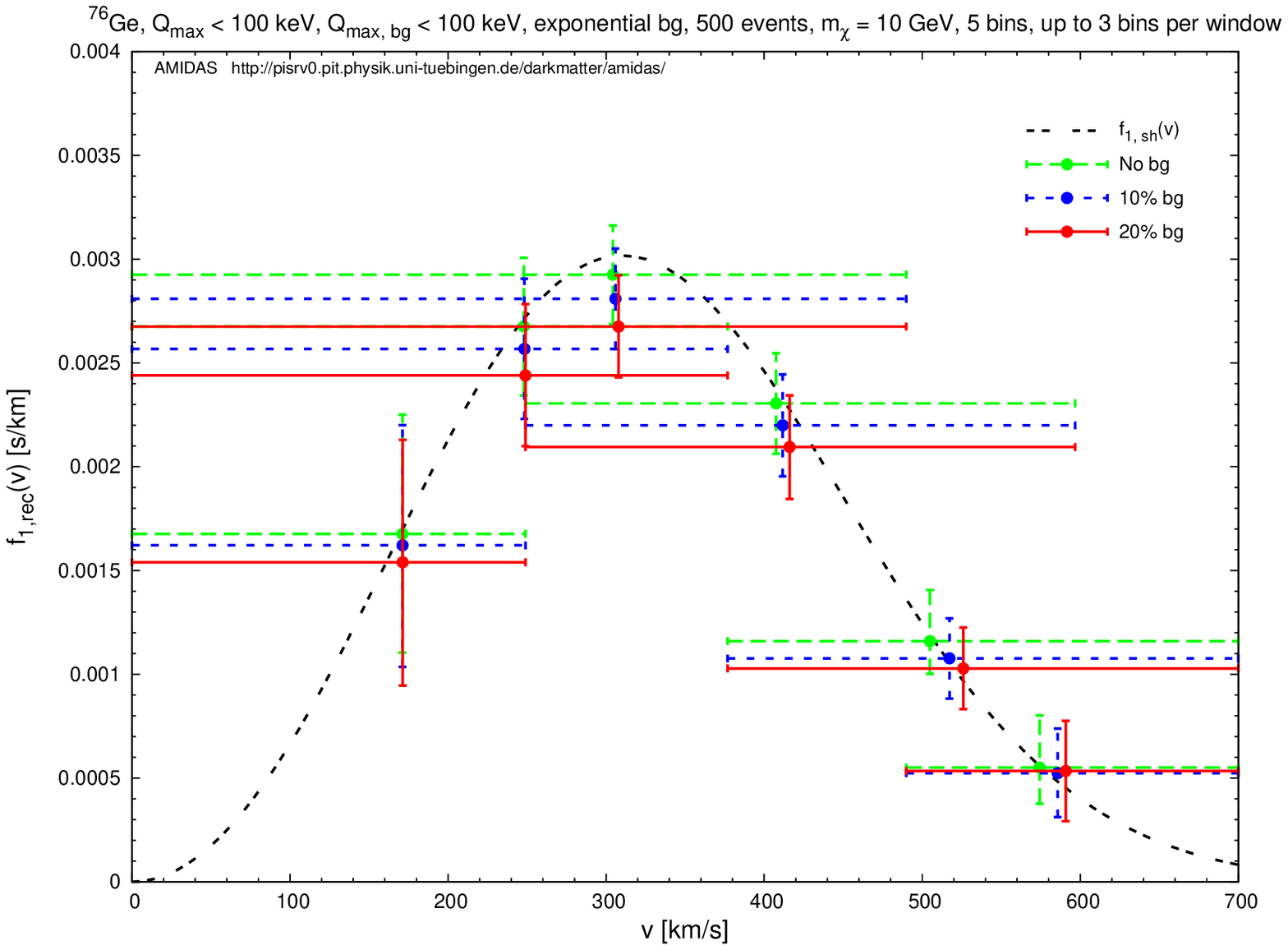} \hspace{-1.1cm}
 \includegraphics[width=9.8cm]{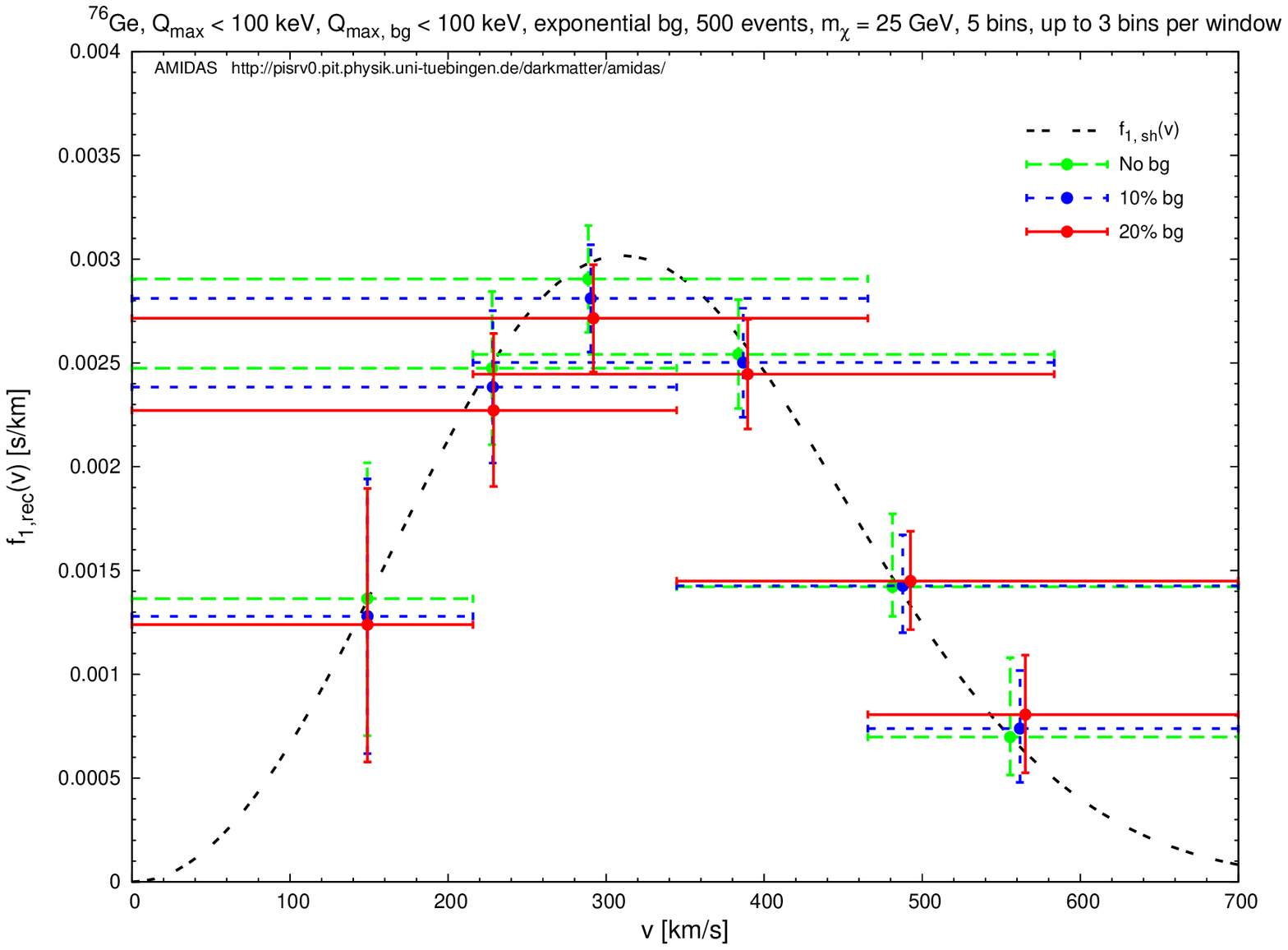} \hspace*{-1.6cm} \\
 \vspace{0.25cm}
 \hspace*{-1.6cm}
 \includegraphics[width=9.8cm]{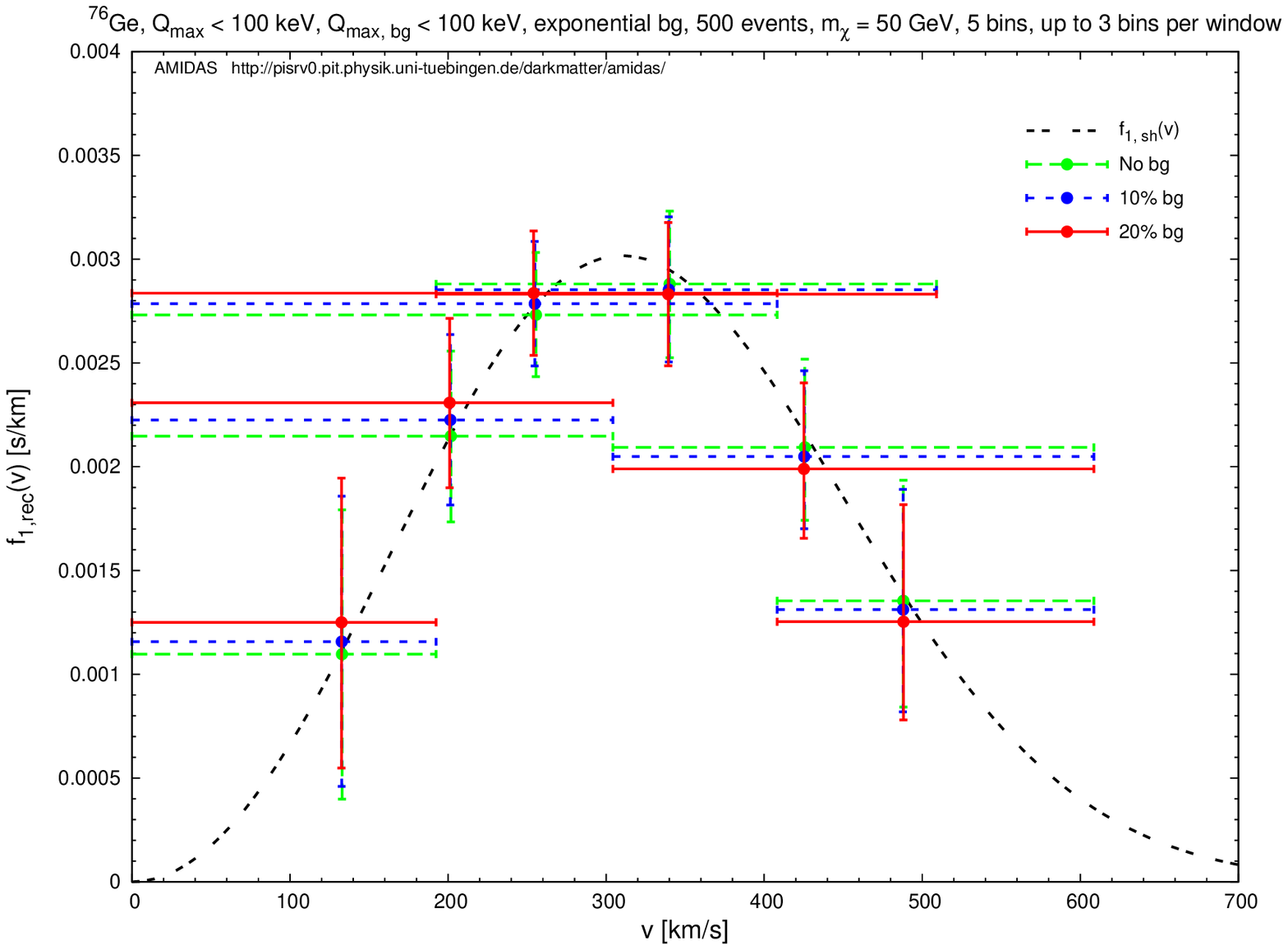} \hspace{-1.1cm}
 \includegraphics[width=9.8cm]{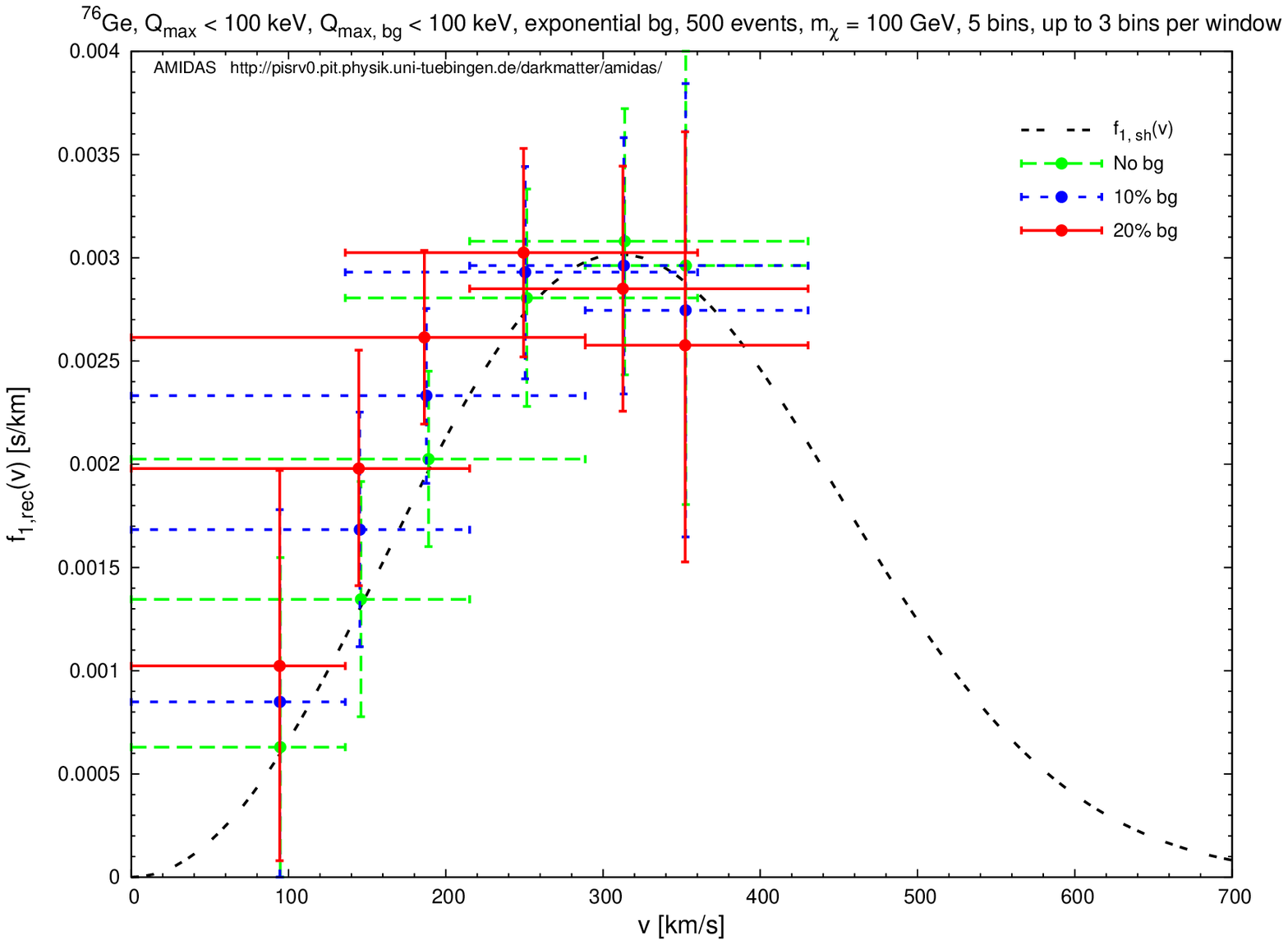} \hspace*{-1.6cm} \\
 \vspace{0.25cm}
 \hspace*{-1.6cm}
 \includegraphics[width=9.8cm]{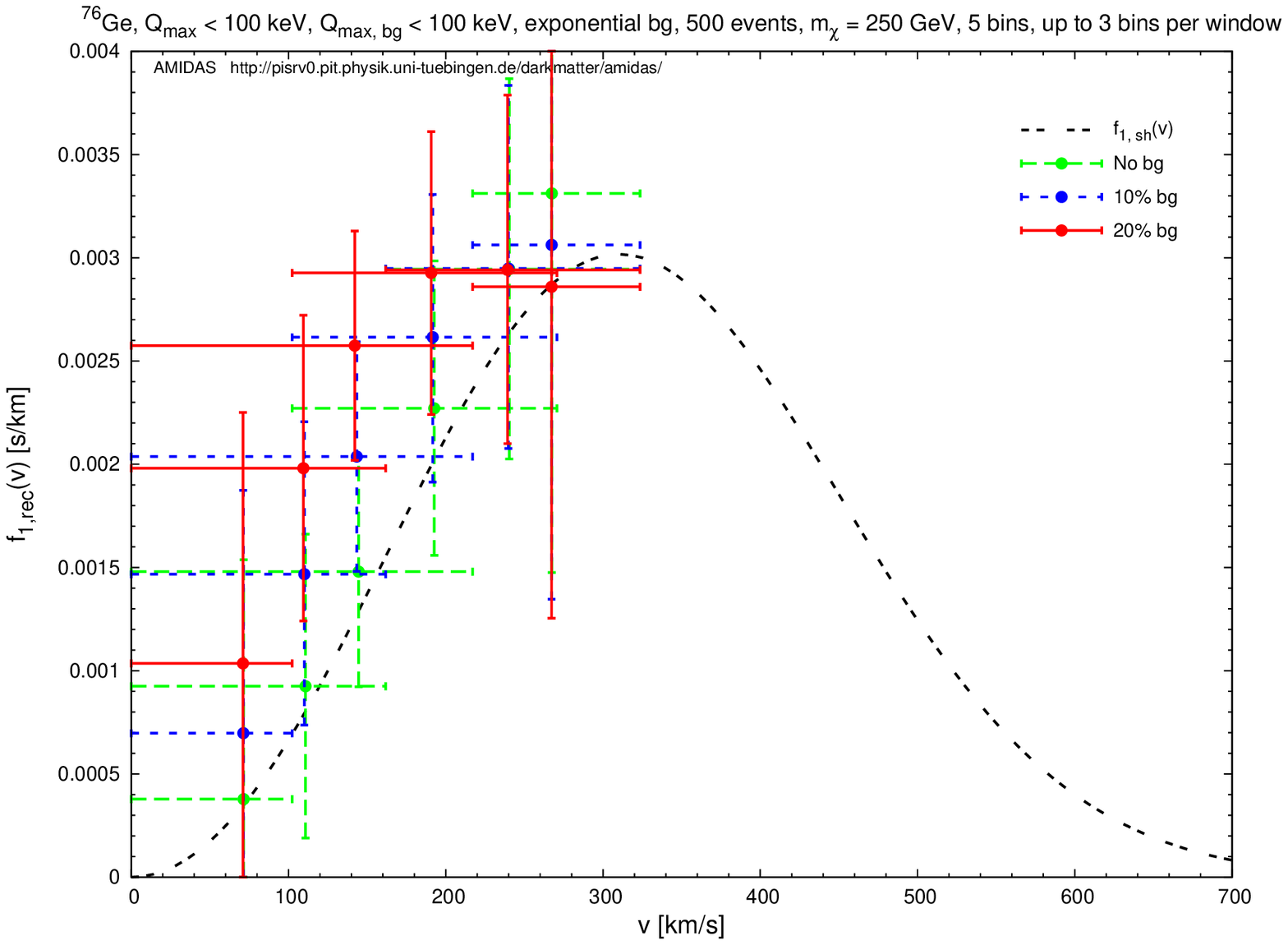} \hspace{-1.1cm}
 \includegraphics[width=9.8cm]{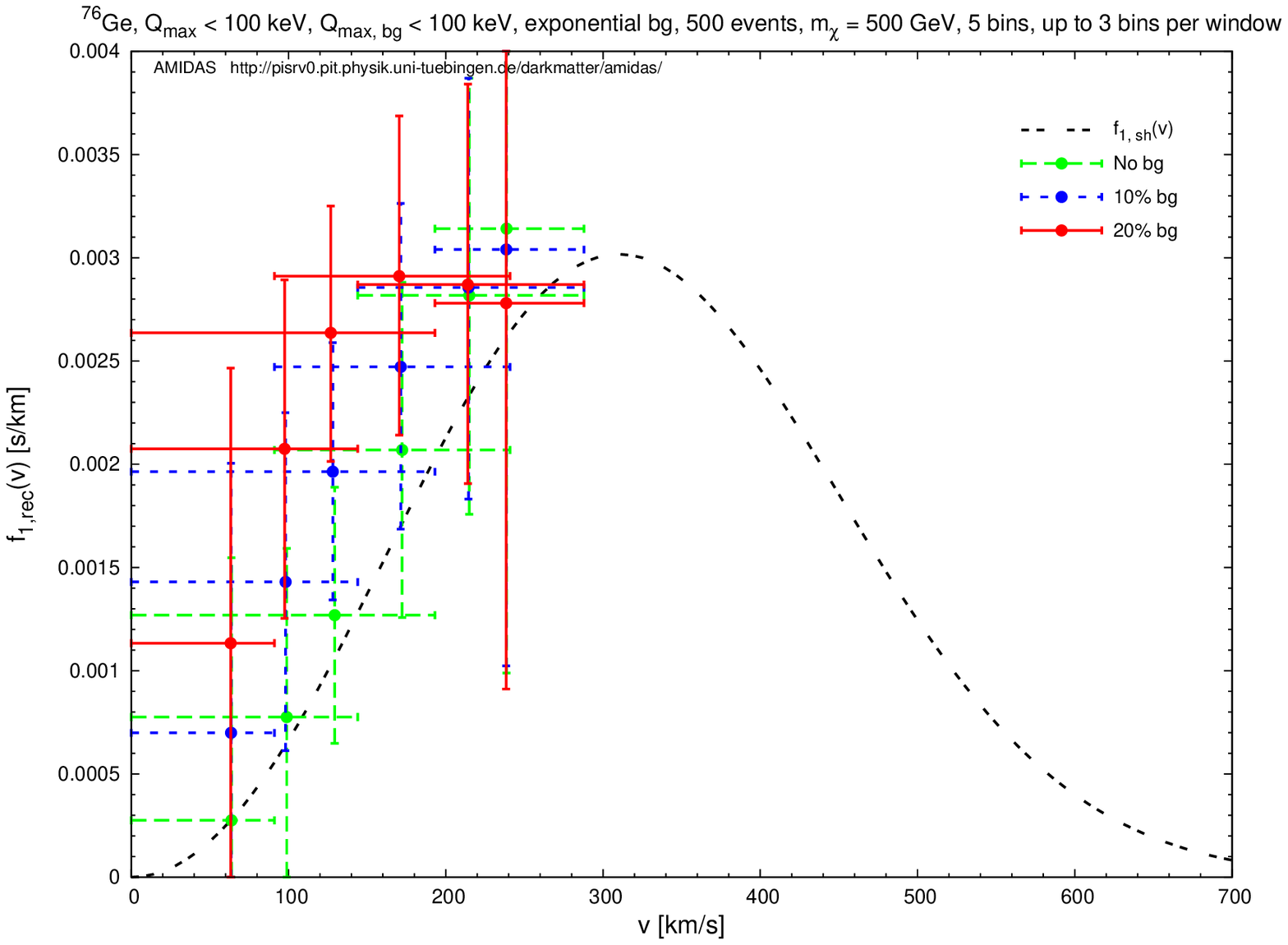} \hspace*{-1.6cm} \\
}
\vspace{-0.5cm}
\end{center}
\caption{
 The one--dimensional WIMP velocity distribution function
 reconstructed by Eq.~(\ref{eqn:f1v_Qsn})
 for six different WIMP masses:
 10, 25, 50, 100, 250, and 500 GeV.
 The double--dotted black curves
 are the input shifted Maxwellian velocity distribution
 in Eq.~(\ref{eqn:f1v_sh}).
 The vertical error bars show
 the square roots of the diagonal entries of the covariance matrix
 given in Eq.~(\ref{eqn:cov_f1v_Qs_mu}),
 while
 the horizontal bars show
 the sizes of the windows used
 for estimating $f_{1, {\rm rec}}(v_{s, \mu})$.
 The background ratios shown here
 are no background (dashed green lines),
 10\% (long--dotted blue lines),
 and 20\% (solid red lines)
 background events in the analyzed data set
 in the background window
 of the entire experimental possible energy range.
 Parameters and notations
 are as in Figs.~\ref{fig:mchi-SiGe-ex-000-100}.
 See the text for further details.
}
\label{fig:f1v-Ge-ex-000-100-0500}
\end{figure}
\begin{figure}[p!]
\begin{center}
\imageswitch{
\begin{picture}(16.5,21.5)
\put(0  ,15  ){\framebox(8,6.5){f1v-Ge-const-000-100-010-0500}}
\put(8.5,15  ){\framebox(8,6.5){f1v-Ge-const-000-100-025-0500}}
\put(0  , 7.5){\framebox(8,6.5){f1v-Ge-const-000-100-050-0500}}
\put(8.5, 7.5){\framebox(8,6.5){f1v-Ge-const-000-100-100-0500}}
\put(0  , 0  ){\framebox(8,6.5){f1v-Ge-const-000-100-250-0500}}
\put(8.5, 0  ){\framebox(8,6.5){f1v-Ge-const-000-100-500-0500}}
\end{picture}}
{\hspace*{-1.6cm}
 \includegraphics[width=9.8cm]{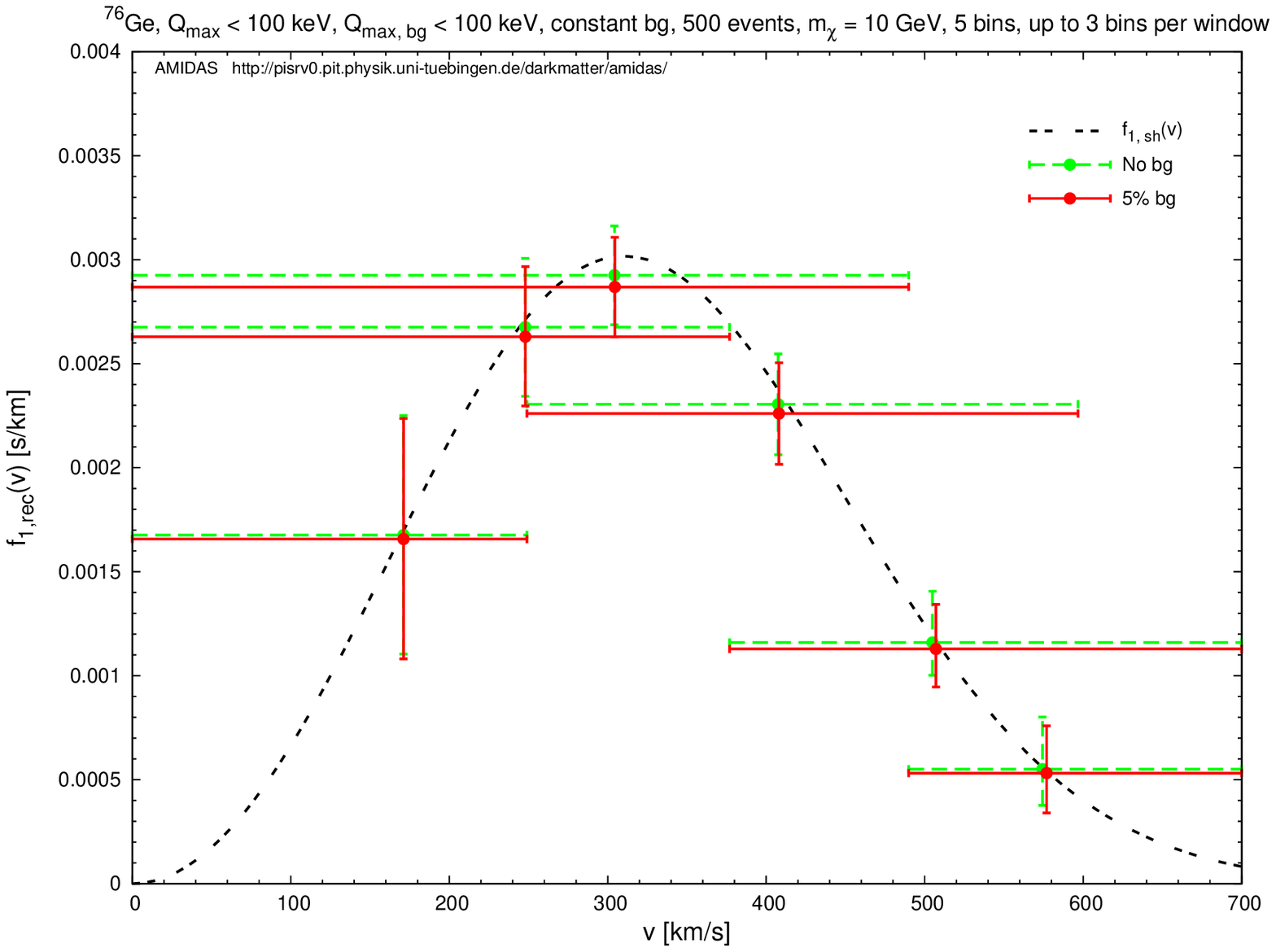} \hspace{-1.1cm}
 \includegraphics[width=9.8cm]{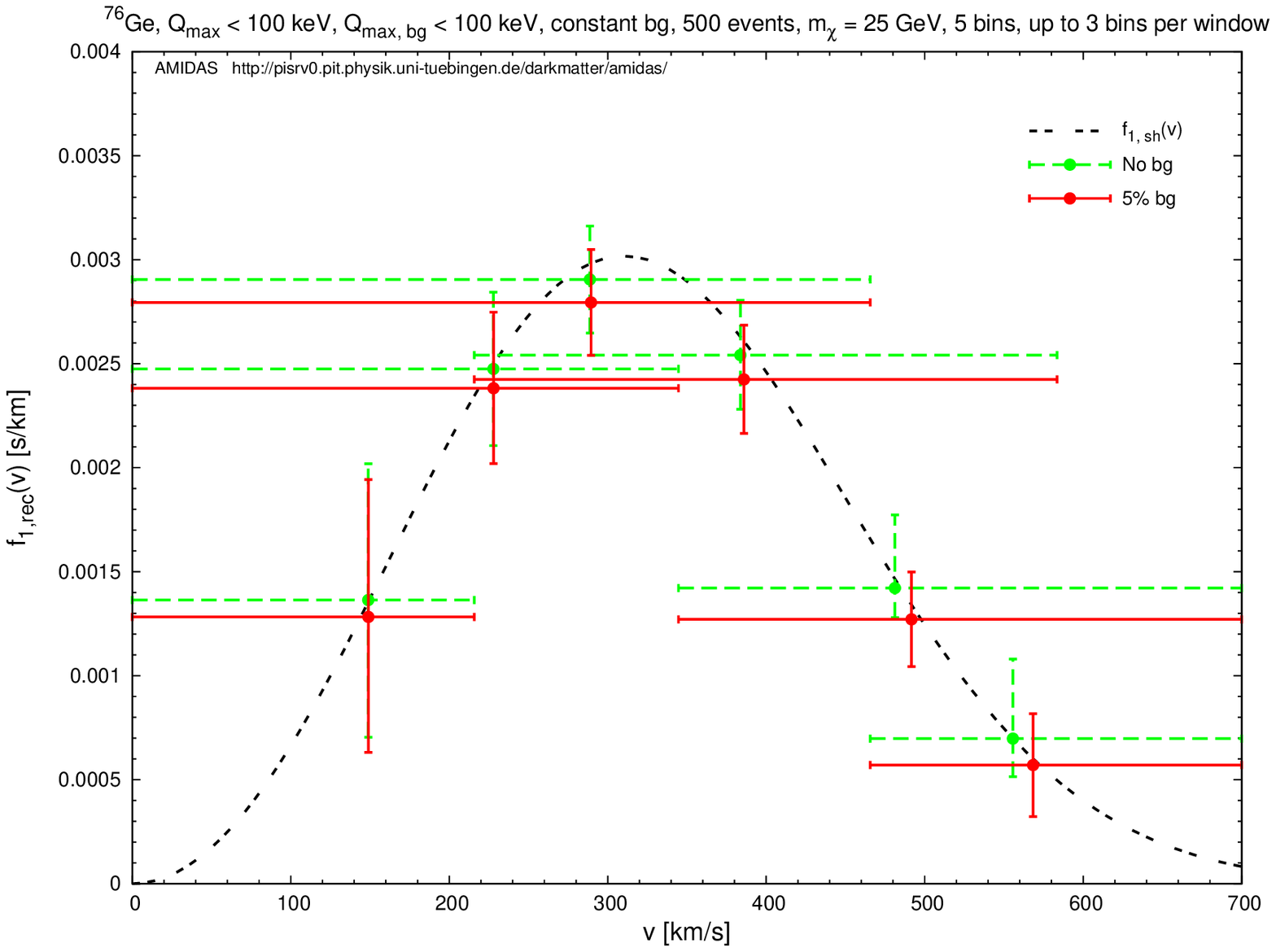} \hspace*{-1.6cm} \\
 \vspace{1cm}
 \hspace*{-1.6cm}
 \includegraphics[width=9.8cm]{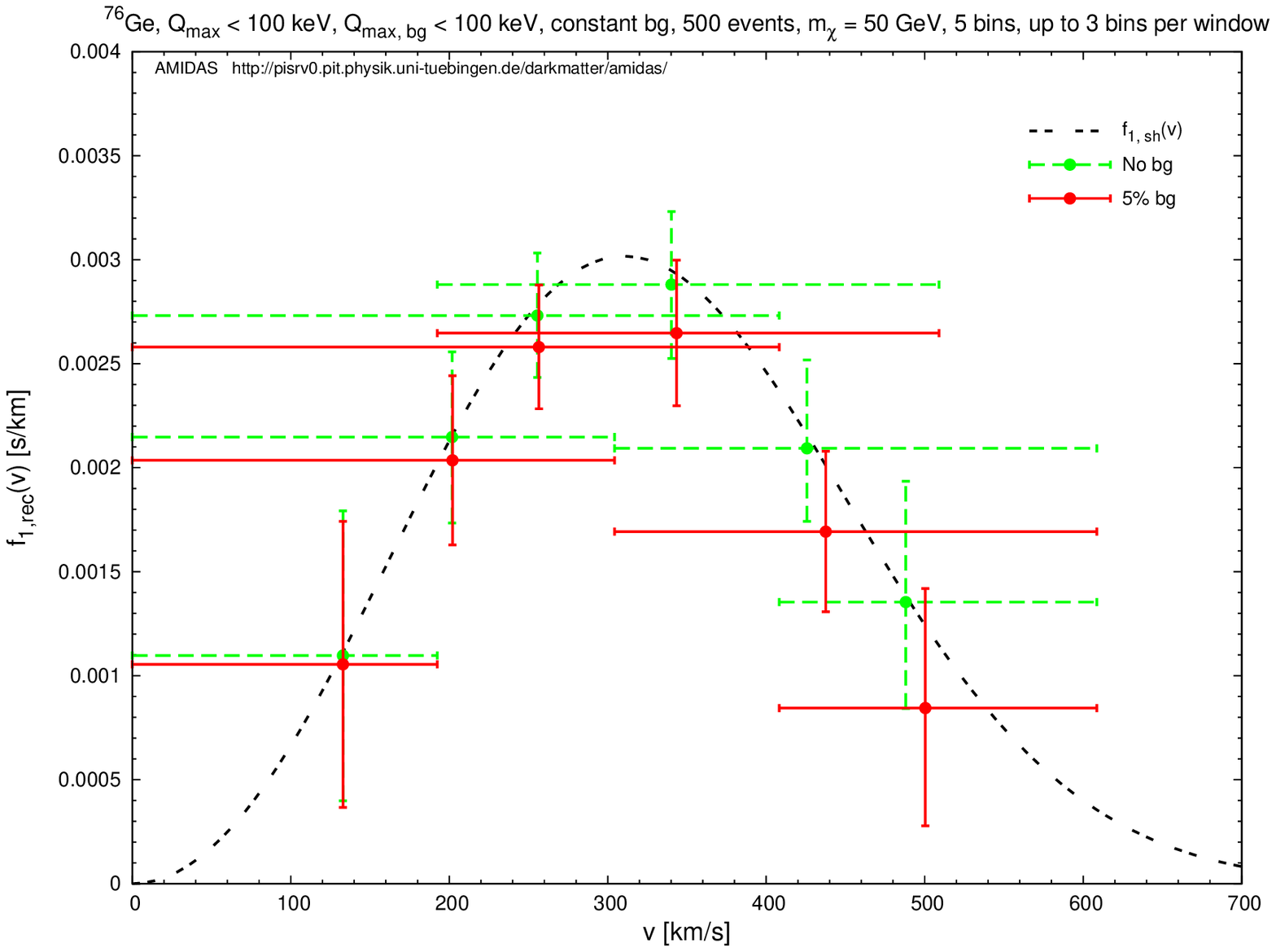} \hspace{-1.1cm}
 \includegraphics[width=9.8cm]{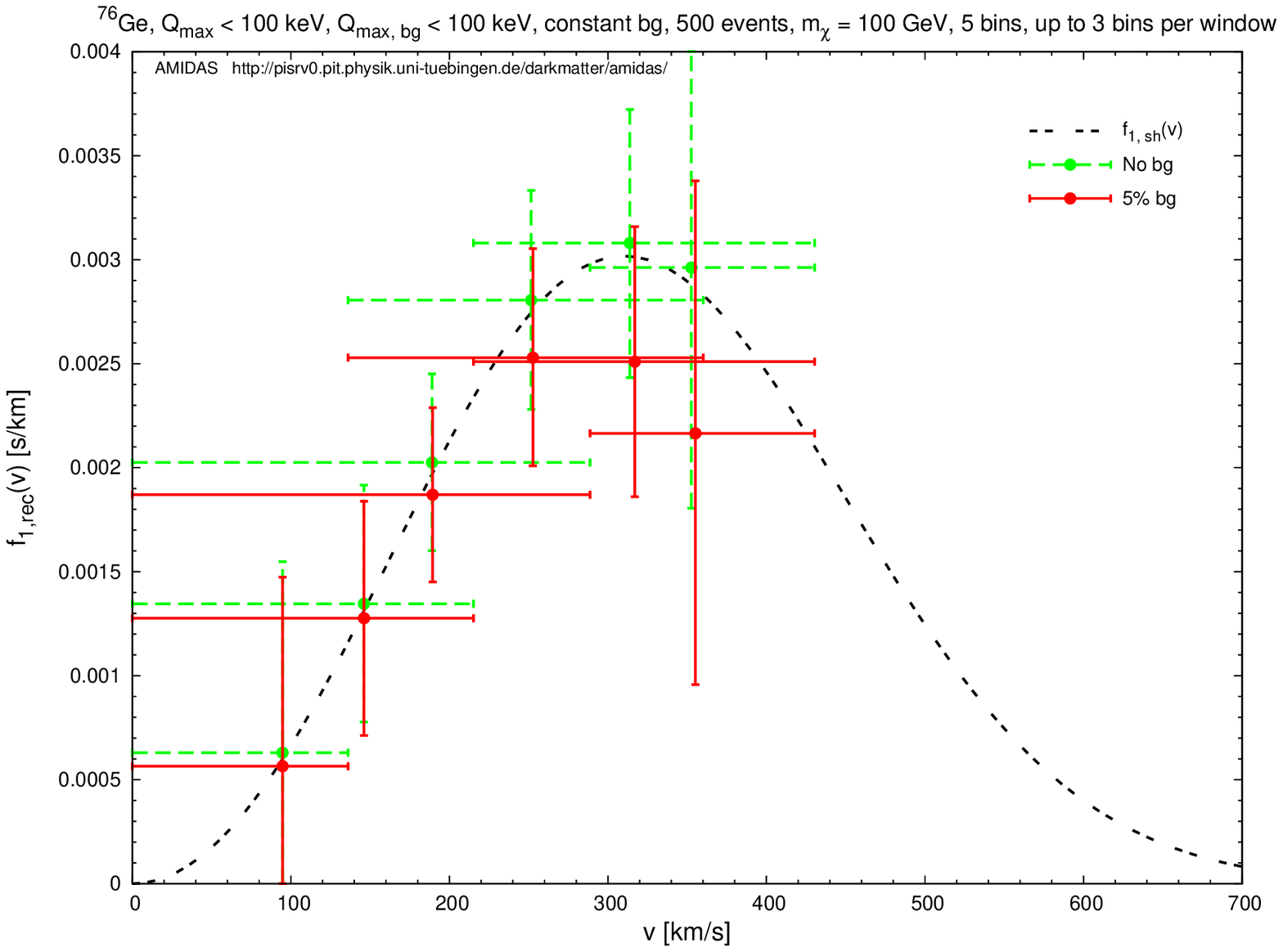} \hspace*{-1.6cm} \\
 \vspace{1cm}
 \hspace*{-1.6cm}
 \includegraphics[width=9.8cm]{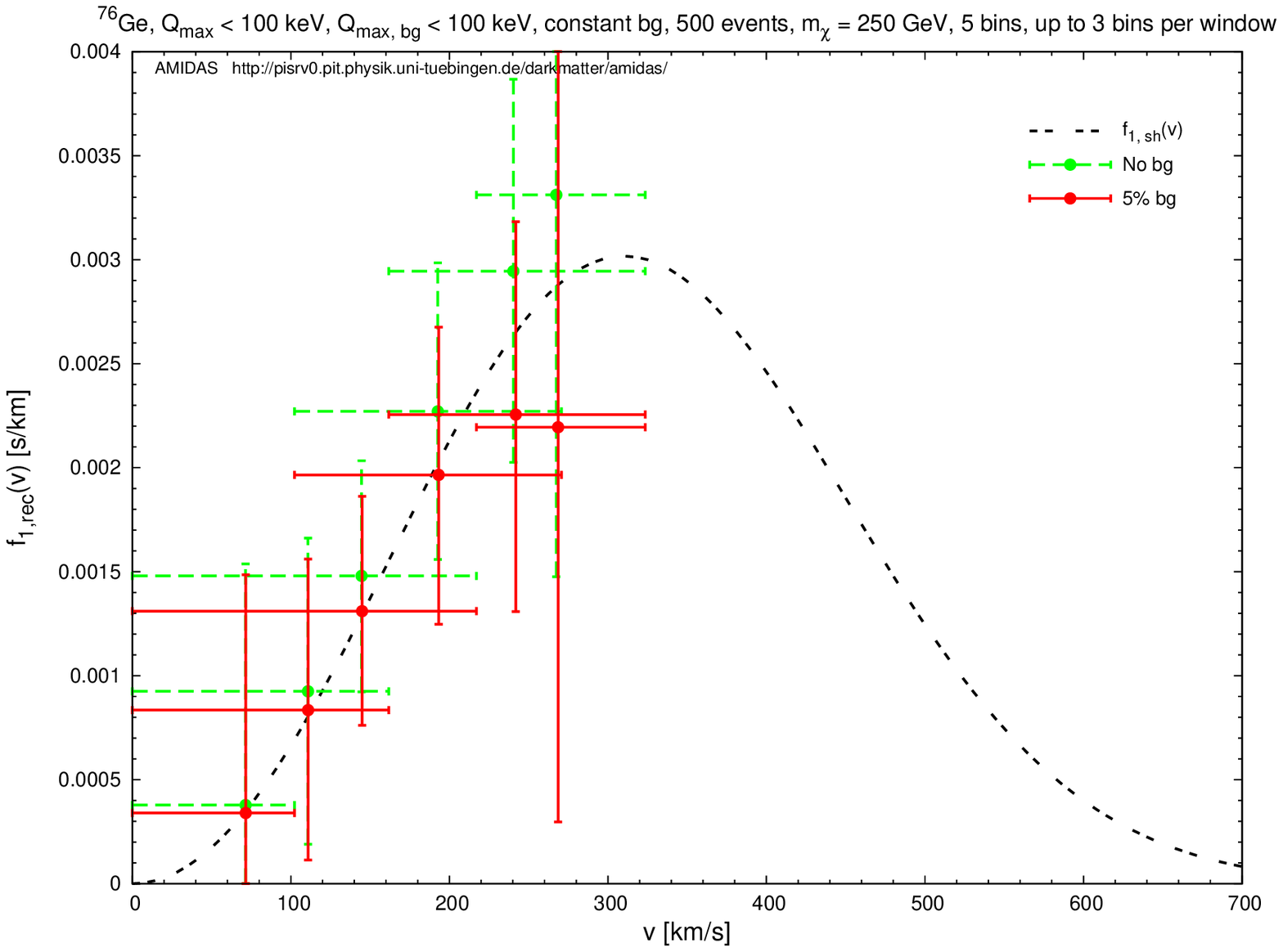} \hspace{-1.1cm}
 \includegraphics[width=9.8cm]{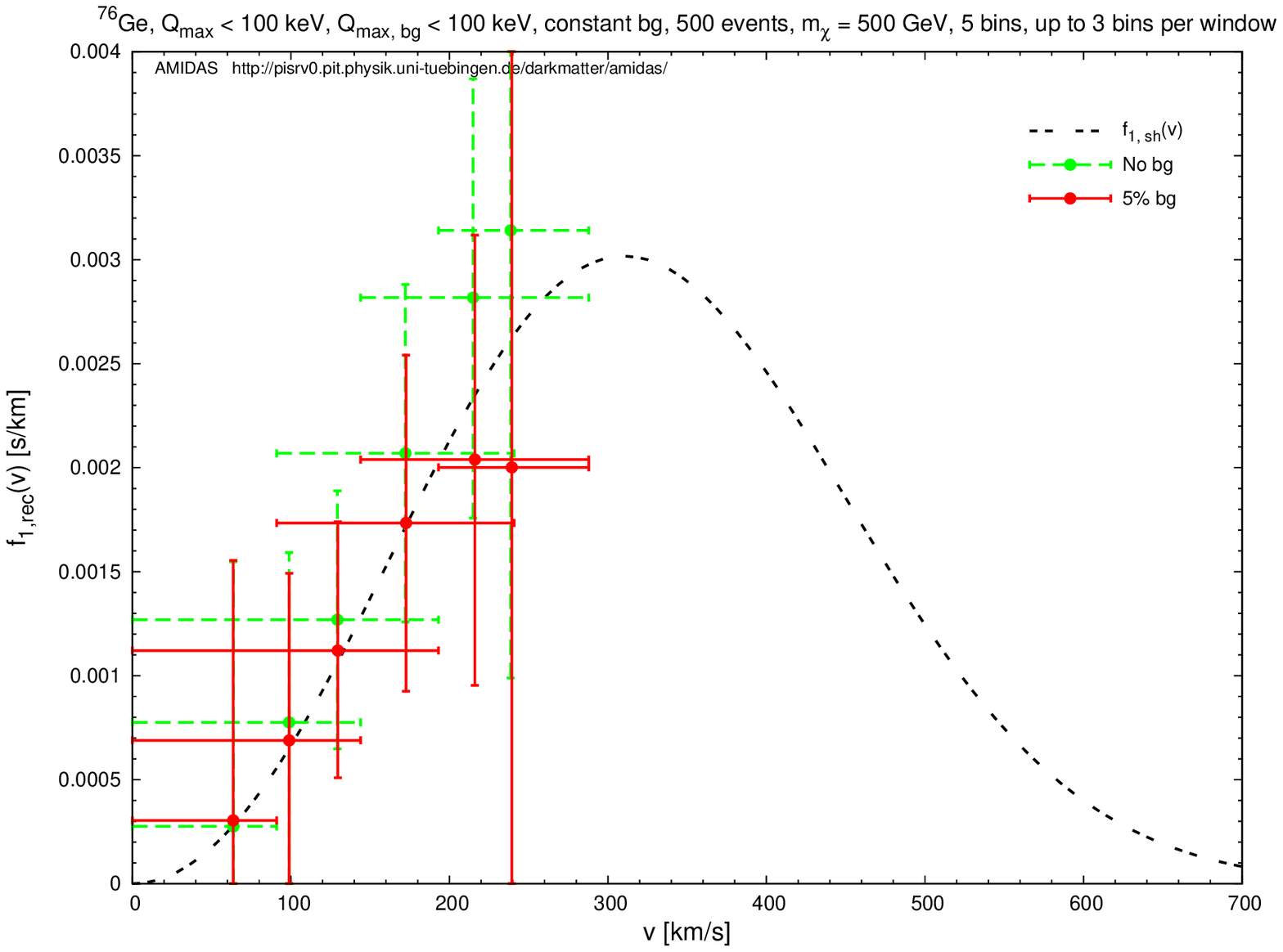} \hspace*{-1.6cm} \\
}
\vspace{-0.25cm}
\end{center}
\caption{
 As in Figs.~\ref{fig:f1v-Ge-ex-000-100-0500},
 except that
 the constant background spectrum
 in Eq.~(\ref{eqn:dRdQ_bg_const})
 has been used.
 Note that
 the solid red lines here
 indicate a background ratio of 5\%.
}
\label{fig:f1v-Ge-const-000-100-0500}
\end{figure}

 Figs.~\ref{fig:f1v-Ge-ex-000-100-0500}
 show the one--dimensional WIMP velocity distribution function
 reconstructed by Eq.~(\ref{eqn:f1v_Qsn})
 with data sets of 500 total events on average
 for six different WIMP masses:
 10, 25, 50, 100, 250, and 500 GeV;
 all events in the data sets
 are treated as WIMP signals.
 The vertical error bars show
 the square roots of the diagonal entries of the covariance matrix%
\footnote{
 Remind that,
 since the neighboring windows overlap,
 the estimates of $f_1$ by Eq.~(\ref{eqn:f1v_Qsn})
 at adjacent values of $v_{s, \mu}$ are correlated.
}
 given in Eq.~(\ref{eqn:cov_f1v_Qs_mu}),
 while
 the horizontal bars show
 the sizes of the windows used
 for estimating $f_{1, {\rm rec}}(v_{s, \mu})$.
 The background ratios shown here
 are no background (dashed green lines),
 10\% (long--dotted blue lines),
 and 20\% (solid red lines)
 background events in the analyzed data set
 in the background window
 of the entire experimental possible energy range.
 Note that,
 since the experimental maximal cut--off energy
 is fixed as 100 keV,
 for heavier input WIMP masses ($\mchi~\gsim~250$ GeV),
 one can reconstruct the velocity distribution function
 only in the velocity range $v~\lsim~300$ km/s.

 It can be seen that,
 as shown in Figs.~\ref{fig:dRdQ-bg-ex-Ge-000-100-20},
 for heavier WIMP masses ($\mchi~\gsim~100$ GeV),
 the relatively sharper background spectrum
 contributes more events to low energy ranges,
 or, equivalently,
 to low velocity ranges.
 This shifts the reconstructed velocity distribution
 to {\em lower} velocities.
 For an input WIMP mass of 100 GeV
 and the background ratio of 10\% (20\%),
 the peak of the reconstructed
 velocity distribution function
 could be shifted by 30 (60) km/s.

 In contrast,
 for lighter WIMP masses ($\mchi~\lsim~50$ GeV),
 the relatively flatter background spectrum
 contributes more events to high energy/velocity ranges.
 This shifts the reconstructed velocity distribution
 to {\em higher} velocities.
 Note that,
 however,
 compared to the cases with the heavy WIMP masses,
 the reconstructed velocity distribution for light WIMPs
 seems not to be shifted to higher velocities very much.
 This should mainly be due to
 the exponential form of the background spectrum:
 its contribution to high energy ranges for light WIMPs
 is relatively not so significant as
 that to low ranges for heavy WIMPs
 (see Figs.~\ref{fig:dRdQ-bg-ex-Ge-000-100-20}).
 One exception should be the case
 with the input WIMP mass of 10 GeV.
 For this case,
 the WIMP scattering spectrum drops very sharp
 in the energy range between 0 and 10 keV,
 while the exponential background spectrum
 extends much wider to 100 keV
 (see Figs.~\ref{fig:dRdQ-bg-ex-Ge-000-100-20}).
 Thus a large part of background events
 contribute to energy ranges higher then 10 keV.
 However,
 because we reconstruct the WIMP velocity distribution
 only in the velocity range below the maximal cut--off velocity,
 i.e., the Galactic escape velocity $\vesc$,
 this leads to a kinematic maximal cut--off of the recoil energy
\beq
   Q_{\rm max, kin}
 = \frac{\vesc^2}{\alpha^2}
\~.
\label{eqn:Qmax_kin}
\eeq
 For a WIMP mass of 10 GeV and
 a $\rmXA{Ge}{76}$ target,
 it can be calculated that
 \mbox{$Q_{\rm max, kin} = 11.8$ keV}.
 Therefore,
 all background events with energies
 larger than $\sim$ 12 keV
 have actually been neglected in the data analysis.

 Not surprisingly,
 the larger the background ratio of our data set,
 the higher/lower the velocities
 to which the reconstructed velocity distribution
 will be shifted.
 But,
 our simulation results
 shown in Figs.~\ref{fig:f1v-Ge-ex-000-100-0500}
 indicate that,
 with an $\sim$ {\em 10\% -- 20\%} background ratio
 (i.e., $\sim$ 50 -- 100 events)
 in the analyzed data set of $\sim$ 500 total events,
 one could in principle still reconstruct
 the one--dimensional velocity distribution function of halo WIMPs
 with an $\sim$ 6.5\%
 (for a 25 GeV WIMP mass, 20\% background events)
 -- $\sim$ 38\%
 (for a 250 GeV WIMP mass, 10\% background events)
 deviation.
 If the mass of halo WIMPs is $\cal O$(50 GeV),
 the maximal acceptable background ratio
 could even be as large as $\sim$ 40\% ($\sim$ 200 events)
 with a deviation of only $\sim$ 14\%.

 In order to check the need of a prior knowledge about
 an (exact) form of the residue background spectrum,
 in Figs.~\ref{fig:f1v-Ge-const-000-100-0500}
 we consider a rather extrem form
 for the residue background spectrum,
 i.e., the constant spectrum
 introduced in Ref.~\cite{DMDDbg-mchi}:
\beq
   \aDd{R}{Q}_{\rm bg, const}
 = 1
\~.
\label{eqn:dRdQ_bg_const}
\eeq
 Here we show only results
 with a background ratio of 5\%
 (solid red lines). 
 It can be seen clearly that,
 although a constant background spectrum
 contributes (much) more events in high energy ranges
 for all six input WIMP masses%
\footnote{
 Illustrations and detailed discussions about
 the effects of the constant form of
 the residue background spectrum
 on the measured energy spectrum
 for different input WIMP masses
 can be found in Ref.~\cite{DMDDbg-mchi}.
},
 taking into account
 the pretty large statistical uncertainty,
 we could at least give a rough outline of
 the WIMP velocity distribution for heavy WIMP masses
 ($\mchi~\gsim~100$ GeV),
 or even reconstruct the distribution pretty well
 for light WIMPs
 ($\mchi~\lsim~100$ GeV),
  thanks to the kinetic maximal cut--off energy $Q_{\rm max, kin}$
  discussed above.
\subsection{With a reconstructed WIMP mass}
\begin{figure}[p!]
\begin{center}
\imageswitch{
\begin{picture}(16.5,21)
\put(0  ,14.5 ){\framebox(8,6.5){f1v-Ge-SiGe-ex-000-100-010-0500}}
\put(8.5,14.5 ){\framebox(8,6.5){f1v-Ge-SiGe-ex-000-100-025-0500}}
\put(0  , 7.25){\framebox(8,6.5){f1v-Ge-SiGe-ex-000-100-050-0500}}
\put(8.5, 7.25){\framebox(8,6.5){f1v-Ge-SiGe-ex-000-100-100-0500}}
\put(0  , 0   ){\framebox(8,6.5){f1v-Ge-SiGe-ex-000-100-250-0500}}
\put(8.5, 0   ){\framebox(8,6.5){f1v-Ge-SiGe-ex-000-100-500-0500}}
\end{picture}}
{\hspace*{-1.6cm}
 \includegraphics[width=9.8cm]{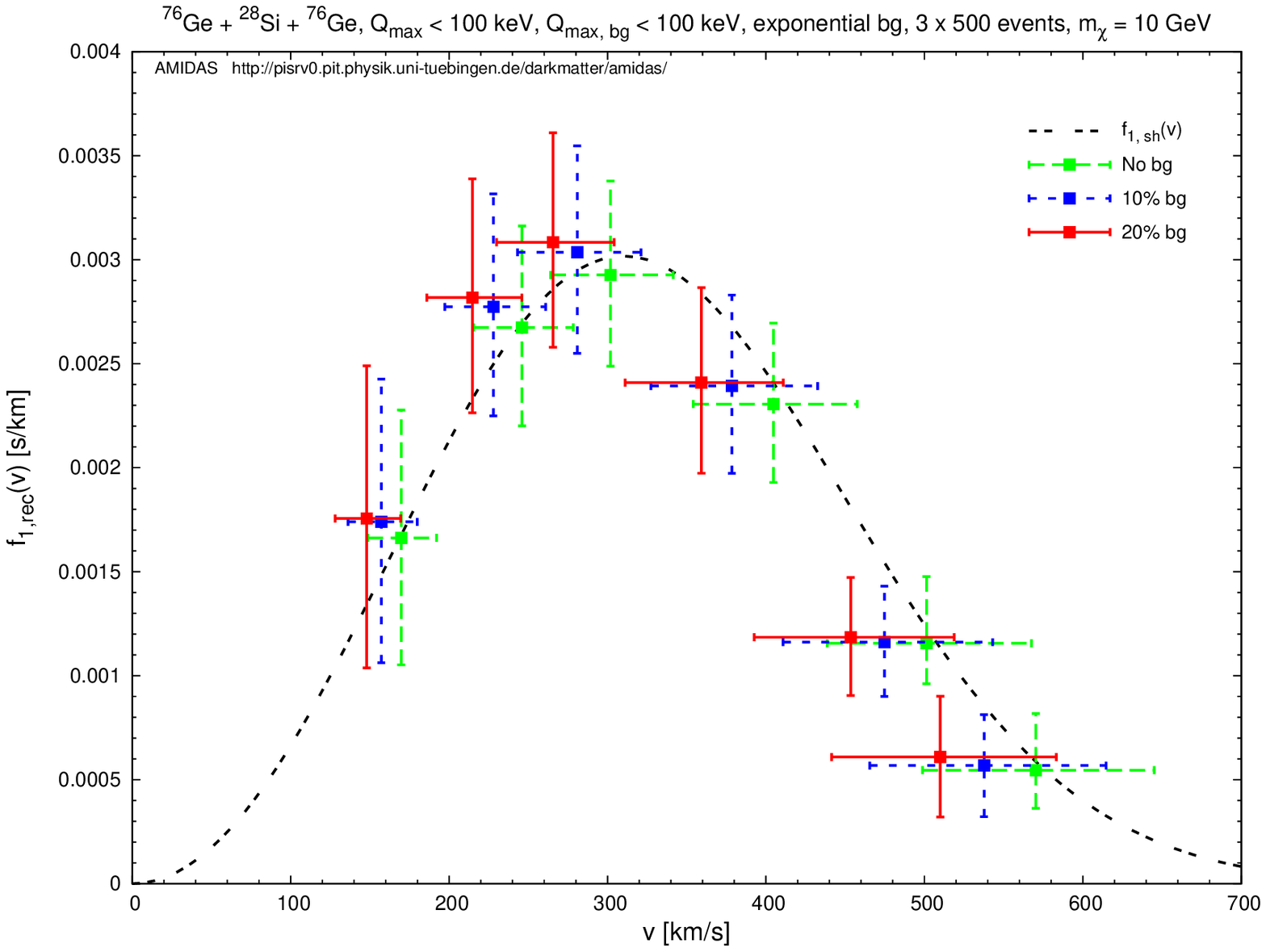} \hspace{-1.1cm}
 \includegraphics[width=9.8cm]{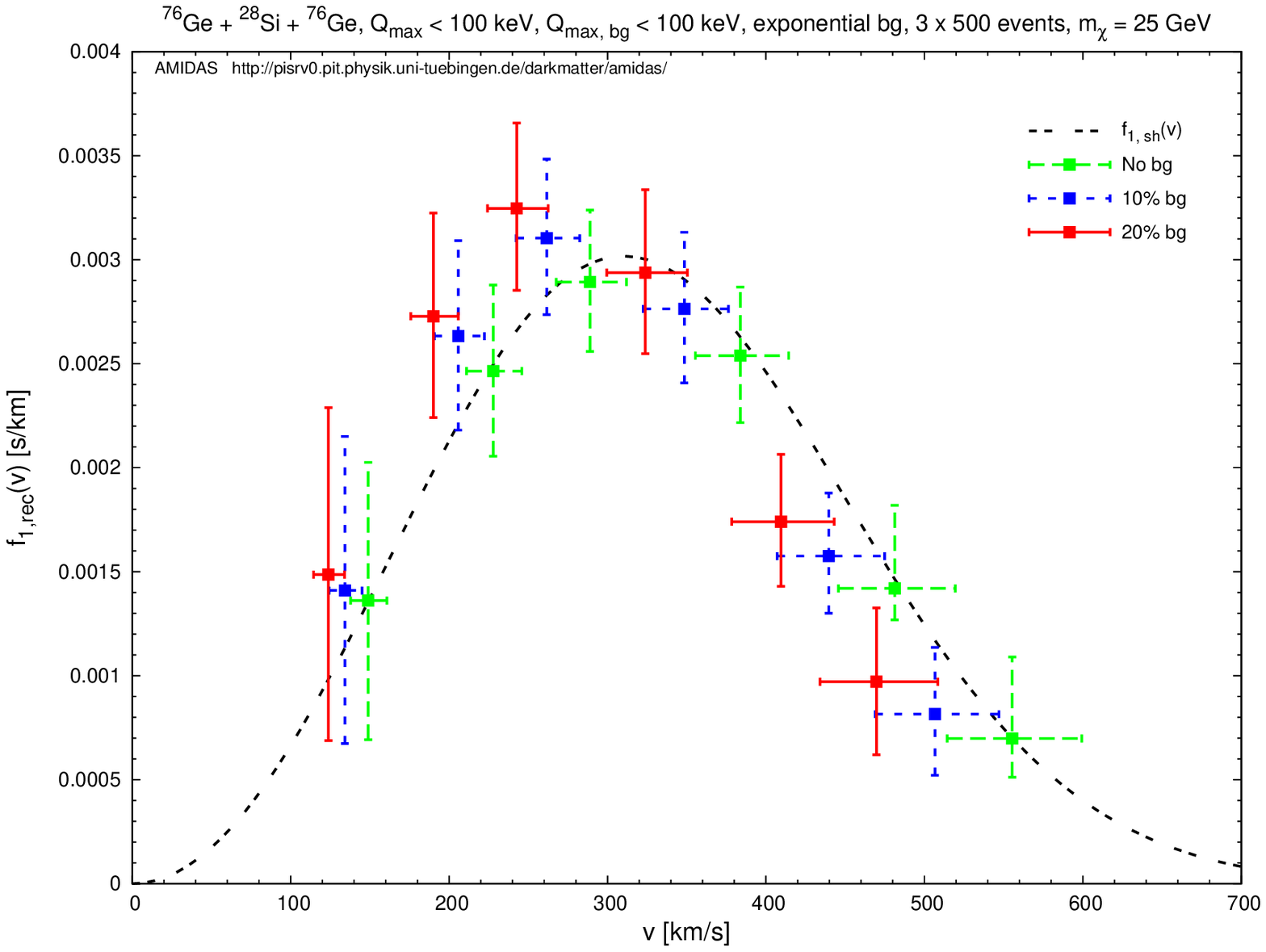} \hspace*{-1.6cm} \\
 \vspace{0.75cm}
 \hspace*{-1.6cm}
 \includegraphics[width=9.8cm]{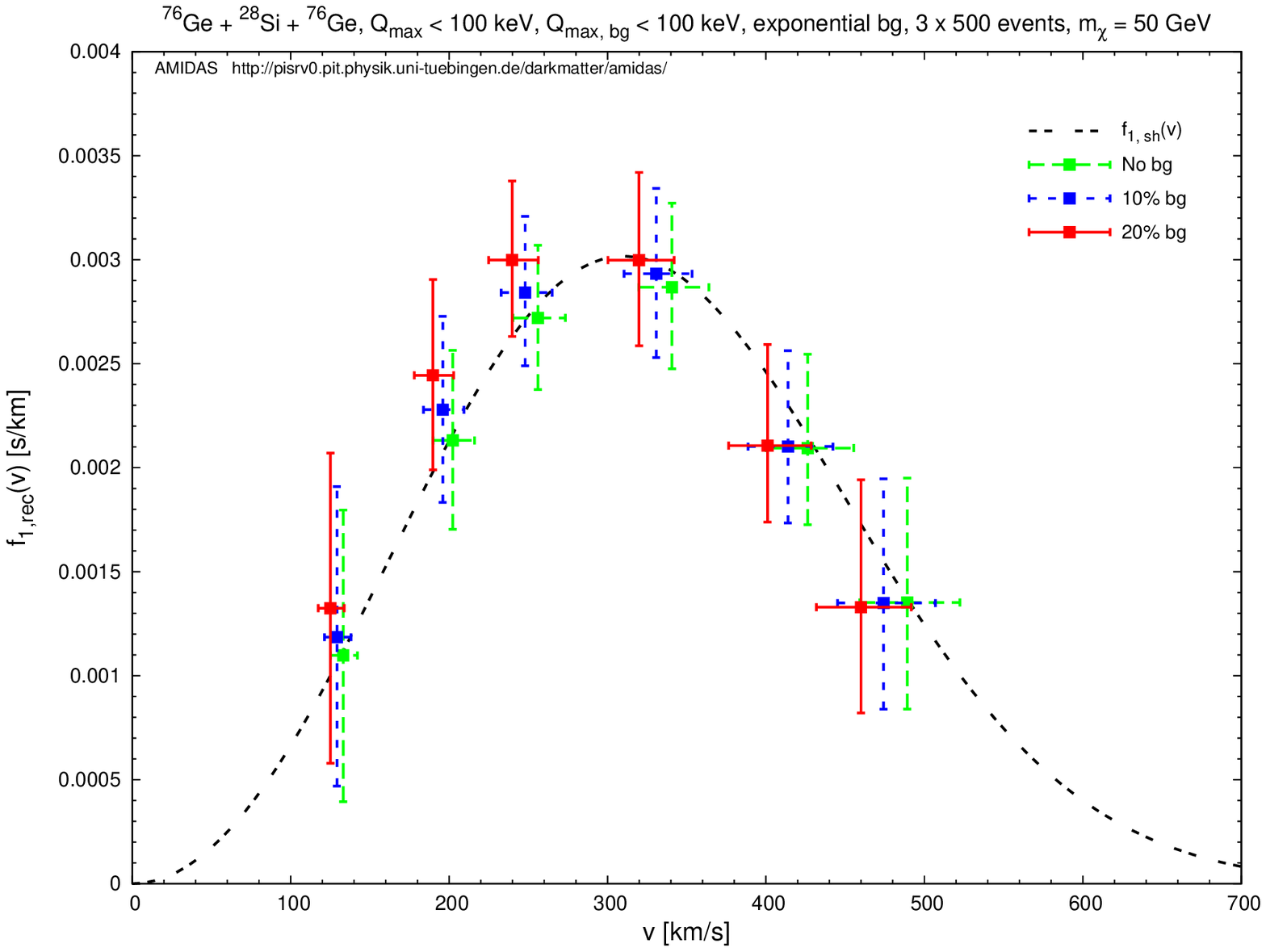} \hspace{-1.1cm}
 \includegraphics[width=9.8cm]{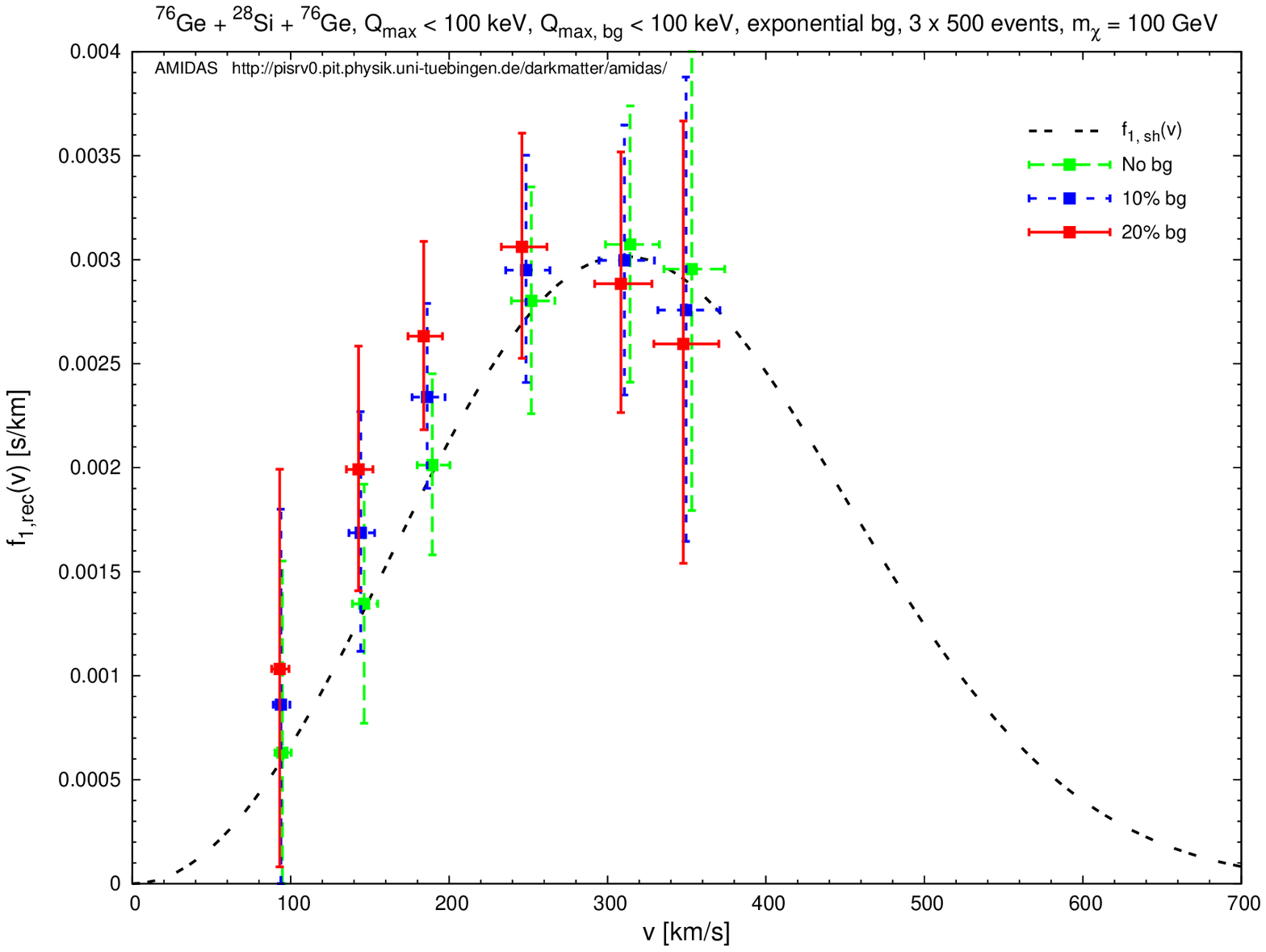} \hspace*{-1.6cm} \\
 \vspace{0.75cm}
 \hspace*{-1.6cm}
 \includegraphics[width=9.8cm]{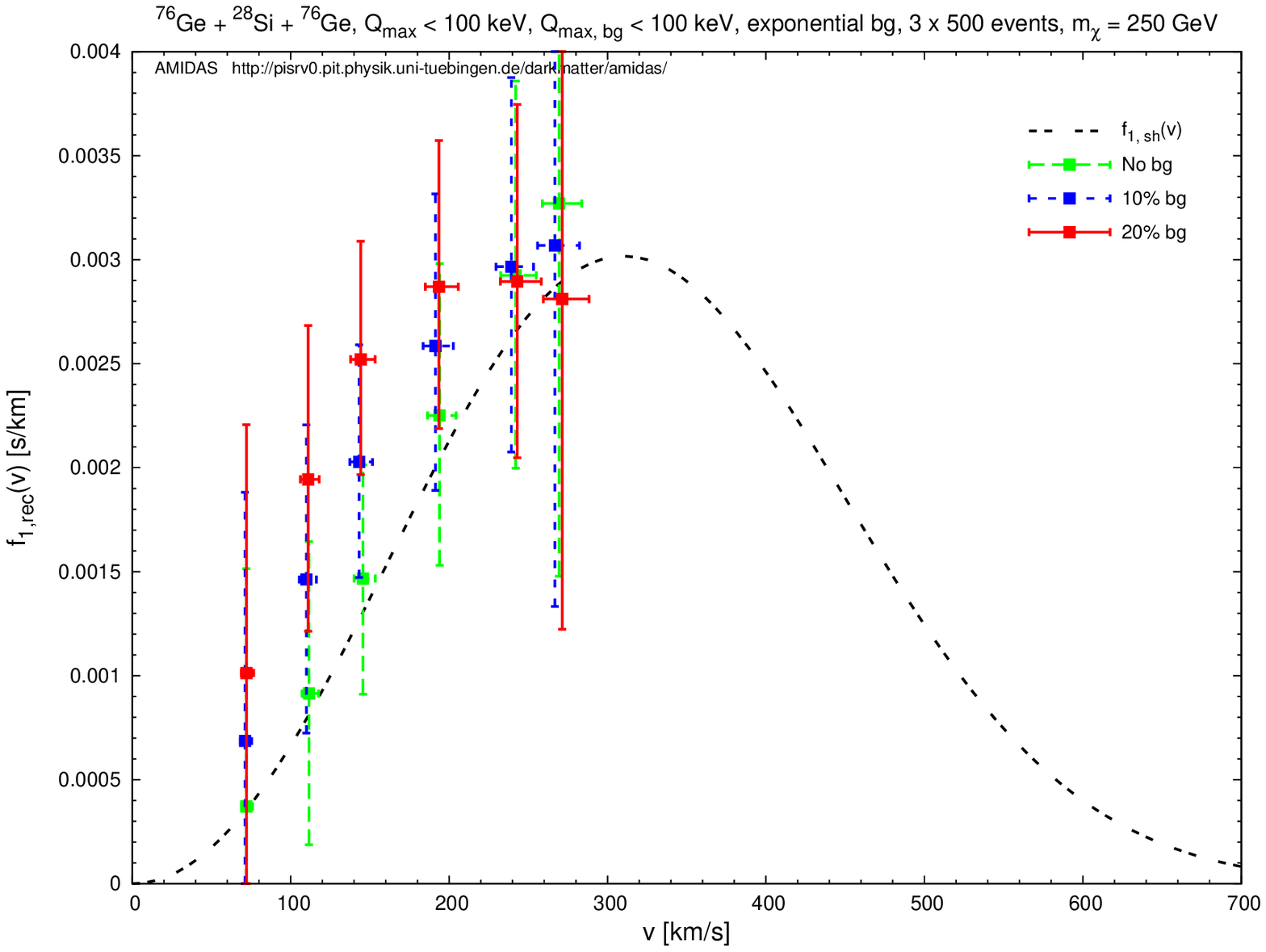} \hspace{-1.1cm}
 \includegraphics[width=9.8cm]{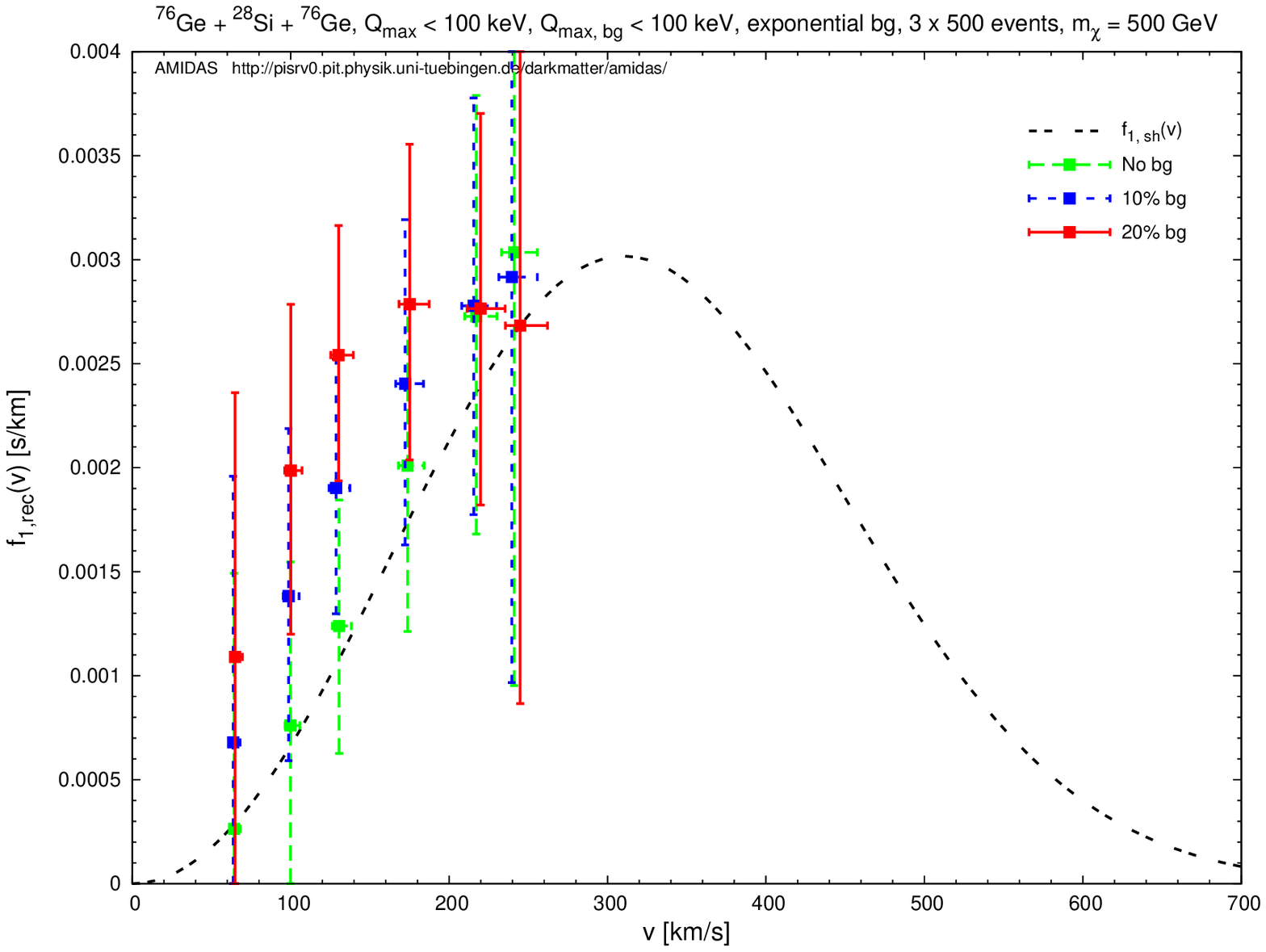} \hspace*{-1.6cm} \\
}
\vspace{-0.5cm}
\end{center}
\caption{
 As in Figs.~\ref{fig:f1v-Ge-ex-000-100-0500},
 except that
 the WIMP masses have been reconstructed
 by means of the procedure introduced
 in Refs.~\cite{DMDDmchi-SUSY07, DMDDmchi}.
 Here the vertical bars show
 the 1$\sigma$ statistical uncertainties
 estimated by Eq.~(\ref{eqn:sigma_f1v_Qsn}),
 while
 the horizontal bars show
 the 1$\sigma$ statistical uncertainties
 on the estimates of $v_{s, \mu}$
 given in Eq.~(\ref{eqn:vsn})
 due to the uncertainty on the reconstructed WIMP mass.
 See the text for further details.
}
\label{fig:f1v-Ge-SiGe-ex-000-100-0500}
\end{figure}
\begin{figure}[p!]
\begin{center}
\imageswitch{
\begin{picture}(16.5,21.5)
\put(0  ,15  ){\framebox(8,6.5){f1v-Ge-SiGe-const-000-100-010-0500}}
\put(8.5,15  ){\framebox(8,6.5){f1v-Ge-SiGe-const-000-100-025-0500}}
\put(0  , 7.5){\framebox(8,6.5){f1v-Ge-SiGe-const-000-100-050-0500}}
\put(8.5, 7.5){\framebox(8,6.5){f1v-Ge-SiGe-const-000-100-100-0500}}
\put(0  , 0  ){\framebox(8,6.5){f1v-Ge-SiGe-const-000-100-250-0500}}
\put(8.5, 0  ){\framebox(8,6.5){f1v-Ge-SiGe-const-000-100-500-0500}}
\end{picture}}
{\hspace*{-1.6cm}
 \includegraphics[width=9.8cm]{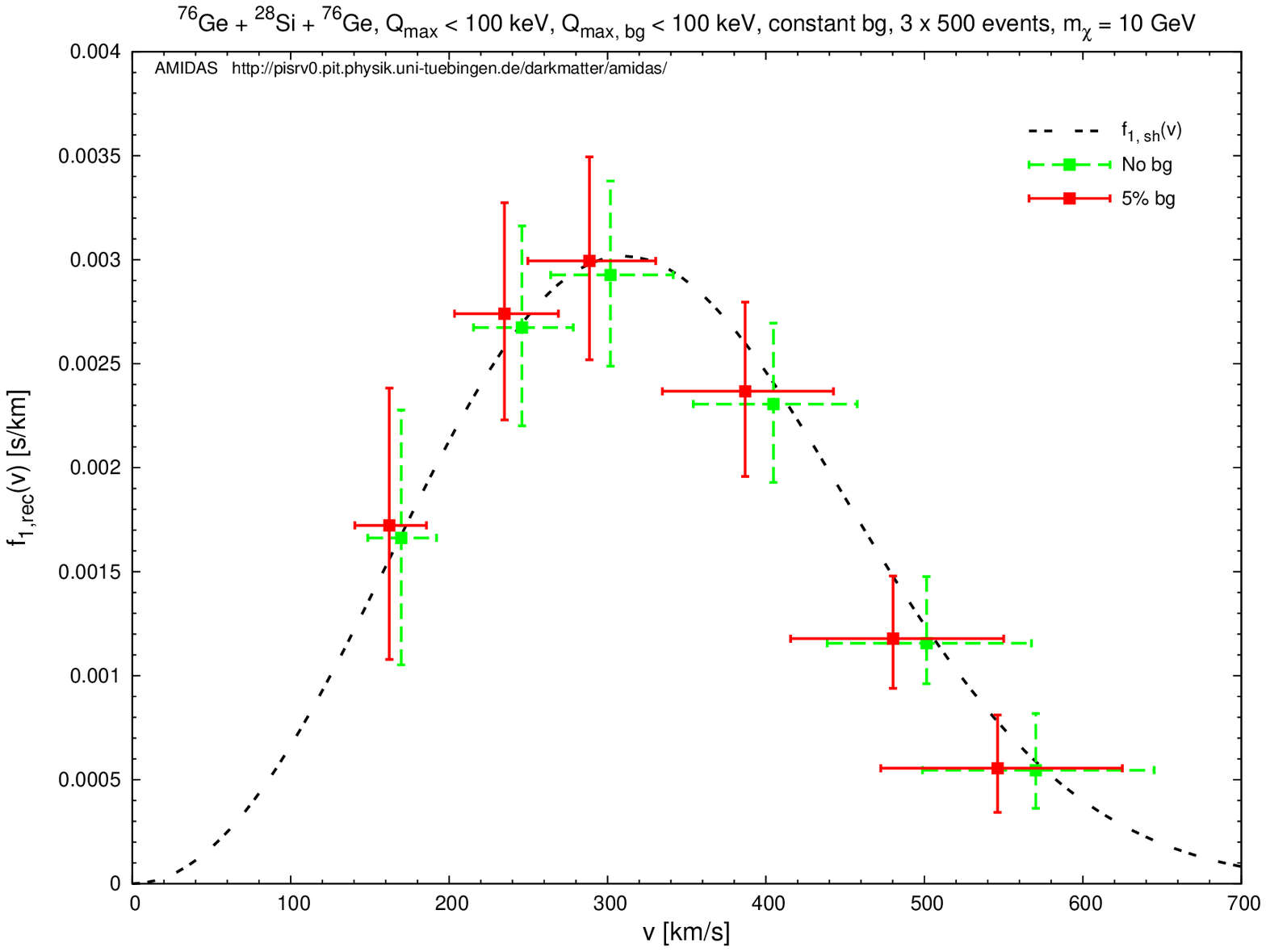} \hspace{-1.1cm}
 \includegraphics[width=9.8cm]{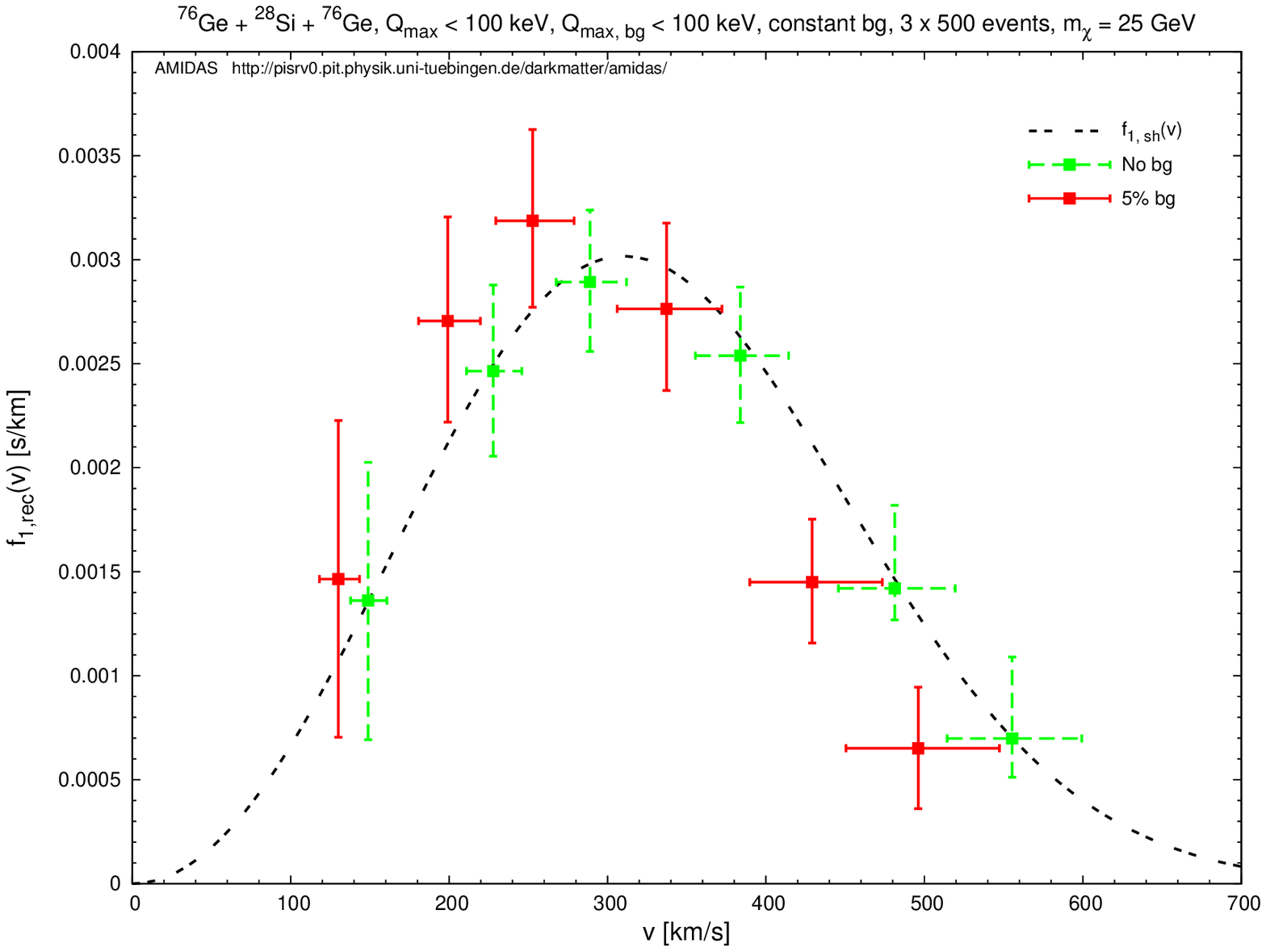} \hspace*{-1.6cm} \\
 \vspace{1cm}
 \hspace*{-1.6cm}
 \includegraphics[width=9.8cm]{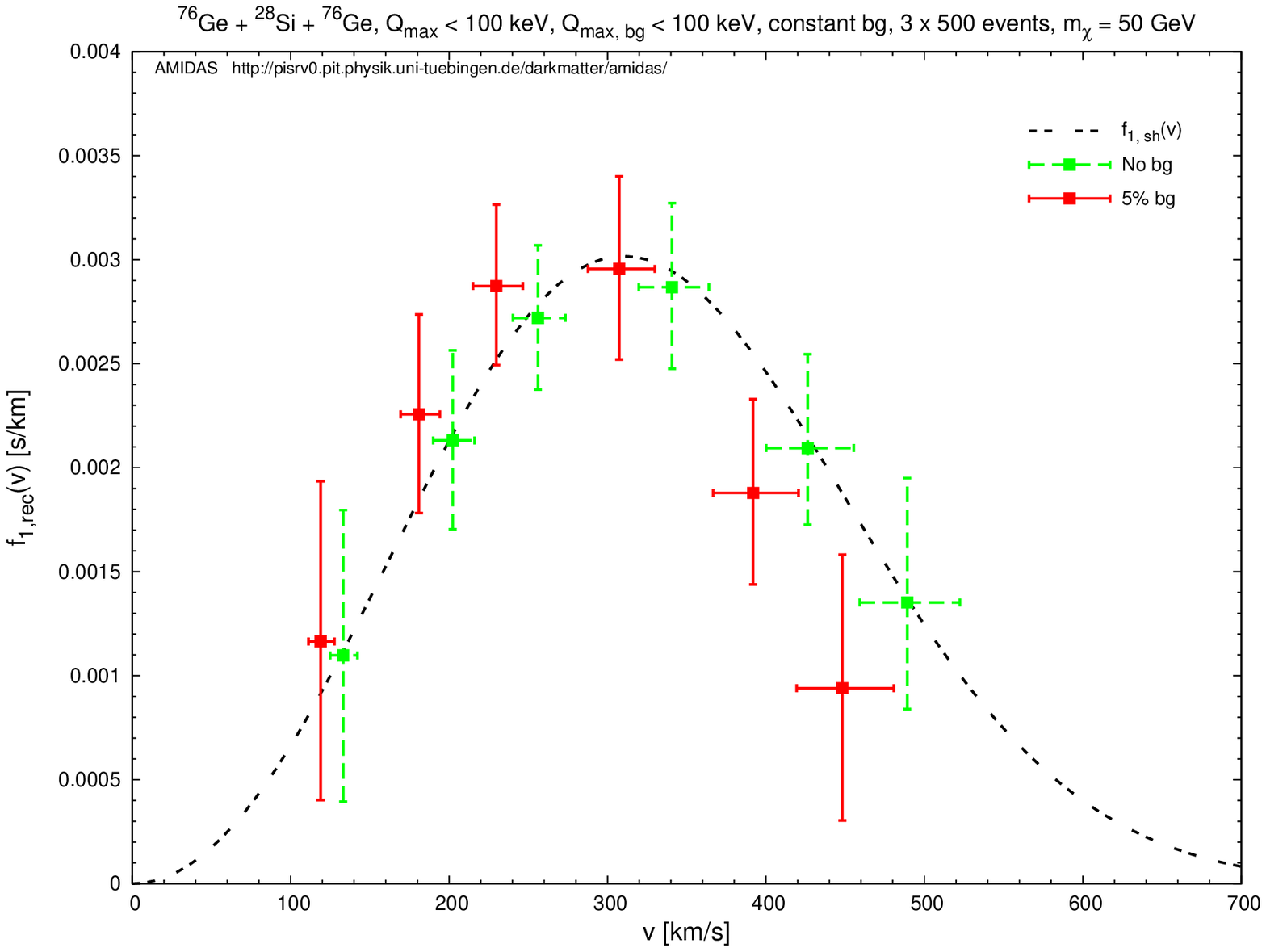} \hspace{-1.1cm}
 \includegraphics[width=9.8cm]{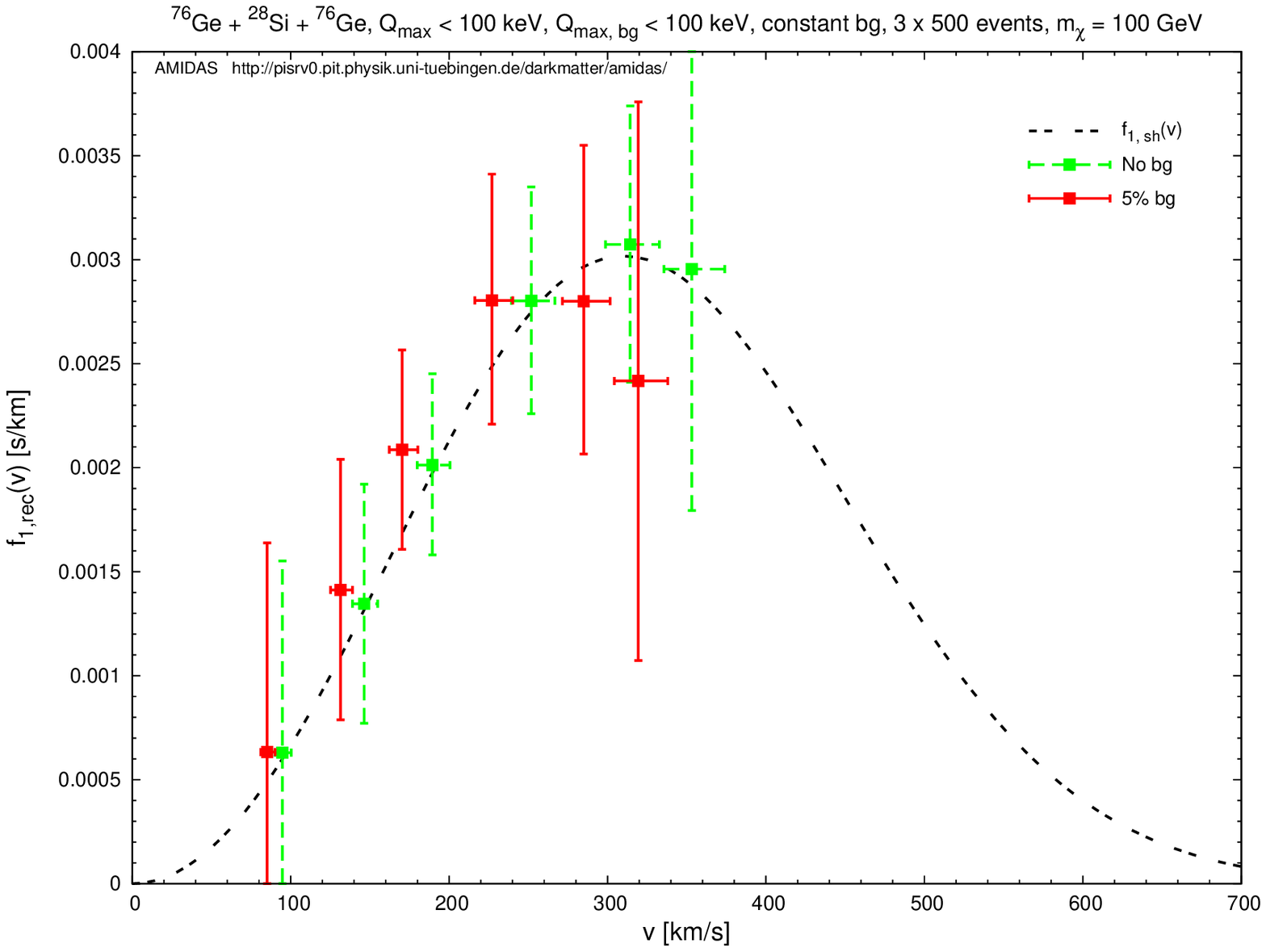} \hspace*{-1.6cm} \\
 \vspace{1cm}
 \hspace*{-1.6cm}
 \includegraphics[width=9.8cm]{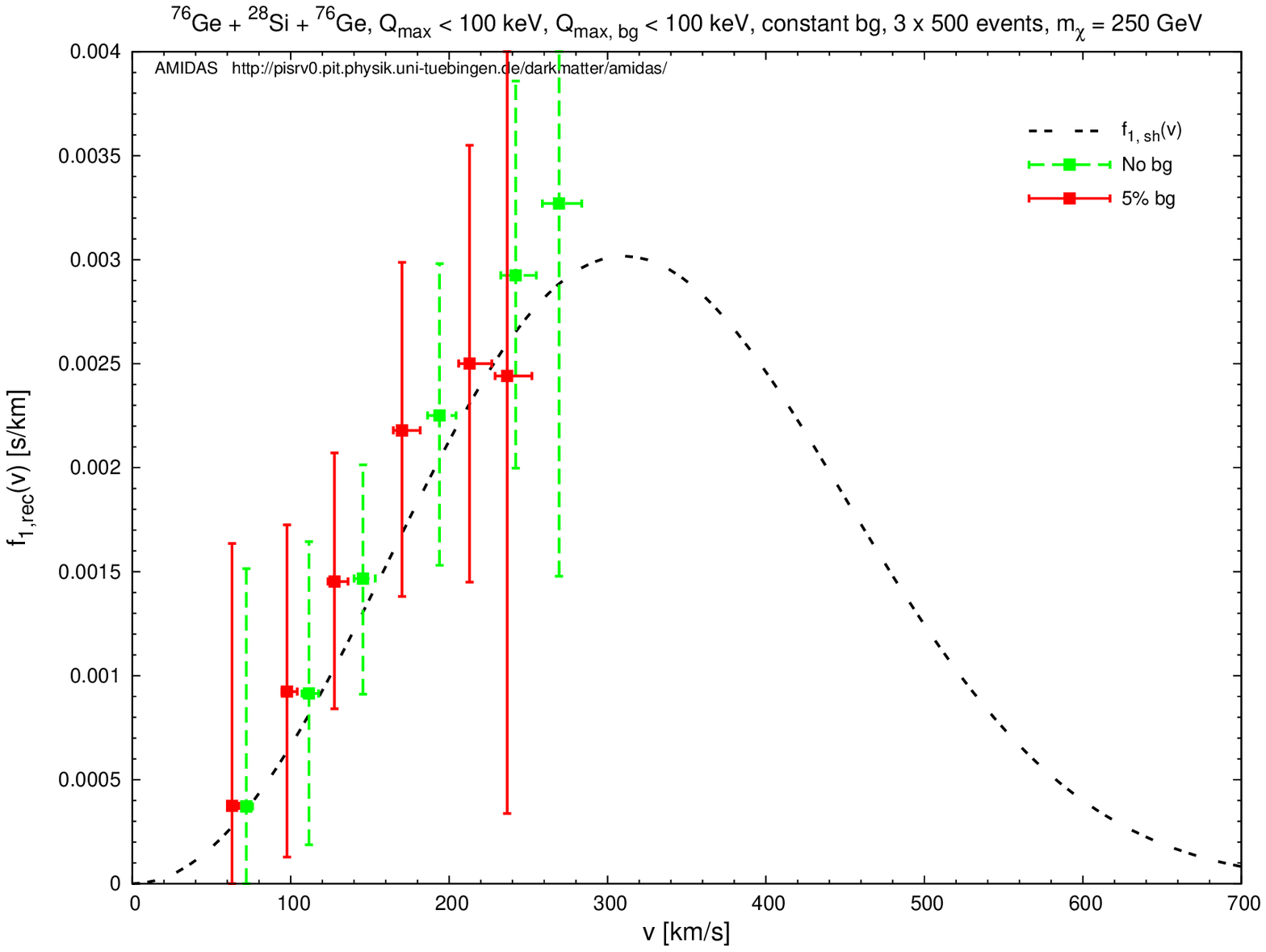} \hspace{-1.1cm}
 \includegraphics[width=9.8cm]{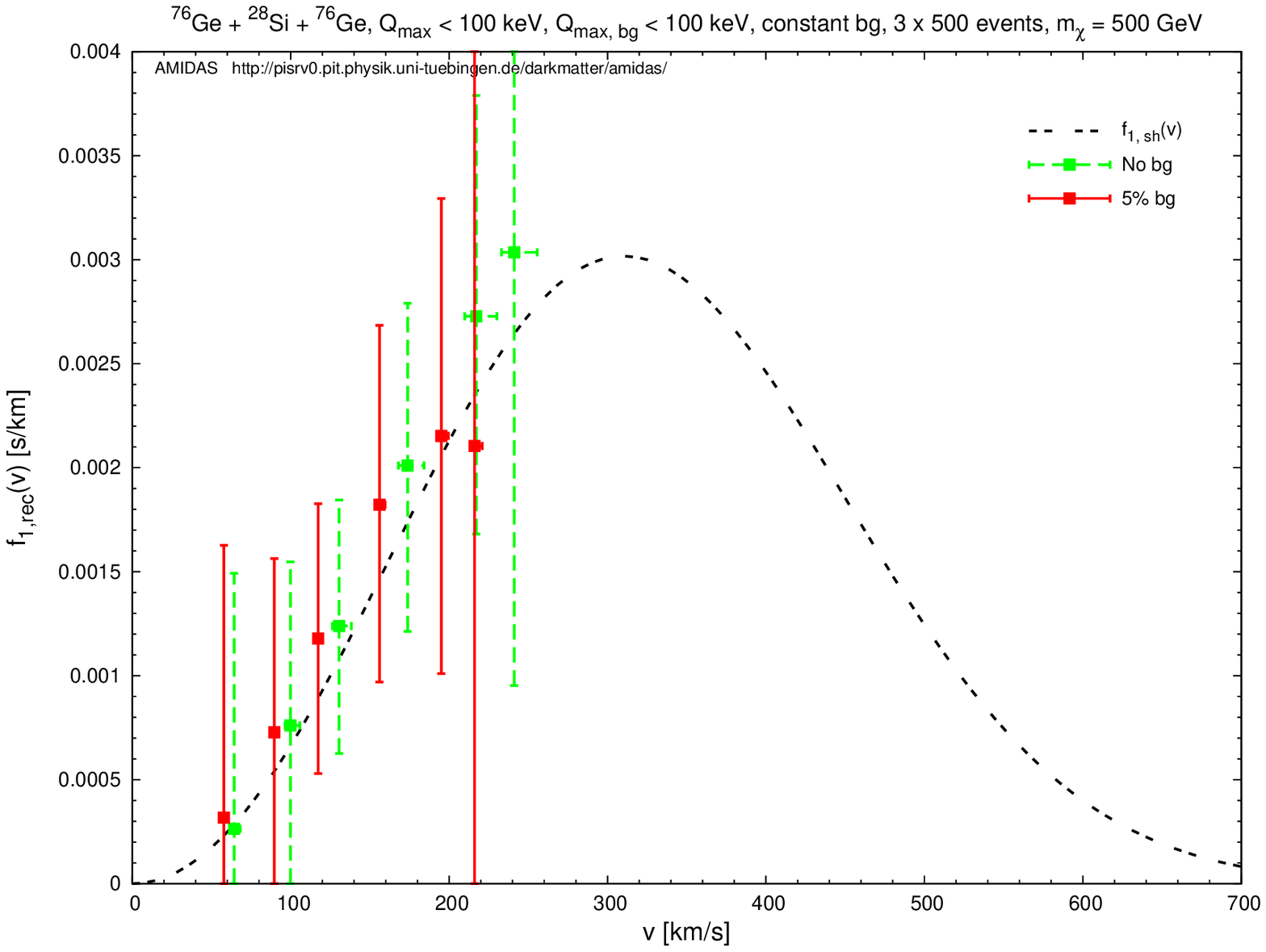} \hspace*{-1.6cm} \\
}
\vspace{-0.25cm}
\end{center}
\caption{
 As in Figs.~\ref{fig:f1v-Ge-SiGe-ex-000-100-0500},
 except that
 the constant background spectrum
 in Eq.~(\ref{eqn:dRdQ_bg_const})
 has been used.
 Note that
 the solid red lines here
 indicate a background ratio of 5\%.
}
\label{fig:f1v-Ge-SiGe-const-000-100-0500}
\end{figure}

 In this subsection,
 the required WIMP mass
 for estimating the reconstructed points $v_{s, \mu}$
 as well as the normalization constant $\calN$
 by Eqs.~(\ref{eqn:vsn}) and (\ref{eqn:calN_sum})
 has been reconstructed
 with other direct detection experiments%
\footnote{
 In order to avoid calculations of
 the correlations between $\mchi$ and $f_{1, {\rm rec}}(v_{s, \mu})$,
 we have assumed here that
 the two data sets using Ge detectors
 are independent of each other.
}.
 Note that
 the statistical uncertainty on $f_{1, {\rm rec}}(v_{s, \mu})$
 estimated as the diagonal entries of the covariance matrix
 given in Eq.~(\ref{eqn:cov_f1v_Qs_mu})
 must thus be modified by taking into account
 the statistical uncertainty on
 the reconstructed WIMP mass $\sigma(\mchi)$ to
\beq
   \sigma^2\aBig{f_{1, {\rm rec}}(v_{s, \mu})}
 = {\rm cov}\aBig{f_{1, {\rm rec}}(v_{s, \mu}), f_{1, {\rm rec}}(v_{s, \mu})}
  + \frac{\mN}{2}
    \bfrac{f_{1, {\rm rec}}(v_{s, \mu})}{\alpha \mchi^2}^2
    \sigma^2(\mchi)
\~.
\label{eqn:sigma_f1v_Qsn}
\eeq
 In Figs.~\ref{fig:f1v-Ge-SiGe-ex-000-100-0500}
 I show the numerical results
 with six different input WIMP masses,
 as shown in Figs.~\ref{fig:f1v-Ge-ex-000-100-0500}.
 Note that,
 while the vertical bars show
 the 1$\sigma$ statistical uncertainties
 estimated by Eq.~(\ref{eqn:sigma_f1v_Qsn}),
 the horizontal bars shown here
 indicate the 1$\sigma$ statistical uncertainties
 on the estimates of $v_{s, \mu}$
 given in Eq.~(\ref{eqn:vsn})
 due to the uncertainty on the reconstructed WIMP mass;
 the statistical and systematic uncertainties
 due to estimating of $Q_{s, \mu}$
 have been neglected here.

 It can be seen that,
 firstly,
 as shown in Figs.~\ref{fig:mchi-SiGe-ex-000-100},
 for an input WIMP mass of 100 GeV,
 the reconstructed mass doesn't differ very much
 from the true value.
 Hence,
 the reconstructed $f_{1, {\rm rec}}(v_{s, \mu})$
 is approximately the same
 for both cases with the input/reconstructed WIMP mass.
 However,
 for light input masses ($\mchi~\lsim~100$ GeV),
 the reconstructed $f_{1, {\rm rec}}(v_{s, \mu})$
 with the reconstructed WIMP mass
 shift to relatively {\em lower} velocities
 compared to the case
 with the input (true) WIMP mass.
 This effect caused directly by
 the overestimate of the reconstructed WIMP mass.
 The coefficient $\alpha$
 defined in Eq.~(\ref{eqn:alpha})
 can be rewritten as
\beq
   \alpha
 = \frac{1}{\sqrt{2 \mN}} \abrac{1 + \frac{\mN}{\mchi}}
\~.
\label{eqn:alpha_mchi}
\eeq
 This implies that,
 once the reconstructed WIMP mass is
 over-/underestimated from the real value,
 the coefficient $\alpha$
 will thus be under-/overestimated.
 Consequently,
 $v_{s, \mu}$ determined by Eq.~(\ref{eqn:vsn})
 will be smaller/larger than the true values.

 Note that,
 firstly,
 this second effect,
 in contrast to the first one
 discussed in the previous subsection,
 draws the reconstructed WIMP velocity distribution
 to the opposite directions;
 i.e., to {\em lower/higher} velocities
 if WIMPs are {\em light/heavy}.
 Secondly,
 for heavier WIMP masses,
 as shown in Figs.~\ref{fig:mchi-SiGe-ex-000-100},
 with a small background fraction,
 the underestimate of the reconstructed WIMP mass
 and thus the shift of the velocity distribution function
 seem not to be significant.
 Nevertheless,
 the simulation results shown
 in Figs.~\ref{fig:f1v-Ge-SiGe-ex-000-100-0500}
 indicate that,
 with an $\sim$ {\em 5\% -- 10\%} background ratio
 (i.e., $\sim$ 25 -- 50 events)
 in the analyzed data sets of $\sim$ 500 total events,
 one could in principle still reconstruct
 the one--dimensional velocity distribution function of halo WIMPs
 with an \mbox{$\sim$ 7\%}
 (for 25 GeV WIMPs, 10\% backgrounds)
 -- $\sim$ 16\%
 (for 250 GeV WIMPs, 5\% backgrounds)
 deviation%
\footnote{
 Remind that,
 as shown in the lower frame of
 Figs.~\ref{fig:mchi-SiGe-ex-000-100},
 with two data sets of $\sim$ 500 total events each
 and a background ratio of $\sim$ 10\%,
 the WIMP mass could in principle be reconstructed
 with a statistical uncertainty of
 $\sim$ 10\% (for  25 GeV WIMPs) --
 $\sim$ 25\% (for 250 GeV WIMPs).
}.
 If the mass of WIMPs is light
 ($\mchi~\lsim~100$ GeV),
 the maximal acceptable background ratio
 could even be as large as $\sim$ 20\%
 ($\sim$ 100 events)
 with a deviation of only $\sim$ 9\%.

\begin{figure}[p!]
\begin{center}
\imageswitch{
\begin{picture}(16.5,21)
\put(0  ,14.5 ){\framebox(8,6.5){f1v-Ge-ex-000-100-010-5000}}
\put(8.5,14.5 ){\framebox(8,6.5){f1v-Ge-ex-000-100-025-5000}}
\put(0  , 7.25){\framebox(8,6.5){f1v-Ge-ex-000-100-050-5000}}
\put(8.5, 7.25){\framebox(8,6.5){f1v-Ge-ex-000-100-100-5000}}
\put(0  , 0   ){\framebox(8,6.5){f1v-Ge-ex-000-100-250-5000}}
\put(8.5, 0   ){\framebox(8,6.5){f1v-Ge-ex-000-100-500-5000}}
\end{picture}}
{\hspace*{-1.6cm}
 \includegraphics[width=9.8cm]{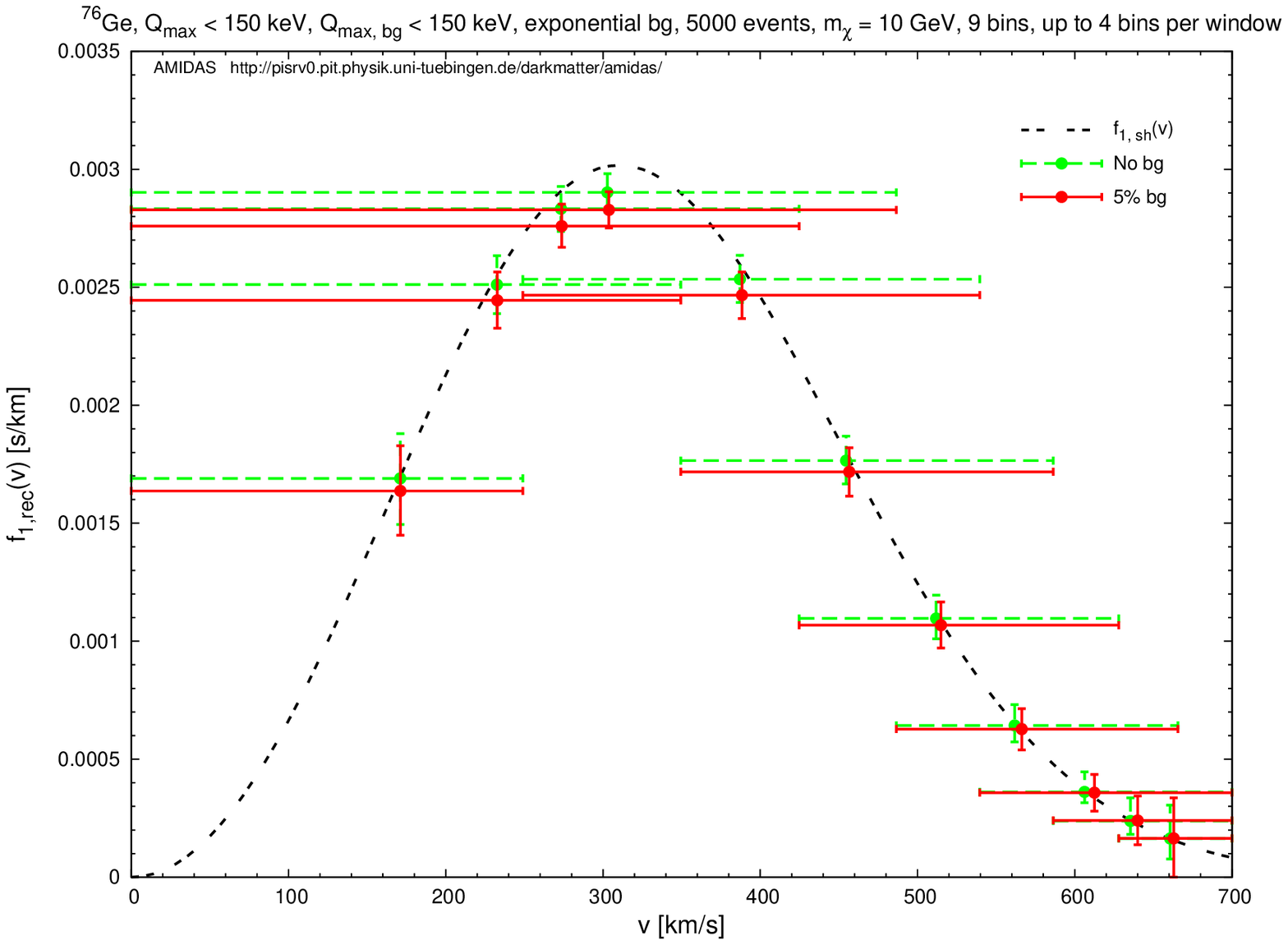} \hspace{-1.1cm}
 \includegraphics[width=9.8cm]{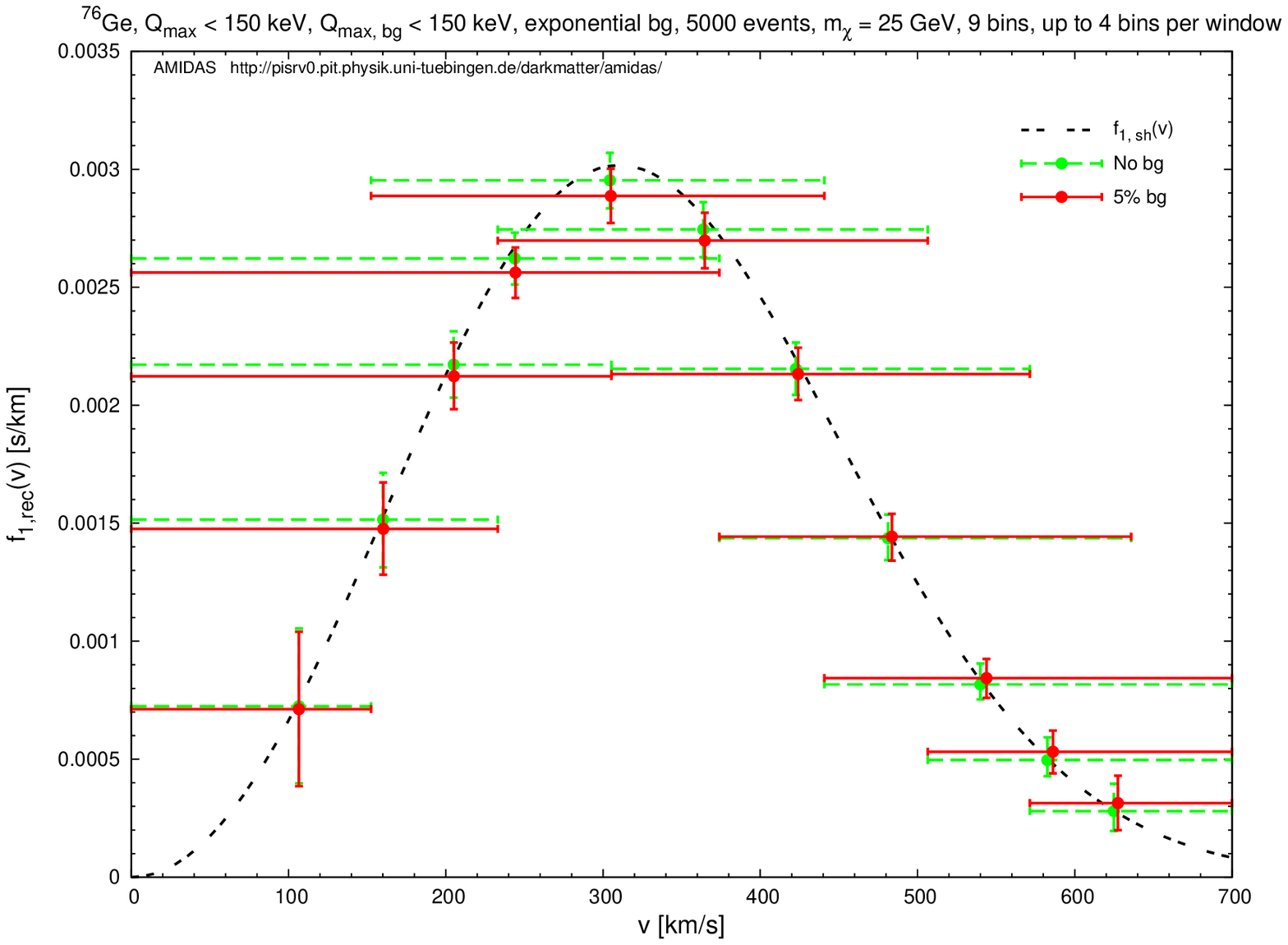} \hspace*{-1.6cm} \\
 \vspace{0.75cm}
 \hspace*{-1.6cm}
 \includegraphics[width=9.8cm]{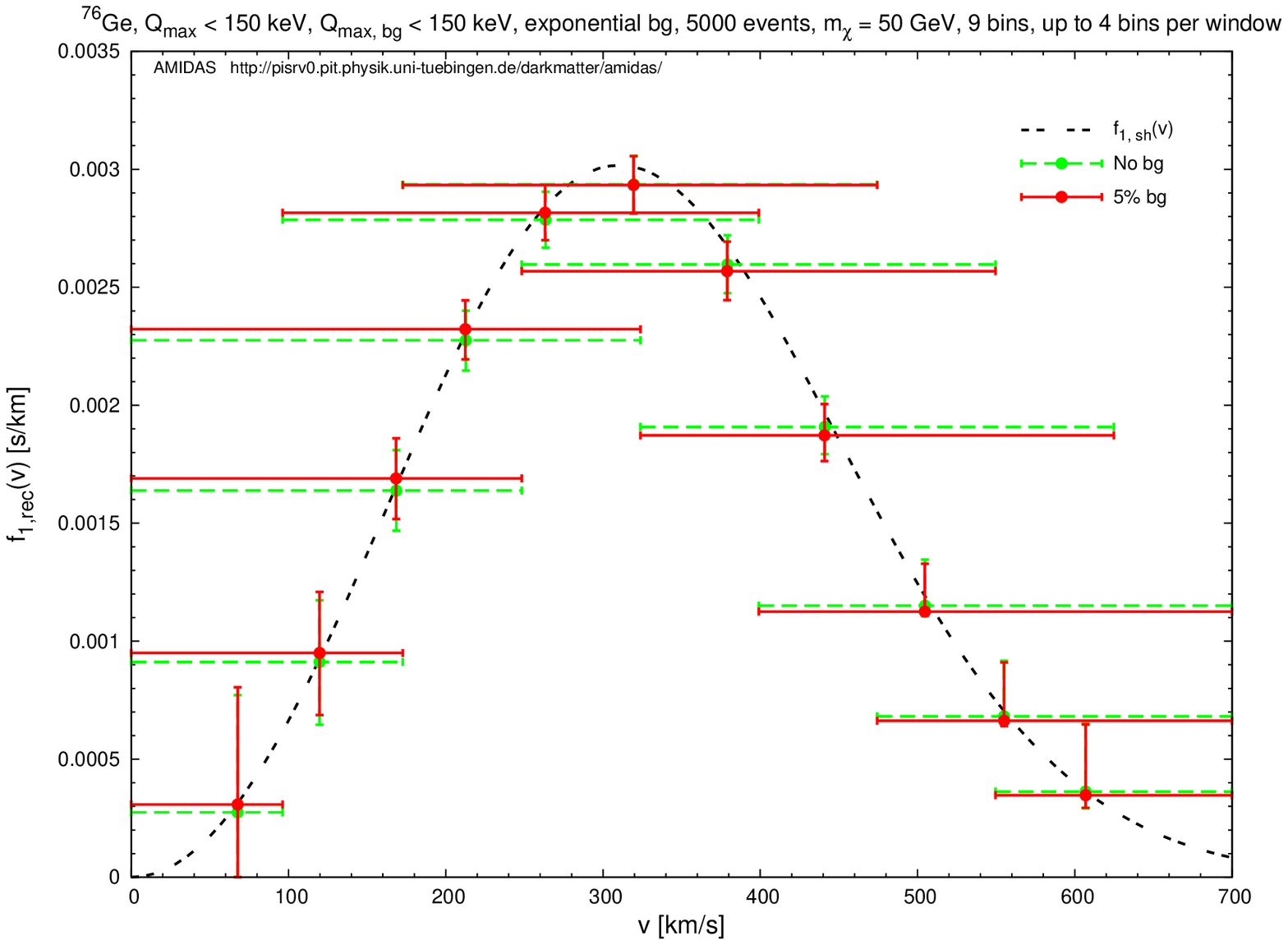} \hspace{-1.1cm}
 \includegraphics[width=9.8cm]{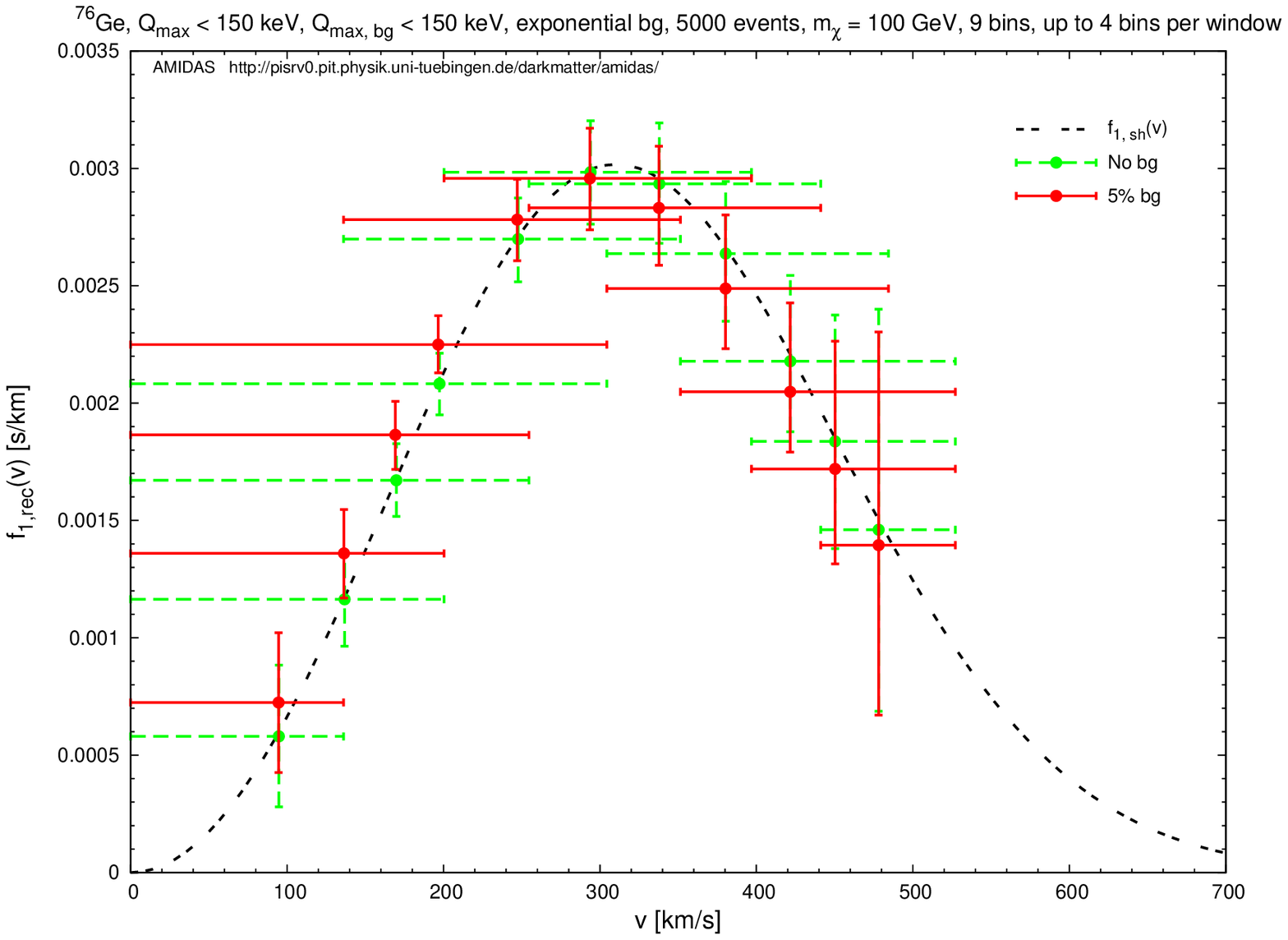} \hspace*{-1.6cm} \\
 \vspace{0.75cm}
 \hspace*{-1.6cm}
 \includegraphics[width=9.8cm]{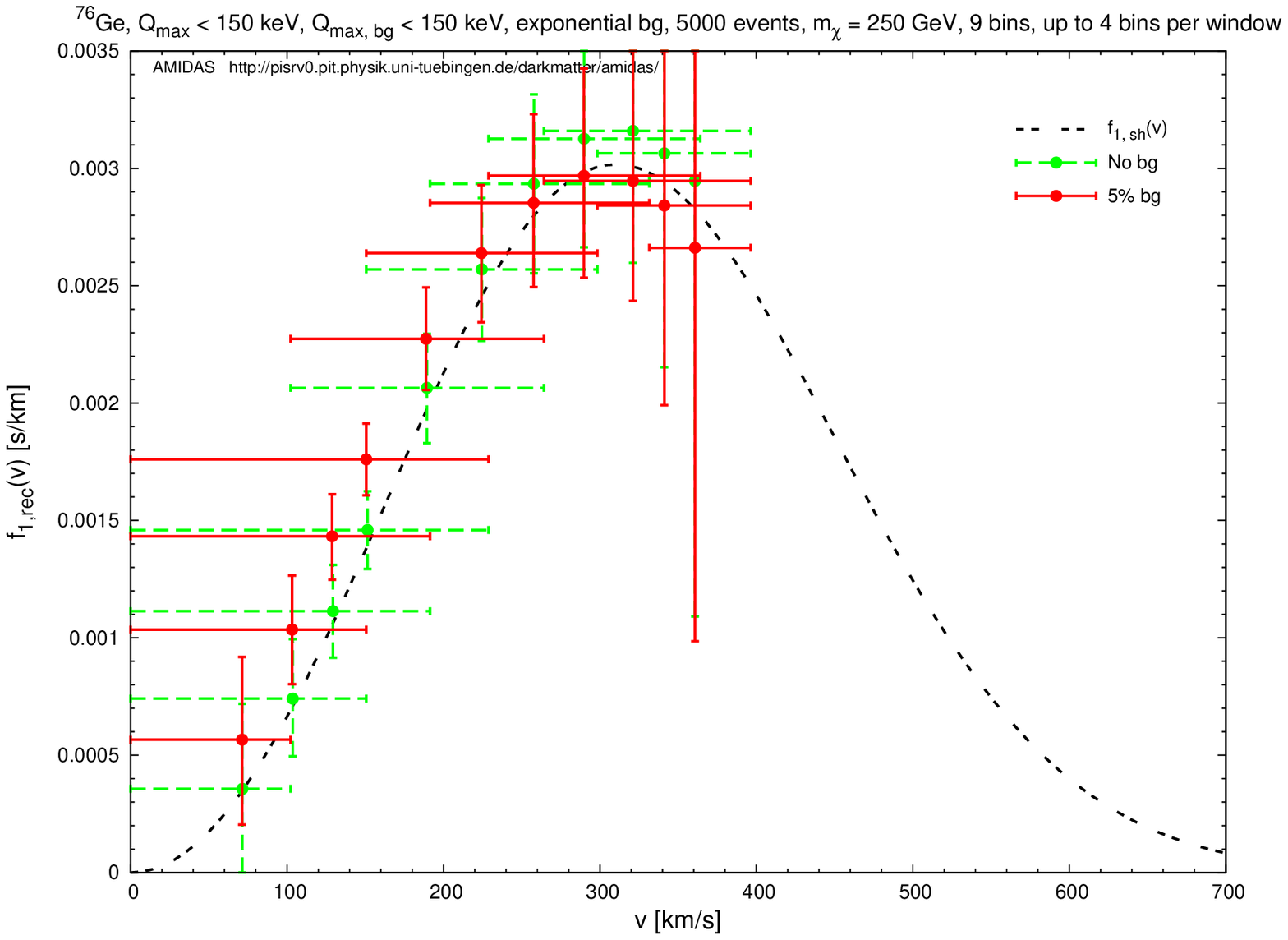} \hspace{-1.1cm}
 \includegraphics[width=9.8cm]{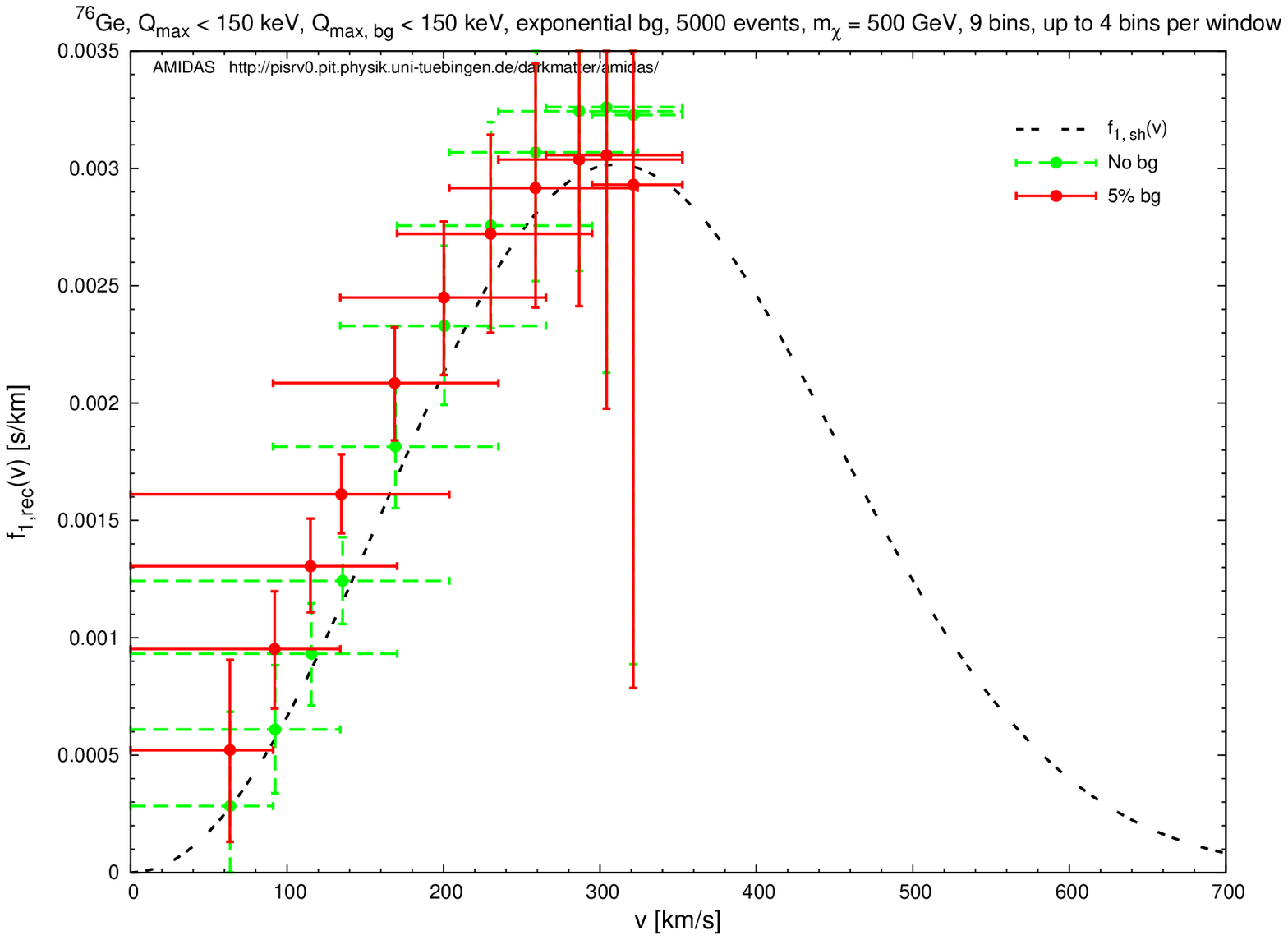} \hspace*{-1.6cm} \\
}
\vspace{-0.25cm}
\end{center}
\caption{
 As in Figs.~\ref{fig:f1v-Ge-ex-000-100-0500},
 except that
 the expected number of total events
 in each experiment
 has been risen to 5,000
 and the experimental maximal cut--off energies
 for WIMP signals and background windows
 have been extended to 150 keV.
 Note that
 the solid red lines here
 indicate a background ratio of 5\%. 
}
\label{fig:f1v-Ge-ex-000-100-5000}
\end{figure}

 In Figs.~\ref{fig:f1v-Ge-SiGe-const-000-100-0500}
 we again use the constant spectrum
 for residue backgrounds
 and the WIMP mass has been reconstructed
 by using other mixed data sets.
 For this case,
 as shown in Ref.~\cite{DMDDbg-mchi},
 the WIMP mass would be overestimated
 for all input masses.
 And, consequently,
 the reconstructed WIMP velocity distribution
 for all six input masses
 are shifted to lower velocity ranges.
 Nevertheless,
 as for the case with a known WIMP mass
 shown in Figs.~\ref{fig:f1v-Ge-const-000-100-0500},
 data sets with background fractions of $\lsim$ {\em 5\%}
 could in principle be used to
 at least give a rough outline of
 the WIMP velocity distribution (for $\mchi~\gsim~100$ GeV),
 or even reconstruct the distribution pretty well
 (for $\mchi~\lsim~100$ GeV).

 Finally,
 we rise the expected event number
 in each experiment a factor of 10,
 to 5,000 events totally.
 The experimental maximal cut--off energies
 for WIMP signals and background windows
 have also been extended to 150 keV.
 Nine bins have been used%
\footnote{
 For the input WIMP masses of 10/25/50 GeV,
 the widths of the first bin
 have been set as 1.5/2.5/2.5 keV.
} and
 up to four bins have been combined to a window.
 Since with 2 $\times$ 5,000 events
 we could in principle determine the WIMP mass
 with an uncertainty of $\lsim~5\%$,
 I show only the results of
 the reconstructed WIMP velocity distribution function
 with the input WIMP mass
 in Figs.~\ref{fig:f1v-Ge-ex-000-100-5000}.
 It can be seen clearly that,
 by using a data set of $\cal O$(5,000) events
 with a maximal background ratio of $\lsim$ {\em 5\%}
 ($\cal O$(250) events),
 we could in principle reconstruct
 the WIMP velocity distribution function
 in the velocity range $v~\lsim~{\cal O}$(500 km/s)
 with a deviation of $\lsim~6\%$
 (for a WIMP mass of 100 GeV).
 Once WIMPs are light ($\mchi~\lsim~{\cal O}$(50 GeV)),
 this deviation could even be
 reduced to $\lsim$ {\em 2.5\%}.
\section{Summary and conclusions}
 In this paper
 we reexamine the model--independent data analysis method
 introduced in Ref.~\cite{DMDDf1v}
 for the reconstruction of
 the one--dimensional velocity distribution function
 of Weakly Interacting Massive Particles
 from data (measured recoil energies) of
 direct Dark Matter detection experiments directly
 by taking into account a fraction of residue background events,
 which pass all discrimination criteria and
 then mix with other real WIMP--induced events
 in the analyzed data sets.
 This method requires neither prior knowledge
 about the WIMP scattering spectrum
 nor about different possible background spectra;
 the unique needed information is the recoil energies
 recorded in direct detection experiments
 and (eventually) the mass of incident WIMPs.

 For the mass of incident WIMPs
 required as an unique input information
 in this method,
 we first assumed that
 it could be known precisely
 with a negligible uncertainty
 from other (e.g., collider) experiments.
 Our simulations show that,
 assuming an exponential form for
 the residue background spectrum,
 with a data set of $\sim$ 500 total events,
 and a background ratio of $\sim$ 10\% -- 20\%,
 the WIMP velocity distribution function
 could in principle be reconstructed
 with an \mbox{$\sim$ 6.5\%}
 (for a 25 GeV WIMP mass, 20\% background events)
 -- $\sim$ 38\%
 (for a 250 GeV WIMP mass, 10\% background events)
 deviation.
 If the WIMP mass is $\cal O$(50 GeV),
 the maximal acceptable background ratio
 could be risen to $\sim$ 40\%
 with a deviation of only $\sim$ 14\%.

 Moreover,
 for lighter/heavier WIMP masses,
 since the relatively flatter/sharper background spectrum
 could contribute relatively more events
 to high/low energy ranges,
 the reconstructed velocity distribution
 could therefore be shifted to higher/lower velocities.
 However,
 since for (very) light WIMPs ($\mchi~\lsim~40$ GeV),
 the kinematic maximal cut--off of the recoil energy
 due to the Galactic escape velocity
 is (much) lower than the experimental cut--off,
 a (large) fraction of background events
 in high energy ranges
 could thus in practice be neglected,
 and the shift could not be very significant
 for WIMPs lighter than $\sim$ 50 GeV.

 Since a model--independent method
 for determining the WIMP mass
 by using two experimental data sets
 with two different target nuclei
 has also been developed
 \cite{DMDDmchi-SUSY07, DMDDmchi},
 we considered in this paper also the case that
 the velocity distribution function is reconstructed
 with a reconstructed WIMP mass
 from other direct detection experiments.
 Our simulations show that,
 since lighter/heavier WIMP masses
 could be over-/underestimated
 by using this method
 with background--mixed data sets
 \cite{DMDDbg-mchi},
 the reconstructed points of the velocity distribution
 would thus be shifted to lower/higher velocities,
 the opposite direction of the shift
 due purely to the background contribution
 to high/low energy ranges.
 Although this second effect shifts
 the reconstructed velocity distribution
 (much) more strongly,
 with $\sim$ 5\% -- 10\% background events
 mixed in the analyzed data sets,
 the WIMP velocity distribution function
 could in principle still be reconstructed
 with an $\sim$ 7\%
 (for 25 GeV WIMPs, 10\% backgrounds)
 -- $\sim$ 16\%
 (for 250 GeV WIMPs, 5\% backgrounds)
 deviation.
 If the WIMP mass is $\lsim~{\cal O}$(100 GeV),
 the maximal acceptable background ratio
 could even be as large as $\sim$ 20\%
 with a deviation of only $\sim$ 9\%.

 Furthermore,
 in order to check the need of a prior knowledge about
 an (exact) form of the residue background spectrum,
 a constant spectrum for residue backgrounds
 has also been considered.
 Since the WIMP mass would
 always be overestimated \cite{DMDDbg-mchi},
 the reconstructed WIMP velocity distribution
 could thus be (strongly) shifted to
 lower velocity ranges.
 However,
 data sets with background fractions of $~\lsim~5\%$
 could in principle be used to
 at least give a rough outline of
 the WIMP velocity distribution (for $\mchi~\gsim~100$ GeV),
 or even reconstruct the distribution pretty well
 (for $\mchi~\lsim~100$ GeV).

 Finally,
 for rather next--to--next generation detectors,
 we considered also the case of 5,000 total events
 and extended the maximal experimental cut--off energies
 for WIMP signals and backgrounds to 150 keV.
 Assuming a maximal background ratio of $\lsim~5\%$,
 our results show that
 the WIMP velocity distribution function
 could in principle be reconstructed
 in the velocity range $v~\lsim~{\cal O}$(500 km/s)
 with a deviation of $\lsim~6\%$
 (for a WIMP mass of 100 GeV).
 Once WIMPs are light ($\mchi~\lsim~{\cal O}$(50 GeV)),
 this deviation could even be reduced to $\lsim~2.5\%$.

 In summary,
 as the second part of
 the study of the effects of residue background events
 in direct Dark Matter detection experiments,
 we considered the reconstruction of
 the velocity distribution function
 of halo WIMPs.
 Our results show that,
 with projected experiments
 using next--generation detectors with
 $10^{-9}$ to $10^{-11}$ pb sensitivities
 \cite{Baudis07a, Drees08, Aprile09a, Gascon09}
 and $< 10^{-6}$ background rejection ability
 \cite{CRESST-bg, 
       EDELWEISS-bg, 
       Lang09b, 
       Ahmed09b}, 
 once one or more experiments with different target nuclei
 could accumulate a few hundreds events
 (in one experiment),
 we could in principle at least give
 a rough outline of
 the WIMP velocity distribution function,
 e.g., an approximate estimate of the location of its peak,
 even though there could be some background events
 mixed in our data sets for the analysis.
 After that,
 by means of
 increased number of observed (WIMP--induced) events
 and improved background discrimination techniques
 \cite{CRESST-bg, 
       EDELWEISS-bg}, 
 the shape and properties of
 the velocity distribution of halo Dark Matter
 should be understood more clearly.
\subsubsection*{Acknowledgments}
 The author would like to thank
 the Physikalisches Institut der Universit\"at T\"ubingen
 for the technical support of the computational work
 demonstrated in this article.
 The author would also like to thank
 the friendly hospitality of the
 Max--Planck--Institut f\"ur Kernphysik in Heidelberg
 where part of this work was completed.
 This work
 was partially supported by
 the National Science Council of R.O.C.~%
 under contract no.~NSC-98-2811-M-006-044
 as well as by
 the Focus Group on Cosmology and Particle Astrophysics,
 National Center of Theoretical Sciences, R.O.C..
\appendix
\setcounter{equation}{0}
\setcounter{figure}{0}
\renewcommand{\theequation}{A\arabic{equation}}
\renewcommand{\thefigure}{A\arabic{figure}}
%
%
%
\section{Formulae needed in Sec.~2}
 Here I list all formulae needed
 for the model--independent method
 described in Sec.~2.
 Detailed derivations and discussions
 can be found in Refs.~\cite{DMDDf1v, PhD}.

 First,
 by using the standard Gaussian error propagation,
 the expression for the error
 on the logarithmic slope $k_n$ can be given
 from Eq.~(\ref{eqn:bQn}) directly as
\beq
   \sigma^2(k_n)
 = k_n^4
   \cbrac{  1
          - \bfrac{k_n b_n / 2}{\sinh (k_n b_n / 2)}^2}^{-2}
            \sigma^2\abrac{\bQn}
\~,
\label{eqn:sigma_kn}
\eeq
 with
\beq
   \sigma^2\abrac{\bQn}
 = \frac{1}{N_n - 1} \bbigg{\bQQn - \bQn^2}
\~.
\label{eqn:sigma_bQn}
\eeq
 For replacing the ``bin'' quantities
 by ``window'' quantities,
 one needs the covariance matrix
 for $\Bar{Q - Q_{\mu}}|_{\mu}$,
 which follows directly from the definition (\ref{eqn:wQ_mu}):
\beqn
 \conti {\rm cov}\abrac{\Bar{Q - Q_{\mu}}|_{\mu}, \Bar{Q - Q_{\nu}}|_{\nu}}
        \non\\
 \=     \frac{1}{N_{\mu} N_{\nu}}
        \sum_{n = n_{\nu-}}^{n_{\mu+}}
        \bbigg{  N_n
                 \abrac{\Bar{Q}|_n - \Bar{Q}|_{\mu}}
                 \abrac{\Bar{Q}|_n - \Bar{Q}|_{\nu}}
               + N_n^2 \sigma^2\abrac{\bQn}}
\~.
\label{eqn:cov_wQ_mu}
\eeqn
 Note that,
 firstly,
 $\mu \leq \nu$ has been assumed here
 and the covariance matrix is, of course, symmetric.
 Secondly,
 the sum is understood to vanish
 if the two windows $\mu$, $\nu$ do not overlap,
 i.e., if $n_{\mu+} < n_{\nu-}$.
 Moreover,
 from Eq.~(\ref{eqn:r_mu}),
 we can get
\beq
   {\rm cov}(r_{\mu}, r_{\nu})
 = \frac{1}{w_{\mu} w_{\nu}} \sum_{n = n_{\nu-}}^{n_{\mu+}} N_n
\~,
\label{eqn:cov_r_mu}
\eeq
 where $\mu \leq \nu$ has again been taken.
 And the mixed covariance matrix can be given by
\beq
   {\rm cov}\abrac{r_{\mu}, \Bar{Q - Q_{\nu}}|_{\nu}}
 = \frac{1}{w_{\mu} N_{\nu}}
   \sum_{n = n_{-}}^{n_{+}} N_n \abrac{\Bar{Q}|_n - \Bar{Q}|_{\nu}}
\~.
\label{eqn:cov_r_mu_wQ_nu}
\eeq
 Note here that
 this sub--matrix is {\em not} symmetric
 under the exchange of $\mu$ and $\nu$.
 In the definition of $n_{-}$ and $n_{+}$
 we therefore have to distinguish two cases:
\beqn
\renewcommand{\arraystretch}{1.6}
\begin{array}{l c l}
   n_{-} = n_{\nu-},~
   n_{+} = n_{\mu+},  &
   ~~~~~~~~~~~~       &
   {\rm if}~\mu \leq \nu\~; \\
   n_{-} = n_{\mu-},~
   n_{+} = n_{\nu+},  &
                      &
   {\rm if}~\mu \geq \nu\~.
\end{array}
\label{eqn:def_n_pm}
\eeqn
 As before,
 the sum in Eq.~(\ref{eqn:cov_r_mu_wQ_nu})
 is understood to vanish if $n_{-} > n_{+}$.

 Furthermore,
 the covariance matrices
 involving the estimators of the logarithmic slopes $k_{\mu}$,
 estimated by Eq.~(\ref{eqn:bQn}) with replacing $n \to \mu$,
 can be given from Eq.~(\ref{eqn:sigma_kn}) as
\beqn
        {\rm cov}\abrac{k_{\mu}, k_{\nu}}
 \=     k_{\mu}^2 k_{\nu}^2
        \cbrac{  1
               - \bfrac{k_{\mu} b_{\mu} / 2}{\sinh (k_{\mu} b_{\mu} / 2)}^2}^{-1}
        \cbrac{  1
               - \bfrac{k_{\nu} b_{\nu} / 2}{\sinh (k_{\nu} b_{\nu} / 2)}^2}^{-1}
        \non\\
        \non\\
 \conti ~~~~~~~~~~~~~~~~ \times 
        {\rm cov}\abrac{\Bar{Q - Q_{\mu}}|_{\mu}, \Bar{Q - Q_{\nu}}|_{\nu}}
\~,
\label{eqn:cov_k_mu}
\eeqn
 and
\beq
   {\rm cov}\abrac{r_{\mu}, k_{\nu}}
 = k_{\nu}^2
   \cbrac{  1
          - \bfrac{k_{\nu} b_{\nu} / 2}{\sinh (k_{\nu} b_{\nu} / 2)}^2}^{-1}
   {\rm cov}\abrac{r_{\mu}, \Bar{Q - Q_{\nu}}|_{\nu}}
\~.
\label{eqn:cov_r_mu_k_nu}
\eeq

\begin{thebibliography}{99}
%
\bibitem{Smith90}
 {P.~F.~Smith and J.~D.~Lewin,
  {\it Phys.~Rep.}~{\bf 187}, 203 (1990).}
%
\bibitem{Lewin96}
 {J.~D.~Lewin and P.~F.~Smith,
  {\it Astropart.~Phys.}~{\bf 6}, 87 (1996).}
%
\bibitem{SUSYDM96}
 {G.~Jungman, M.~Kamionkowski and K.~Griest,
  {\it Phys.~Rep.}~{\bf 267}, 195 (1996).}
%
\bibitem{Bertone05}
 {G.~Bertone, D.~Hooper and J.~Silk,
  {\it Phys.~Rep.}~{\bf 405}, 279 (2005).}
%
%
\bibitem{DMDDf1v}
 {M.~Drees and C.~L.~Shan,
  {\it J.~Cosmol.~Astropart.~Phys.}~{\bf 0706}, 011 (2007).}
%
\bibitem{DMDDmchi-SUSY07}
 {C.~L.~Shan and M.~Drees,
  {\tt arXiv:0710.4296 [hep-ph]} (2007).}
%
\bibitem{DMDDmchi}
 {M.~Drees and C.~L.~Shan,
  {\it J.~Cosmol.~Astropart.~Phys.}~{\bf 0806}, 012 (2008).}
%
\bibitem{Ahmed09b}
 {CDMS Collab., Z.~Ahmed {\it et al.},
  {\tt arXiv:0912.3592 [astro-ph.CO]} (2009).}
%
\bibitem{CRESST-bg}
 {CRESST Collab., R.~F.~Lang {\it et al.},
  {\it Astropart.~Phys.}~{\bf 33}, 60 (2010);}
%
 {CRESST Collab., J.~Schmaler {\it et al.},
  {\it AIP Conf.~Proc.}~{\bf 1185}, 631 (2009).}
%
\bibitem{Aprile09a}
 {E.~Aprile and L. Baudis, for the XENON100 Collab.,
  {\it PoS} {\bf IDM2008}, 018 (2008).}
%
\bibitem{EDELWEISS-bg}
 {EDELWEISS Collab., A.~Broniatowski {\it et al.},
  {\it Phys.~Lett.}~{\bf B 681}, 305 (2009);}
%
 {EDELWEISS Collab., E.~Armengaud {\it et al.},
  {\it Phys.~Lett.}~{\bf B 687}, 294 (2010).}
%
\bibitem{Lang09b}
 {CRESST Collab., R.~F.~Lang {\it et al.},
  {\it Astropart.~Phys.}~{\bf 32}, 318 (2010).}
%
\bibitem{DMDDbg-mchi}
 {Y.~T.~Chou and C.~L.~Shan,
  {\tt arXiv:1003.5277 [hep-ph]} (2010).}
%
\bibitem{PhD}
 {C.~L.~Shan,
  Ph.D.~Thesis,
  {\tt arXiv:0707.0488 [astro-ph]} (2007).}
%
\bibitem{Freese88}
 {K.~Freese, J.~Frieman and A.~Gould,
  {\it Phys.~Rev.}~{\bf D 37}, 3388 (1988).}
%
\bibitem{Engel91}
 {J.~Engel,
  {\it Phys.~Lett.}~{\bf B 264}, 114 (1991).}
%
\bibitem{DMDDmchi-NJP}
 {C.~L.~Shan,
  {\it New J.~Phys.}~{\bf 11}, 105013 (2009).}
%
\bibitem{AMIDAS-web}
 {
  {\tt http://pisrv0.pit.physik.uni-tuebingen.de/darkmatter/amidas/}.}
%
\bibitem{AMIDAS-eprints}
 {C.~L.~Shan,
  {\it AIP Conf.~Proc.}~{\bf 1200}, 1031 (2010);
%
  {\tt arXiv:0910.1971 [astro-ph.IM]} (2009).}
%
\bibitem{Baudis07a}
 {L.~Baudis,
  {\tt arXiv:0711.3788 [astro-ph]} (2007).}
%
\bibitem{Drees08}
 {M.~Drees and G.~Gerbier,
  contribution to
  {\it ``The Review of Particle Physics 2008''},
  C.~Amsler {\it et al.}, {\it Phys.~Lett.}~{\bf B 667}, 1 (2008).}
%
\bibitem{Gascon09}
 {J.~Gascon,
  {\tt arXiv:0906.4232 [astro-ph.HE]} (2009).}
%
\end{thebibliography}
\end{document}